\documentclass[parskip=half, twoside=false]{scrartcl}

\usepackage[left=2cm, right=2cm, top=2cm, bottom=3cm]{geometry}
\usepackage{scrextend}
\usepackage{afterpage}
\usepackage[round,authoryear,sort,comma,numbers,elide]{natbib}
\usepackage{selinput}
\usepackage[english]{babel}
\usepackage[T1]{fontenc}
\usepackage[
	anchorcolor=.,
	citecolor=blue,
	linkcolor=.,
	filecolor=., 
	menucolor=.,
	runcolor=.,
	urlcolor=black,
	colorlinks=true]{hyperref}
\usepackage{amsmath}
\usepackage{amsfonts}
\usepackage{bbm}
\usepackage{graphicx}
\usepackage[dvipsnames]{xcolor}
\usepackage{color}
\usepackage{colortbl}
\usepackage{tikz}
\usetikzlibrary{quotes, angles, positioning, calc} 
\usetikzlibrary{arrows, decorations.markings}
\usetikzlibrary{decorations.pathreplacing}
\usetikzlibrary{arrows.meta}

\usepackage{caption}
\usepackage{subcaption}
\usepackage[
	nogroupskip,
	nonumberlist, 
	acronym]{glossaries}
\newglossary[slg]{symbolslist}{syi}{syg}{List of Symbols}

\usepackage{setspace}
\usepackage{algpseudocode}
\usepackage{placeins}
\usepackage{booktabs}
\usepackage{tabularx}
\usepackage{rotating}
\usepackage{diagbox}
\usepackage{multirow}
\usepackage{enumitem}
\usepackage{mathtools}
\usepackage{float}

\allowdisplaybreaks
\addtokomafont{disposition}{\rmfamily\normalfont}
\addtokomafont{subject}{\Large\normalfont}
\addtokomafont{title}{\LARGE\normalfont}
\addtokomafont{subtitle}{\normalsize\normalfont}
\addtokomafont{author}{\normalfont}
\addtokomafont{date}{\normalsize\normalfont}
\addtokomafont{publishers}{\normalsize\normalfont}

\title{Simulation-based Forecasting for Intraday Power Markets: Modelling Fundamental Drivers for Location, Shape and Scale of the Price Distribution}
\author{Simon Hirsch\textsuperscript{1,2} \and Florian Ziel\textsuperscript{1}}
\date{%
    \textsuperscript{1}House of Energy Markets and Finance, University of Duisburg-Essen\\%
    \textsuperscript{2}Statkraft Trading GmbH\\[4ex]%
    \today\\[4ex]%
    Working Paper 
}

\newacronym{SIDC}{SIDC}{Single IntraDay Coupling}
\newacronym{XBID}{XBID}{Cross Border Intraday}
\newacronym{RES}{RES}{Renewable Energy Sources}
\newacronym{GAM}{GAM}{Generalized Additive Models}
\newacronym{GAMLSS}{GAMLSS}{Generalized Additive Models for Location, Shape and Scale}
\newacronym{MAE}{MAE}{Mean Absolute Error}
\newacronym{RMSE}{RMSE}{Root Mean Squared Error}
\newacronym{CRPS}{CRPS}{Continuously Ranked Probability Score}
\newacronym{ES}{ES}{Energy Score}
\newacronym{DM}{DM}{Diebold-Mariano}
\newacronym{PB}{PB}{Pinball-Score}
\newacronym{WS}{WS}{Winkler-Score}
\newacronym{CR}{CR}{Coverage Ratio}
\newacronym{REMIT}{REMIT}{{Regulation on wholesale Energy Market Integrity and Transparency}}
\newacronym{LASSO}{LASSO}{{Least Absolute Shrinkage and Selection Operator}}
\newacronym{CDF}{CDF}{{Cumulative Distribution Function}}
\newacronym{PDF}{PDF}{{Probability Distribution Function}}
\newacronym{EEX}{EEX}{{European Energy Exchange}}
\newacronym{OTC}{OTC}{Over-The-Counter}
\newacronym{BIC}{BIC}{Bayesian Information Criterion}
\newacronym{MEH}{MEH}{Market Efficiency Hypothesis}
\newacronym{CET}{CET}{Central European Time}
\newacronym{CEST}{CEST}{Central European Summer Time}
\newacronym{TSO}{TSO}{Transmission System Operator}
\newacronym{AR}{AR}{Auto-Regressive}
\newacronym{ARX}{ARX}{Auto-Regressive with eXogenous input}
\newacronym{ARMA}{ARMA}{AutoRegressive Moving Average}
\newacronym{ARIMA}{ARIMA}{AutoRegressive Integrated Moving Average}
\newacronym{VWAP}{VWAP}{Volume-Weighted Average Price}
\newacronym{ACER}{ACER}{Agency for the Cooperation of Energy Regulators}
\newacronym{DA}{DA}{day-ahead}
\newacronym{ID}{ID}{intraday}
\newacronym{SDE}{SDE}{Stochastic Differential Equation}
\newacronym{PI}{PI}{Prediction Interval}
\newacronym{GARCH}{GARCH}{Generalized AutoRegressive Conditional Heteroscedasticity}

\newglossaryentry{symb:Pi}{name=$\pi^{d,s}_t$, description={Probability of at least one trade in the 5 minute interval $t$ on delivery day $d$ and hour $s$},sort=pi, type=symbolslist}
\newglossaryentry{symb:P}{name={$P$}, description={A price}, sort=P, type=symbolslist}
\newglossaryentry{symb:delta_P}{name={$\Delta P$}, description={First difference of $P$}, sort=P_delta, type=symbolslist}
\newglossaryentry{symb:P_d_s}{name={$P^{d,s}_t$}, description={The \gls{VWAP} at the 5-minute interval $t$}, sort=P_d_s, type=symbolslist}
\newglossaryentry{symb:delta_P_d_S}{name={$\Delta P$}, description={First difference of $P^{d,s}_t$}, sort=P_delta, type=symbolslist}

\newglossaryentry{symb:d}{name=$d$, description={Delivery day},	sort=time_d, type=symbolslist}
\newglossaryentry{symb:s}{name=$s$,	description={Delivery hour in $s = \{0, ... ,S\}$ with $S=23$},	sort=time_s, type=symbolslist}
\newglossaryentry{symb:t}{name=$t$,	description={Simulation step in $t = \{1, ... ,T\}$},	sort=time_t, type=symbolslist}

\newglossaryentry{symb:F}{name=$F$,	description={Some continuous distribution $F$},	sort=dist_F, type=symbolslist}
\newglossaryentry{symb:dirac}{name=$D_0$, description={The Dirac Distribution $D_0$}, sort=dist_dirac, type=symbolslist}
\newglossaryentry{symb:mu}{name=$\mu^F$, description={Location parameter of a distribtion $F$. Simplified as $\mu$},sort=dist_mu_F, type=symbolslist}
\newglossaryentry{symb:sigma}{name=$\sigma^F$, description={Scale parameter of a distribution $F$. Simplified as $\sigma$},	sort=dist_sigma, type=symbolslist}
\newglossaryentry{symb:nu}{name=$\nu^F$, description={First shape parameter of a distribtion $F$. Simplified as $\nu$},	sort=dist_nu, type=symbolslist}
\newglossaryentry{symb:tau}{name=$\tau^F$, description={Second shape parameter of a distribtion $F$. Simplified as $\tau$}, sort=dist_tau, type=symbolslist}
\newglossaryentry{symb:theta}{name=$\boldsymbol{\theta}^F$,	description={Parameter vector of a distribtion $F$},sort=dist_theta, type=symbolslist}

\newglossaryentry{symb:MO}{name=$\textit{MO}^{d,s}_{q}$,	description={Slope of the merit order for $d,s$ for the price level at $t$ and $\pm q$ MW.},sort=var_mo, type=symbolslist}
\newglossaryentry{symb:W}{name=$\hat{W}^{d,s}_{t}$,	description={Wind production forecast for $d,s$ issued at $t$.},sort=var_w, type=symbolslist}
\newglossaryentry{symb:L}{name=$\hat{L}^{d,s}_{t}$,	description={Load forecast for $d,s$ issued at $t$.},sort=var_l, type=symbolslist}
\newglossaryentry{symb:S}{name=$\hat{S}^{d,s}_{t}$,	description={Solar production forecast for $d,s$ issued at $t$.},sort=var_s, type=symbolslist}

\newcommand{\review}[1]{\textcolor{Black}{#1}}

\begin{document}
\setcounter{tocdepth}{2}
\maketitle

\begin{abstract}
\textit{Abstract:} During the last years, European intraday power markets have gained importance for balancing forecast errors  due to the rising volumes of intermittent renewable generation. However, compared to day-ahead markets, the drivers for the intraday price process are still sparsely researched. In this paper, we propose a modelling strategy for the location, shape and scale parameters of the return distribution in intraday markets, based on fundamental variables. We consider wind and solar forecasts \review{and their intraday updates}, outages, price information and a novel measure for the shape of the merit-order, derived from spot auction curves as explanatory variables. We validate our modelling by simulating price paths and compare the probabilistic forecasting performance of our model to benchmark models in a forecasting study for the German market. The approach yields significant improvements in the forecasting performance, especially in the tails of the distribution. At the same time, we are able to derive the contribution of the driving variables. \review{We find that, apart from the first lag of the price changes, none of our fundamental variables have explanatory power for the expected value of the intraday returns. This implies weak-form market efficiency as renewable forecast changes and outage information seems to be priced in by the market.} We find that the volatility is driven by the merit-order regime, the time to delivery and the closure of cross-border order books. The tail of the distribution is mainly influenced by past price differences and trading activity. Our approach is directly transferable to other continuous intraday markets in Europe.
\end{abstract}

\deffootnote[1.5em]{0em}{1.5em}{\makebox[1.5em][l]{\thefootnotemark )\ }}


\newcommand{\modelshorthand}[2]{\boldsymbol{\beta}^\text{#1}_{#2} \boldsymbol{X'}^\text{#1}_{#2}}
\newcommand{\iddrei}{$\text{ID}_3$}

\newcommand{\deltaPID}[1]{\ensuremath{\Delta P^{d,s}_{\text{ID},#1}}}
\newcommand{\PDA}{\ensuremath{P^{d,s}_\text{DA}}}
\newcommand{\PID}[1]{\ensuremath{P^{d,s}_{\text{ID},#1}}}
\newcommand{\ALPHA}[1]{\ensuremath{\alpha^{d,s}_#1}}

\section{Introduction} \label{ch:introduction}

Intraday power markets gained significant importance throughout the past few years, most visible in sharply increased trading volumes. In the European power market structure, they provide traders, asset owners and marketers of intermittent \gls{RES} the opportunity to balance forecast errors arising after the day-ahead auction until five minutes before the beginning of the delivery period \citep{hirth2019}. Increasingly, this balancing action is taken over by algorithmic trading strategies, for which reliable short-term price and volatility forecasts are necessary. \review{At the same time, our results shed light on the influence of idiosyncratic features of intraday markets such as \gls{SIDC} on the price process and are thus valuable for policy makers concerned with short-term markets.} 

The recent and still sparse literature on probabilistic forecasting in intraday markets \citep{janke2019, ziel2020b, uniejewski2019} and the markets' driving fundamentals has, so far, focussed on modelling the impact of renewable forecast \citep{ziel2017, kath2019, pape2016, gurtler2018, balardy2018} and forecast errors \citep{ziel2017, kulakov2020, kuppelwieser2021}. A different strand of literature emerged around modelling of the merit-order effect for price changes and price elasticity \citep{kiesel2017, kiesel2020a, kiesel2020b, kulakov2019, balardy2018}. To the best knowledge of the authors, only \cite{ziel2020a} and \cite{baule2021} focus on modelling the volatility in intraday markets. This paper aims to generalize the above research by investigating the fundamental drivers of the location, scale and shape parameters of the intraday price return distribution. We use the \gls{GAMLSS} framework to model the distribution moments in a parametric and explainable fashion. Our contribution is thus two-fold: We are able to significantly improve forecasting performance compared to benchmarks and qualitatively analyse the impact of fundamental drivers for the distribution moments. While this paper focuses on the German intraday market, our methodology is transferable to any continuous intraday market such as France, Great Britain, Spain or Turkey. 

This paper builds on the work of \cite{ziel2020a} to develop a simulation-based probabilistic forecasting model for the path of the five-minute-volume-weighted price \gls{symb:P} between 185 and 30 minutes before delivery. Instead of directly modelling the price $P$ as it is common in day-ahead forecasting and is done in other forecasting studies on the intraday market \citep{uniejewski2019, janke2019}, the first differences $\Delta P$ will be modelled. 
The path of $P$ is thus the cumulative sum of the initial price $P_0$ and all price differences in the forecasting period. Following the suggestions of \cite{ziel2020a}, we assume the first differences \gls{symb:delta_P} follow a mixture distribution of the Dirac distribution $\delta_0$ \review{with an atom at 0} and a continuous distribution \gls{symb:F}. In \cite{ziel2020a}, the latter is assumed to be $t$-distributed without any detailed justification, except the observation that $\Delta P$ tends to be heavy tailed. This manuscript extends the approach in four dimensions:

\begin{enumerate}
\item We study in more detail the distribution assumption for \gls{symb:F}. We investigate more distributions, including those with potential skewness.
The distribution for $F$ is selected from the skew-$t$ and the Johnson's $S_U$. These distributions have been used successfully to model asset returns or and have been applied to forecasting in energy markets \citep{serinaldi2011, bunn2018, bunn2018a}.
\item One natural starting point to improve the models is adding intra-daily updated forecasts for wind and solar generation to reflect the information set available to market participants better. High forecast updates indicate that market participants with \gls{RES} assets need to solve larger positions in the intraday market. Capturing this effect aims at improving the modelling of the location and volatility of $\Delta P$. The findings will be discussed in the light of the market efficiency hypothesis indicated by \cite{ziel2020b, kuppelwieser2021}. Additionally, the variance of individual forecast updates is used to model the volatility of $\Delta P$. This is motivated by the assumption that uncertainty regarding the weather situations is transmitted to the price formation.
\item A measure for the intraday price elasticity is derived from the day-ahead auction curves, which are used as proxy for the intraday merit-order and included in the modelling of the volatility \citep{kulakov2019, kulakov2020}. This is motivated by \cite{kiesel2020a, kiesel2020b}, who show that the price impact of forecast errors depends on whether the market is in a flat or steep merit-order 'regime' and the work of \cite{balardy2018}, who shows that the intraday bid-ask spread can be explained by the price elasticity derived from the day-ahead auction curves. Additionally, the predictive power of the day-ahead level of outages and the change in planned and unplanned outages between day-ahead and intraday is tested.
\item As in \cite{ziel2020a}, we utilize the GAMLSS framework for parameter estimation. However, we allow for a more flexible parameter training approach utilizing an automatic variable selection for all distribution moments by utilising the adaptive \gls{LASSO} estimation technique for the \gls{GAMLSS} model as well. This procedure was similarly used in \cite{ziel2021m5} in the context of the M5 forecasting competition. 
\end{enumerate}

The extended models are tested in a forecasting study and compared to the benchmark models. We evaluate the probabilistic forecasting performance by utilizing established probabilistic scoring rules and calibration measures. Statistical significance is evaluated by the widely used \gls{DM}-test \citep{ziel2020a, ziel2019b, weron2018, diebold2002}. \review{In our forecasting study, the GAMLSS-based model assuming Johnson's $S_U$ significantly outperforms all proposed benchmark models as well as the GAMLSS-based model assuming the popular skew-$t$ distribution. The GAMLSS-based model assuming the skew-$t$ distribution exhibits stark sensitivity towards outliers.} Qualitatively, our results indicate that price changes $\Delta P$ in the intraday market are influenced by the first lag, while other explanatory variables have little predictive power for the expected value of $\Delta P$. This result supports the notion of weak-form efficient markets already indicated by \cite{ziel2020b, kuppelwieser2021}. We find evidence for a merit-order effect in the volatility and kurtosis of the distribution $\Delta P$. A steeper merit-order implies higher volatility and heavier tails. Additionally, the volatility rises with decreasing time to delivery and with the gate closure of XBID/SIDC, while kurtosis is more driven by trading-related variables such as lagged absolute price differences. We find that none of the included explanatory variables has predictive power for the skewness of the distribution.

The presented models and methodology are also of interest to practitioners in intraday markets. Path-based forecasts allow to price short-term asset optionality using Asian option valuation. Additionally, the explicit modelling of the volatility provides a starting point to introduce time-varying volatility to mathematical finance models for market making and position solving \citep{luckner2017, glas2020, aid2016, kath2020}. 

The remainder of this paper is structured as follows: Section \ref{ch:market_structure} gives a short introduction to the structure of the German short-term power markets. Section \ref{ch:data} presents the data preparation of the intraday trade data, the forecast and outage data sets and the transformation of the day-ahead auction curves and related assumptions. Also, some exploratory data analysis is carried out in this Section. Section \ref{ch:models} introduces the used models. The forecasting study design and scoring rules are discussed in Section \ref{ch:forecasting_study_evaluation}. Finally, Sections \ref{ch:results} and \ref{ch:discussion_conclusion} present the results and conclude this paper.

\FloatBarrier 
 
\section{Structure of the German Power Market} \label{ch:market_structure}

\begin{figure} 
\caption[Daily procedure in the German short-term power markets.]{Daily procedure in the German short-term power markets \citep[based on][]{epex2018, ziel2020a, nordpool2018}.} \label{fig:market_structure}
\begin{center}
\begin{tikzpicture}
\begin{footnotesize}

\coordinate (A) at (0,0);
\coordinate (B) at (14, 0); 

\coordinate[label={[above, align=center, font=\scriptsize] Day-ahead \\ Spot Auction}] (11) at (0.5,2);
\coordinate[label={[below, align=center] $d-1$, \\ 12.00 }] (12) at (0.5,-0.5);

\coordinate[label={[above, align=center, font=\scriptsize] First Spot Auction \\ Results $P_\text{DA}^{d,s}$}] (31) at (2, 1.1); 
\coordinate[label={[below, align=center] $d-1$, \\ 12.45 }] (32) at (2,-0.5);

\coordinate[label={[above, align=center, font=\scriptsize] Hourly Intraday \\ opens}] (21) at (3.5,2);
\coordinate[label={[below, align=center] $d-1$, \\ 15.00 }] (22) at (3.5,-0.5);



\coordinate[label={[above, align=center, font=\scriptsize] Start Simulation \\ with $P_{\text{ID},0}^{d,s}$}] (41) at (8.5,1.1); 
\coordinate[label={[below, align=center] $d$, \\ $s-185$ \\ min}] (42) at (8.5,-0.5);

\coordinate[label={[above, align=center, font=\scriptsize] SIDC \\ opens}] (51) at (6,2);
\coordinate[label={[below, align=center] $d-1$, \\ 18.00 }] (52) at (6,-0.5);

\coordinate[label={[above, align=center, font=\scriptsize] SIDC \\ closes}] (61) at (10.5,2);
\coordinate[label={[below, align=center] $d$, \\ $s-60$ \\ min}] (62) at (10.5,-0.5);

\coordinate[label={[above, align=center, font=\scriptsize] Market \\ closes}] (71) at (11.5,1.25);
\coordinate[label={[below, align=center] $d$, \\ $s-30$ \\ min}] (72) at (11.5,-0.5);

\coordinate[label={[above, align=center, font=\scriptsize] Control zones \\ close}] (81) at (12.5,0.5);
\coordinate[label={[below, align=center] $d$, \\ $s-5$ \\ min}] (82) at (12.5,-0.5);

\coordinate[label={[above, align=center, font=\scriptsize] Delivery $b(d,s)$}] (91) at (13.5,2);
\coordinate[label={[below, align=center] $d$, $s$}] (92) at (13.5,-0.5);

\draw [->,line width=1.25pt] (A) -- (B);
\draw [-] (11) -- (12);
\draw [-] (21) -- (22);
\draw [-] (31) -- (32);
\draw [-] (41) -- (42);
\draw [-] (51) -- (52);
\draw [-] (61) -- (62);
\draw [-] (71) -- (72);
\draw [-] (81) -- (82);
\draw [-,line width=1.25pt] (91) -- (92);

\end{footnotesize}
\end{tikzpicture}
\end{center}
\end{figure}
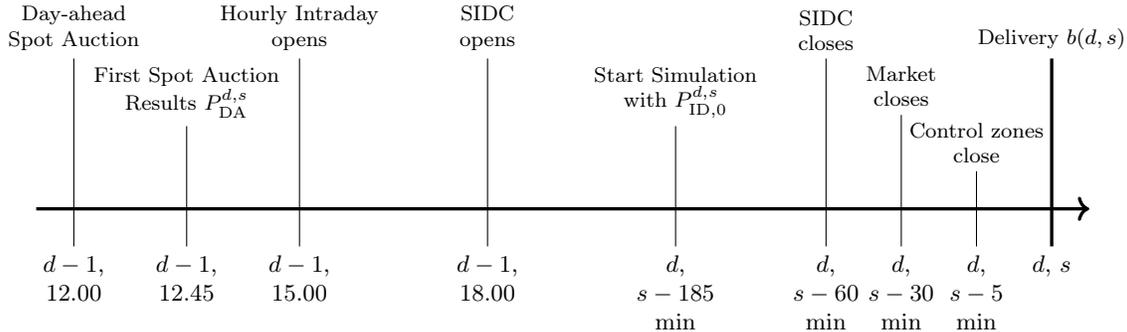

This section briefly introduces the relevant structure of the German power market. As we work with data from the day-ahead auction and the intraday market, the description focuses on these markets. Generally, denote the delivery day as \gls{symb:d} and the delivery hour as \gls{symb:s} for $s = 0, ..., S$ and $S=23$. Times are usually expressed in local time unless otherwise noted.\footnote{\gls{CET} respectively \gls{CEST} for Germany.} Electric power markets generally follow the structure of a forward market, where different delivery periods in the future can be traded almost up to the start of actual physical delivery. Figure \ref{fig:market_structure} shows the time line of the German short-term markets, for more details see \cite{viehmann2017}. \review{Let us generally note here that we place indices referring to the delivery periods $d,s$ as superscript, while placing indices relating to the time where the price is determined as subscript. The same holds for other variables as e.g. production forecasts.}

The spot market is organized as a pay-as-cleared auction. The order book closes on $d-1$ at 12:00 and first auction results are published around 12:42 on $d-1$. Official results shall be published at latest at 14:00 on $d-1$. From the bids submitted by the market participants, EPEX Spot calculates aggregated supply and demand curves for each delivery period. The intersection between supply and demand curves is the market clearing price $\PDA$. Additionally to normal bids, market participants can submit special bids such as block bids spanning more than one delivery period and linked bids, where execution is linked to neighbouring bids. The day-ahead spot price also serves as reference price for cascading financial futures. \cite{epex2020c} publishes aggregate curves together with the official market results. \review{The lower price level is set to -500 EUR/MWh, the upper level is set to 3000 EUR/MWh}. 

\review{The intraday market is structured as continuous pay-as-bid auction similar to financial markets. However, contrary to equity or currency markets, the individual trading sessions of the intraday electricity markets are not part of a larger process, as the intraday trading session ends with the physical delivery of power. Hence, trading sessions for the same delivery period on different delivery days might be driven by fundamentally different circumstances and need to be viewed separately.} Trading starts at 15:00, 15:30, 16:00  on $d-1$ for hourly, half-hourly and quarter-hourly products with delivery on day $d$. At 18:00 on $d-1$, cross-border trading within the \gls{SIDC} system, formerly known as \gls{XBID}, starts in Germany, Denmark, Netherlands, Norway and Poland.\footnote{For the sake of consistency, both the \gls{XBID} and \gls{SIDC} are referred to as \gls{SIDC} throughout this paper.} At 22:00 on $d-1$, the remaining countries of the core market area follow. Here, the intraday order books of all participating countries are shared and orders can be matched internationally as long as sufficient transmission capacity is available. For each product, the cross-border shared order books close one hour before delivery. \gls{SIDC} went live on June 18, 2018 \citep{nordpool2018}. 30 minutes before delivery, the Germany-wide order book closes and trading resumes in local (control zone) products up to five minutes before delivery. \review{Note that all open delivery periods are traded in parallel.} The market price limits  are at  $\pm9999$ EUR/MWh. The smallest possible price tick changed multiple times throughout the last few years and is currently set to 0.01 EUR/MWh. The smallest possible volume tick is 0.1 MW \citep{epex2020b, epex2018, viehmann2017}.

\FloatBarrier
\section{Data and Exploratory Analysis} \label{ch:data}
\subsection{Intraday Trade Data} \label{sec:data_trades}

\review{On the intraday market, trading happens continuously. Hence, the transactions are irregularly spaced and need to be aggregated. The following paragraphs and Figure \ref{fig:data_prep_process} give a brief overview of the aggregation. A detailed description can be found in Appendix \ref{app:trade_aggregation}.} Trade data is obtained from \cite{epex2020a}. The data consists of all hourly trades on the continuous intraday market between January 1st, 2016 and July 31st, 2020.

\begin{figure}
\begin{center}
\includegraphics[width=\textwidth]{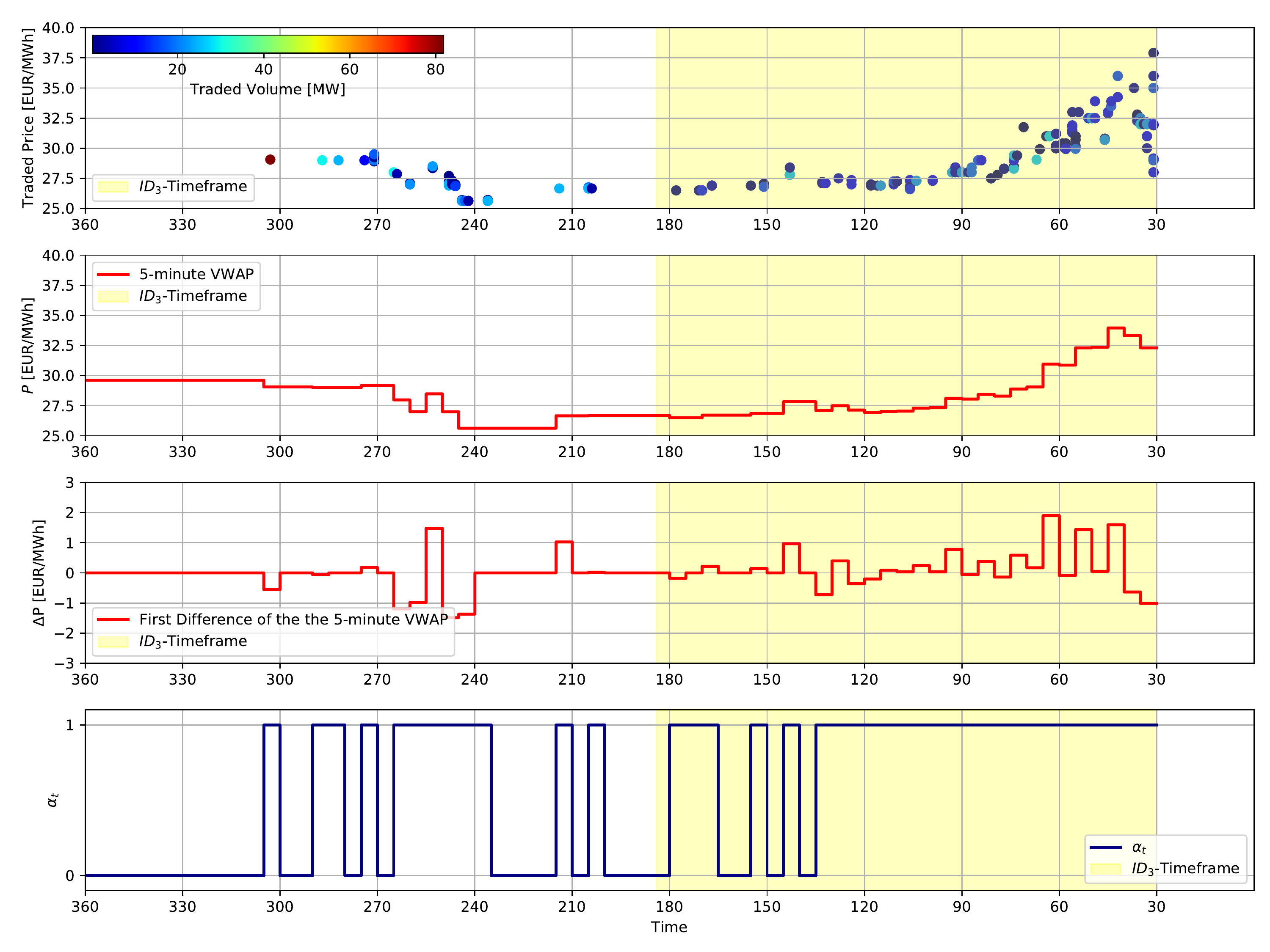}
\caption[Overview of the data preparation.]{Overview of the data preparation. The first figure shows the raw trade data. The color refers to the traded volume. Below, the 5-minute \gls{VWAP} $P$ and their first differences $\deltaPID{t}$ are shown. Lastly, $\ALPHA{t}$ is shown. \review{The plots show the data for $d$ = January 1st, 2016 for delivery period $s=12$ in the last 6 hours to delivery.}} \label{fig:data_prep_process}
\end{center}
\end{figure}

For each delivery period $d, s$ we aggregate all trades on an equidistant 5-minute grid by taking the volume-weighted average price within each bucket, denoted by $\PID{t}$, where $t$ denotes the 5-minute interval (see panel 2 in Figure \ref{fig:data_prep_process}). We then take first differences $\deltaPID{t} = \PID{t} - \PID{t-1}$ (see panel 3). Lastly, we define a boolean variable $\ALPHA{t}$, which takes the value 1 if there has been at least one trade within the 5-minute interval (see panel 4). As the trading sessions in the intraday market are of varying length for the different delivery periods and our simulation concerns the last 185 minutes of trading for each product, we define $t$ relative to the start of the physical delivery. $t=1$ denotes the first 5-minute interval in the simulation window, thus 185 to 180 minutes before the start of physical delivery and $t = 31 = T$ denotes the last 5-minute interval in the simulation window, 35 to 30 minutes before the start of physical delivery. Similar aggregation methods have been used by \cite{ziel2020a, ziel2020b} and \cite{serafin2022}.

\begin{figure}[H]
\includegraphics[width=\textwidth]{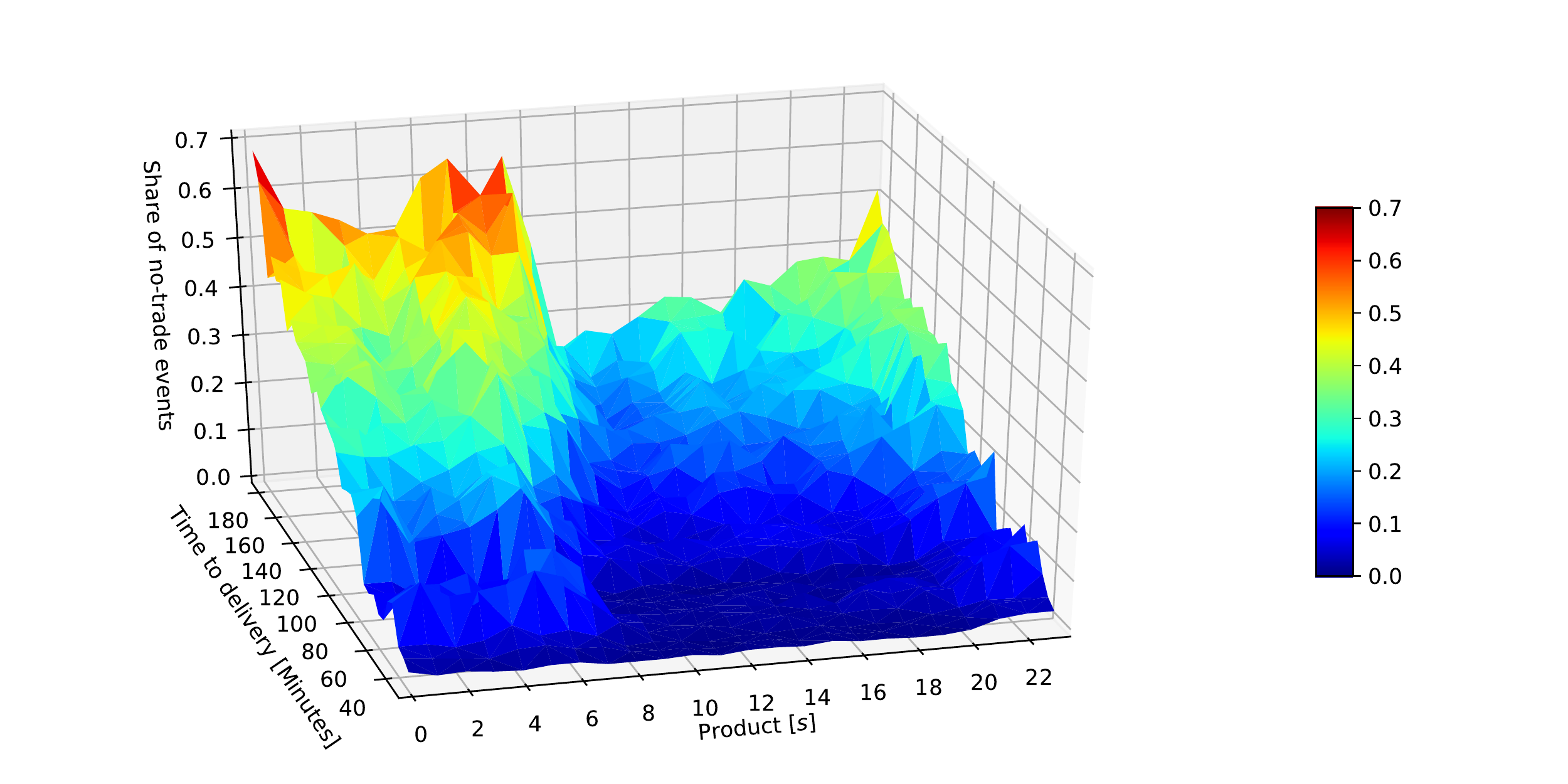}
\caption[Share of no-trade events over time to delivery and delivery hour.]{Share of no-trade events over time to delivery and delivery hour for the initial one-year  training set from January 1st to December 31st, 2016. Colour corresponds to the $z$-axis. Low values correspond to few no-trade events, high values correspond to many no-trade events.} \label{fig:alpha_timetodelivery}
\end{figure}

\review{Figure \ref{fig:alpha_timetodelivery} shows the relationship between the share of no-trade events, i.e. 5-minute intervals where $\alpha_t^{d,s} = 0$, relative to the time to delivery on the initial training set. With decreasing time to delivery, the probability of no-trade events decreases in a non-linear fashion. For periods close to 30 minutes to delivery, the share of no-trade events in the initial training data set is close to 0, while further away from delivery, there are more periods without trades. For products with delivery in the peak hours, there are less no-trade events at the beginning of the $\text{ID}_3$ period already.} Additionally, Table \ref{tab:summary_statistics_alpha_delta_p} presents summary statistics for $\alpha_t$ and $\deltaPID{t}$ for all 5-minute intervals with at least one trade,  grouped by year. The share of 5-minute intervals where $\alpha_t = 1$, i.e. at least one trade happens happens, increases throughout the years. It is almost 1 from  2018 onwards, implying that there are barely any periods without trades. Accordingly, the number of observations for $\alpha_t$ and $\deltaPID{t} \mid \alpha_t = 1$ converge. \review{We can thus identify two levels of time-varying behaviour of $\alpha^{d,s}_t$, first across the multiple years of the data set, but also second within each trading session. While we explicitly model the latter, the first will be coped with due to the set-up of a rolling window forecasting study.}

\review{The mean and median values of $\deltaPID{t}$ are close to 0 across all years in the dataset. However the standard deviation is rather high and the extreme minima and maxima already hint at a leptokurtic distribution of $\deltaPID{t}$.} The minima and maxima increase throughout the data set, \review{while the 5\% respectively 95\% and the 10\% respectively 90\% quantiles are roughly constant.} Especially for 2020, the minima and maxima of more than 2000 respectively less than -2000 EUR/MWh are noteworthy. Driven by these larger outliers in 2020, the standard deviation of $\deltaPID{t}$ rises fourfold between 2019 and  2020, while staying roughly constant before. \review{The more robust dispersion measures median absolute deviation (MAD) and the interquartile range (IQR) support this notion.} 

\begin{table}
\caption[Summary statistics for $\ALPHA{t}$ and $\deltaPID{t}$.]{Summary statistics for $\ALPHA{t}$ and $\deltaPID{t}$ for 185 to 30 minutes before delivery. $Q_\tau$ denotes the empirical $\tau \cdot 100\%$-quantile. Note that all $\deltaPID{t} \;|\; \ALPHA{t} = 0$ are 0 by definition.} \label{tab:summary_statistics_alpha_delta_p}
\begin{center}
\begin{footnotesize}\label{tab:tab_summary_statistics_delta_p}
\begin{tabular}{llrrrrr}
\toprule
                              &     &       2016 &       2017 &       2018 &       2019 &       2020 \\
\midrule
$\alpha_t$ 					  	& Count ($n$) 		&  257424 &  252960 &  268584 &  267096 &  158472 \\
                              	& Mean ($\mu$) &       0.82 &       0.87 &       0.95 &       0.98 &       0.98 \\
                              	& Std. ($\sigma$) &       0.39 &       0.34 &       0.23 &       0.14 &       0.15 \\ \midrule
$\Delta P_t \mid \alpha_t = 1$ 	& Count ($n$) &  209995 &  219258 &  253931 &  261721 &  154597 \\
                              	& Mean ($\mu$) &       0.02 &       0.02 &      -0.00 &      -0.00 &      -0.01 \\
                              	& Std. ($\sigma$) &       1.57 &       2.06 &       1.94 &       2.15 &       8.13 \\
		                        & MAD   			& 0.94 &       1.11 &       1.03 &       0.91 &       1.10 \\
								& IQR   			& 1.17 &       1.33 &       1.23 &       1.01 &       1.07 \\
								& Skewness     &       1.07 &       1.94 &       8.02 &      12.25 &       0.65 \\
								& Kurtosis &     131.50 &     322.14 &     902.99 &    2544.57 &   64968.44 \\
                              	& Min 				& -76.38 &    -127.70 &    -108.22 &    -214.16 &   -2161.80 \\
                              	& $Q_{0.10}$ &      -1.40 &      -1.61 &      -1.52 &      -1.29 &      -1.45 \\
                               & $Q_{0.25}$ &      -0.57 &      -0.65 &      -0.62 &      -0.51 &      -0.54 \\
                               & $Q_{0.50}$ &       0.00 &       0.00 &       0.00 &      -0.01 &       0.00 \\
                               & $Q_{0.75}$ &       0.60 &       0.68 &       0.61 &       0.50 &       0.53 \\
                               & $Q_{0.90}$ &       1.43 &       1.64 &       1.49 &       1.26 &       1.40 \\
                              & Max &      82.16 &     119.88 &     183.94 &     226.26 &    2164.23 \\
\bottomrule
\end{tabular}

\end{footnotesize}
\end{center}
\end{table}

\begin{figure}
\begin{subfigure}[c]{0.35\textwidth}
\includegraphics[width=\textwidth]{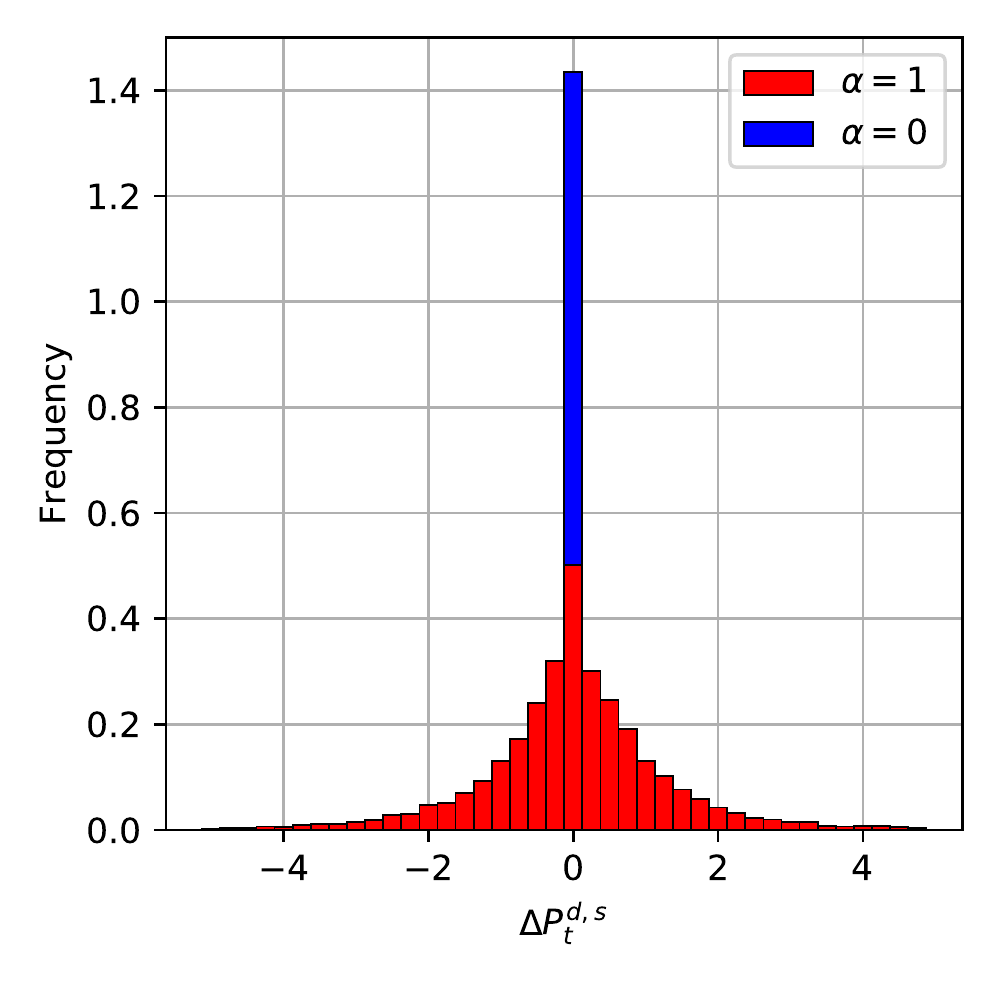}
\includegraphics[width=\textwidth]{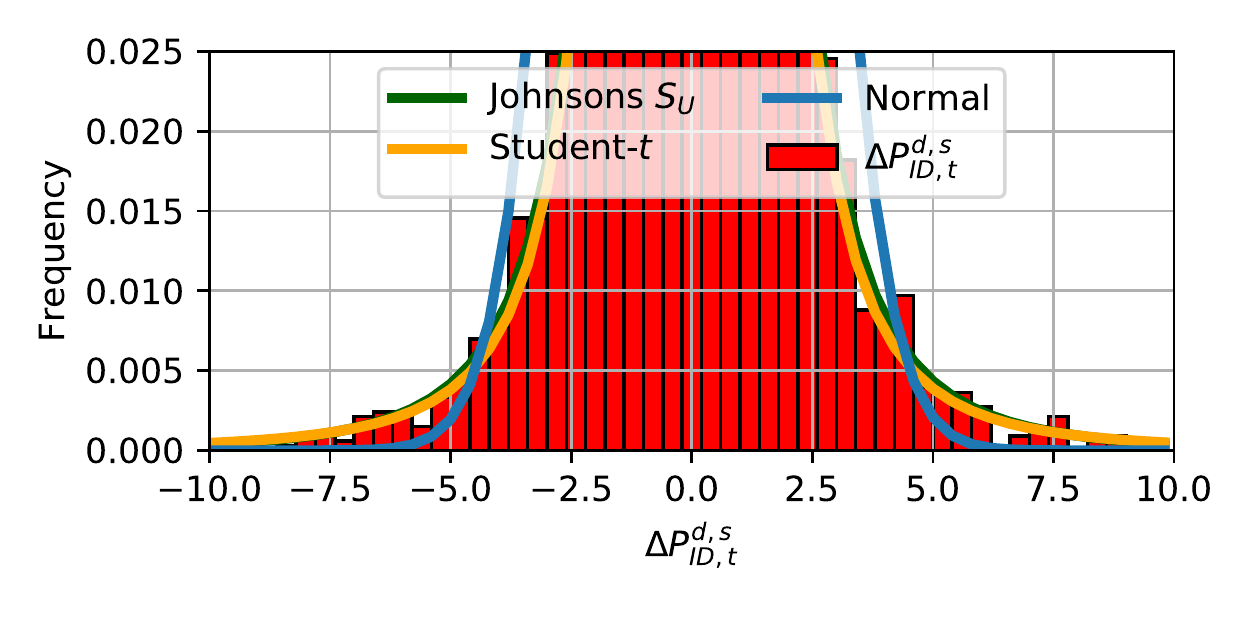}
\subcaption{Hour $s=4$.} 
\end{subfigure}%
\begin{subfigure}[c]{0.35\textwidth}
\includegraphics[width=\textwidth]{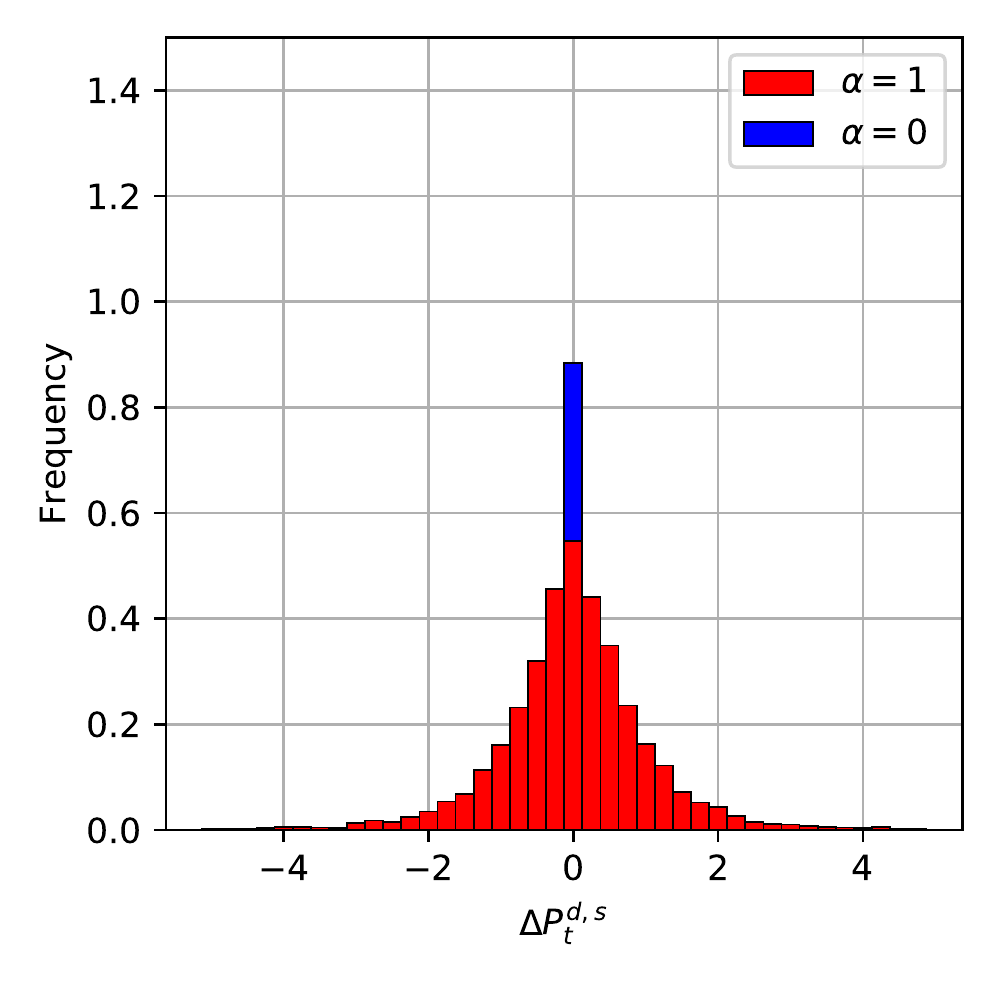}
\includegraphics[width=\textwidth]{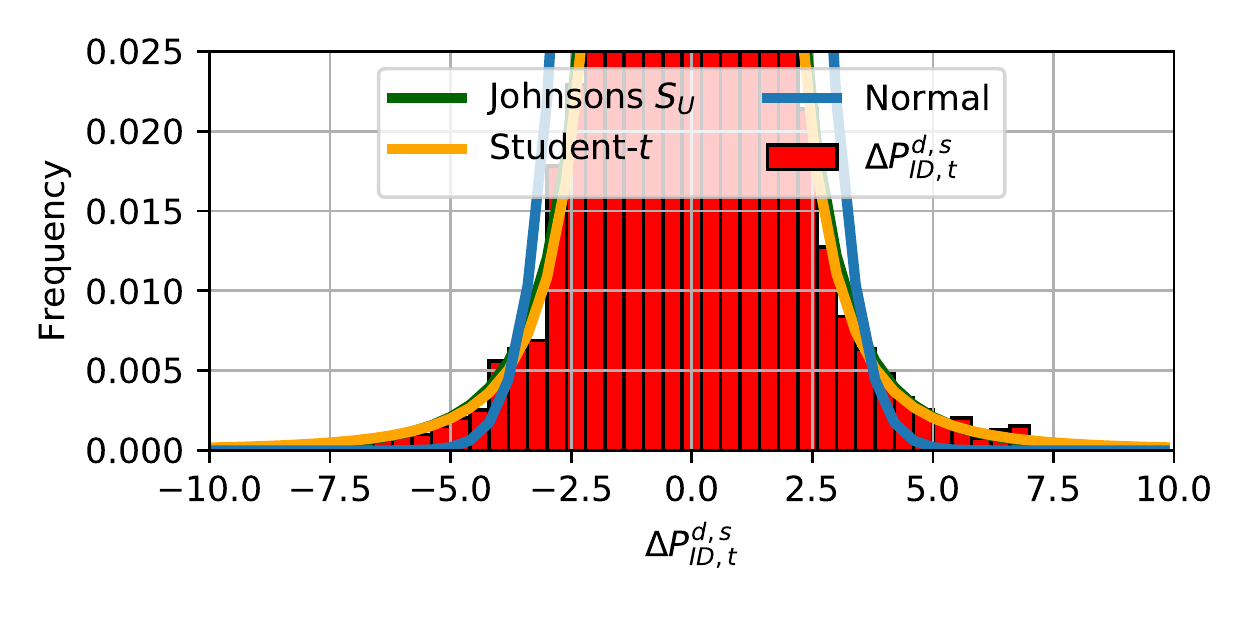}
\subcaption{Hour $s=12$.} 
\end{subfigure}%
\begin{subfigure}[c]{0.35\textwidth}
\includegraphics[width=\textwidth]{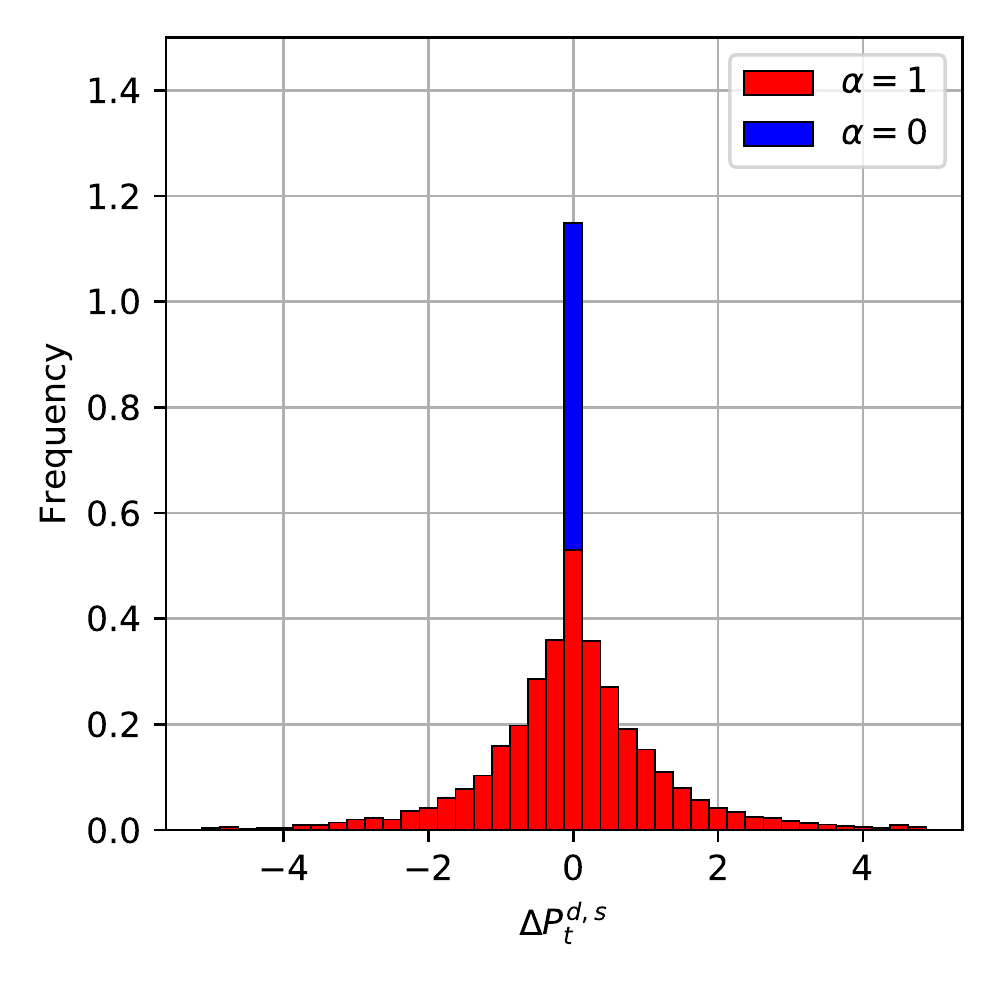}
\includegraphics[width=\textwidth]{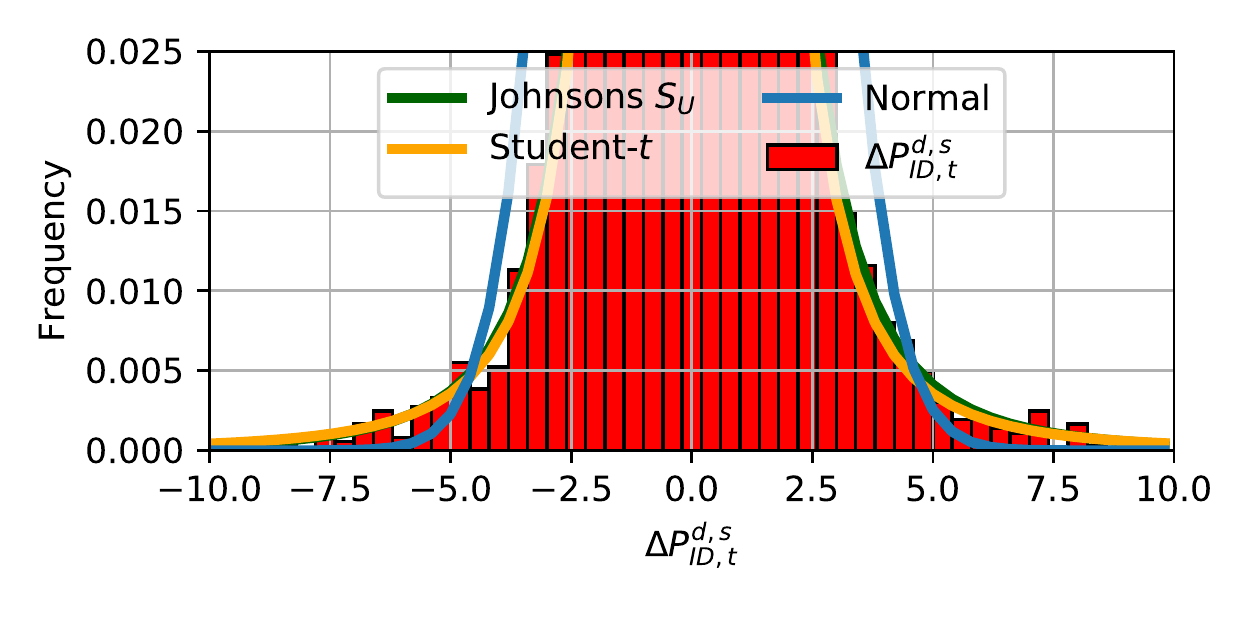}
\subcaption{Hour $s=20$.} 
\end{subfigure}%
\caption[Histograms of $\deltaPID{t}$.]{Histograms of $\deltaPID{t}$ for $s = 4, 12, 20$ in the initial one-year training set.} \label{fig:histogram_delta_P}
\end{figure}

\begin{figure}
\begin{subfigure}[c]{\textwidth}
\includegraphics[width=\textwidth]{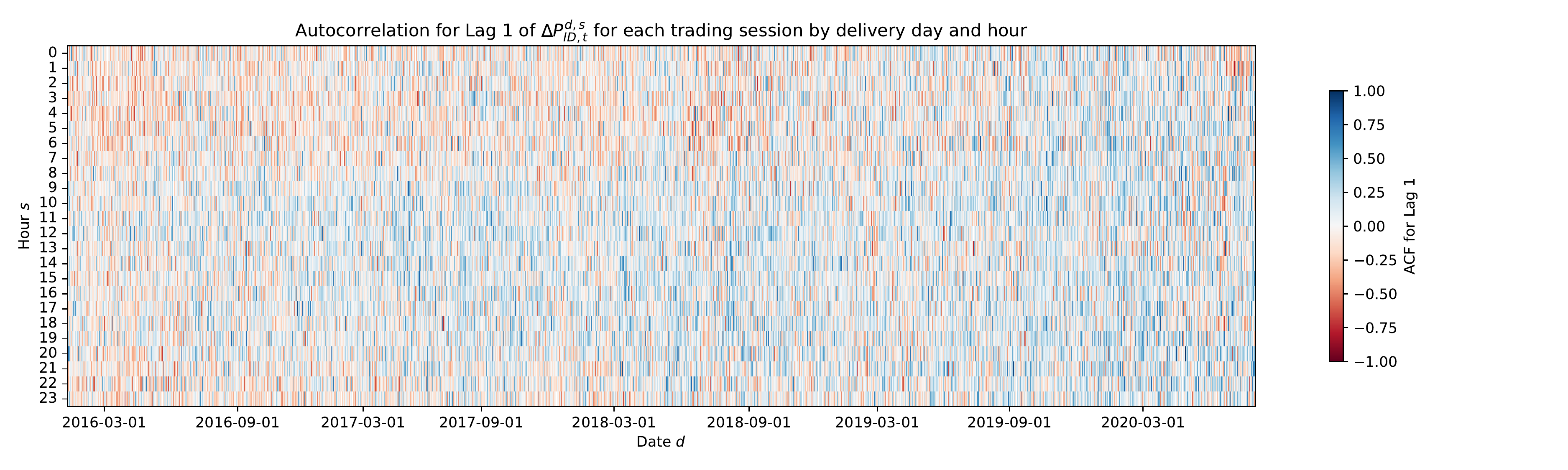}
\subcaption{Lag 1 autocorrelation of $\deltaPID{t}$.} 
\end{subfigure}
\begin{subfigure}[c]{\textwidth}
\includegraphics[width=\textwidth]{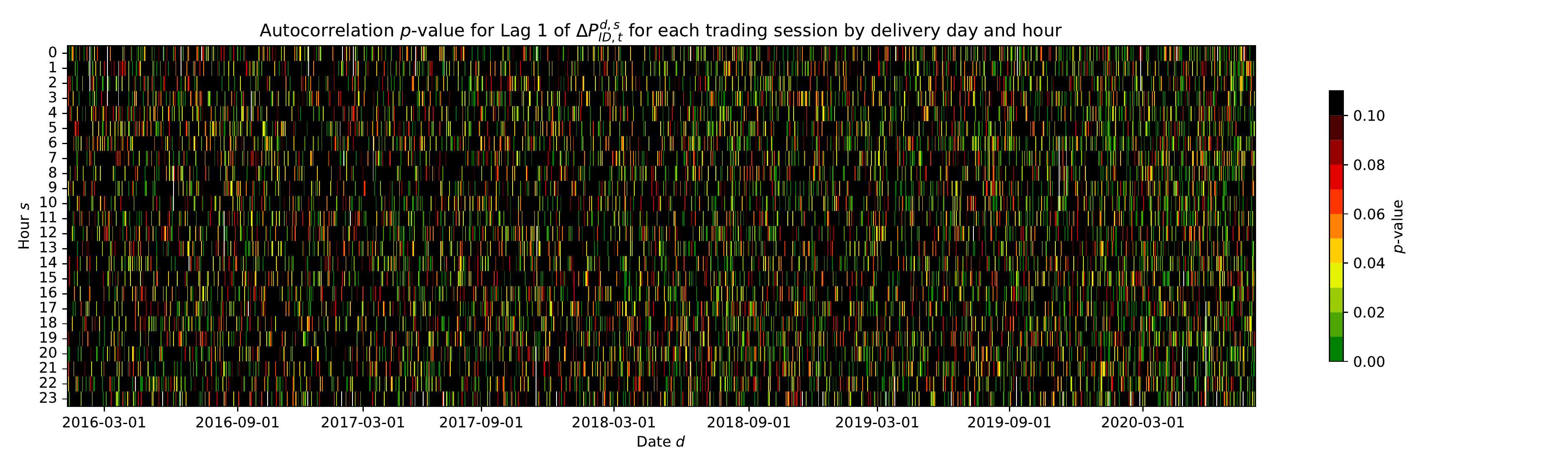}
\subcaption{$p$-values for lag 1 autocorrelation of $\deltaPID{t}$.} 
\end{subfigure}
\caption{Autocorrelation of $\deltaPID{t}$ per trading window for lags 1 and according $p$-values. The first heat maps show the size of the correlation coefficient by delivery day $d$ and hour $s$, second shows according $p$-values. Lags 2 and 3 can be found in Appendix \ref{app:plots}.} \label{fig:autocorrelation}
\end{figure}

Figure \ref{fig:histogram_delta_P} plots histograms for $\deltaPID{t}$ in the initial training set in the hours $s \in \{4, 12, 20\}$ exemplary. These delivery hours represent the typical night, noon and afternoon peak load hours. The first plot focuses on the general shape of the distribution as well as the relation between intervals with and without trades. \review{The center bar shows the relative weight of 5-minute intervals without trades (i.e. $\ALPHA{t} = 0$) and 5-minute intervals with at least one trade (i.e. $\ALPHA{t} = 1$), but small or no price changes. As visible already in Figure \ref{fig:alpha_timetodelivery}, the share of 5-minute intervals with $\ALPHA{t} = 0$ decreases strongly for delivery hours after 8.} In the second Figure, the tails of the distribution are shown together with fitted normal, student-$t$, and Johnson's $S_U$ distributions. 
\review{Additionally, Figure \ref{fig:autocorrelation} (a) plots the pearson autocorrelation coefficients $r$ for $\deltaPID{t}$ for each trading session for the first lag. Colour intensity corresponds to the coefficient size. For the first lag, slight positive autocorrelation is present for the morning hours, while some negative correlation is visible for noon and evening hours. Figure \ref{fig:autocorrelation} (b) shows the according $p$-values for the test statistic $r \cdot \sqrt{n-2} / \sqrt{1 - r^2}$ where $n$ is the number of 5-minute intervals in the trading session. We find that for around one-fifth of all trading sessions, the lag 1 autocorrelation coefficient is significant at the 10\%{}-level and for only 15\%{} of all trading sesions, the lag 1 autocorrelation is significant at the 5\%{}-level. For lags 2 and 3, we find even less significant autocorrelation (see Figures \ref{fig:autocorrelation_lag2} and \ref{fig:autocorrelation_lag3} in Appendix \ref{app:plots}).}

\begin{figure}
\includegraphics[width=\textwidth]{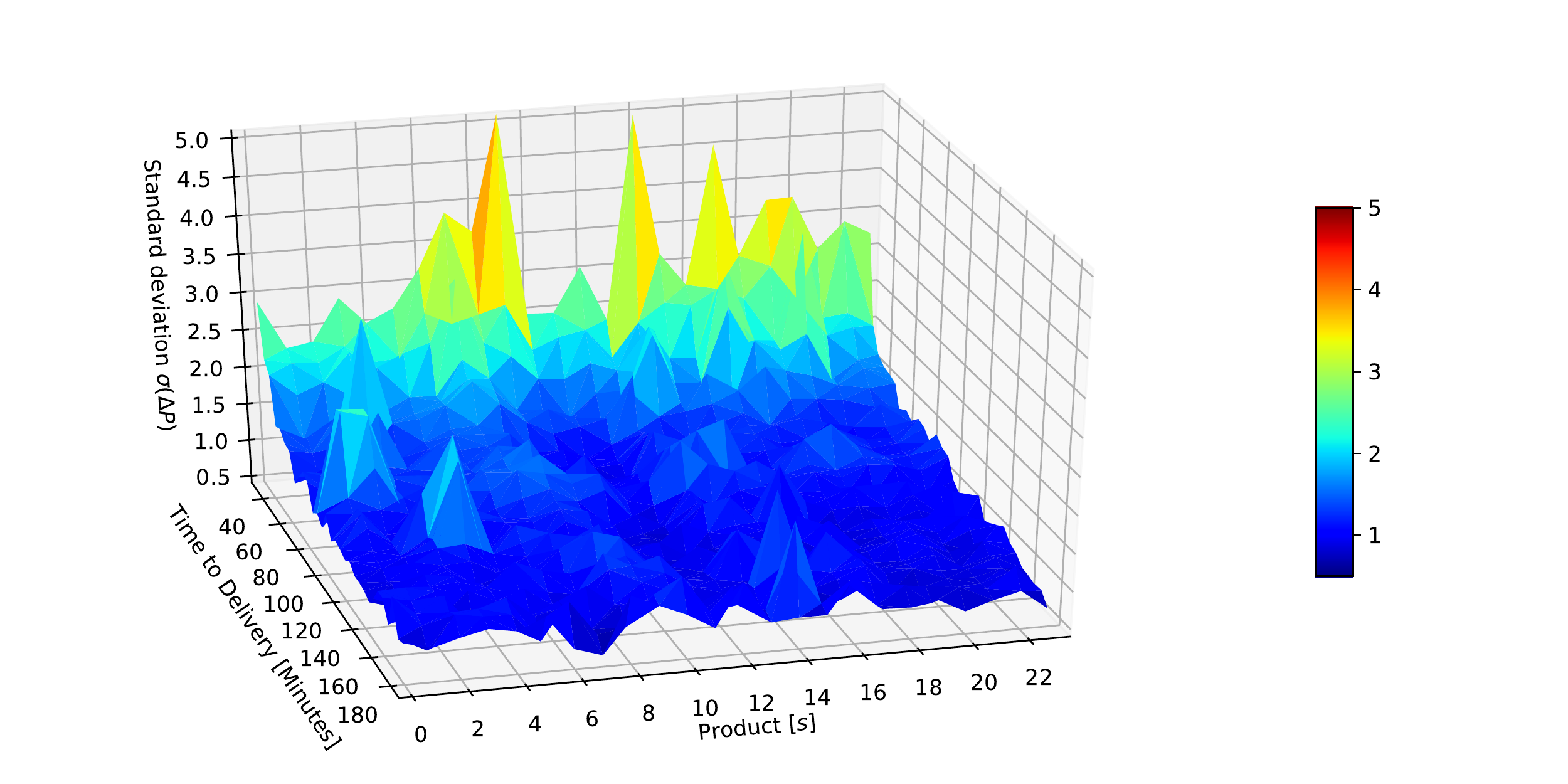}
\caption[Volatility development over time to delivery and delivery hour.]{Volatility development over time to delivery and delivery hour for the initial training one-year set from January 1st to December 31st, 2016. Colour corresponds to the $z$-axis. Note the inverted $y$-axis (Time to delivery) compared to Figure \ref{fig:alpha_timetodelivery} to ease visualisation.}  \label{fig:volatility_timetodelivery}
\end{figure}
\FloatBarrier

The relationship between the realized variance and the time to delivery in the initial training data is shown in Figure \ref{fig:volatility_timetodelivery}. The volatility increases slightly until 60 minutes before the start of the delivery and rises sharply between 60 and 30 minutes before the start the delivery. \review{Here, we note three levels of time-varying behaviour: first, across the full data set, volatility is increasing. Second, within each day, the volatility moves with the peak/off-peak hours. Third, within each trading session, volatility increases towards the end of the trading session.}

\review{To analyse the stationarity properties of the differenced and un-differenced price series, we apply the augmented Dickey-Fuller test to each simulation window individually and report aggregate results in Table \ref{tab:stationarity_tests}. For the majority of the trading windows, we find stationarity of the price differences and unit-root behaviour in the prices. We note though, that due to the heteroskedasticity present in the individual trading windows, the underlying assumptions of the ADF-test might be violated. Together with the results of \cite{lohndorf2022value}, who aggregate trades in the intraday market on a 1-hour grid and report similar results for the ADF-test at the 10\%{}-level, we conclude that the price changes in the intraday market are stationary.}

\begin{table}[H]
\caption{Aggregate results of the augmented Dickey-Fuller unit-root tests. Tests are applied to each simulation window individually. The table reports the share of simulation windows where the test result implies stationarity, i.e. rejection of the $H_0$ of a unit root.} \label{tab:stationarity_tests}
\begin{center}
\begin{tabular}{lllrrrrr}
\toprule
          							& Year 				&   2016 &   2017 &   2018 &   2019 &   2020 \\
Variable 							& Significance 		&        &        &        &        &        \\
\midrule
$\Delta P_{\text{ID},t}^{d,s}$ 		& $\alpha = 0.01$ &  0.675 &  0.658 &  0.633 &  0.596 &  0.555 \\
    	      						& $\alpha = 0.05$ &  0.750 &  0.744 &  0.717 &  0.684 &  0.641 \\
$P_{\text{ID},t}^{d,s}$ 			& $\alpha = 0.01$ &  0.059 &  0.055 &  0.055 &  0.059 &  0.088 \\
    	     						& $\alpha = 0.05$ &  0.128 &  0.125 &  0.122 &  0.117 &  0.157 \\
\bottomrule
\end{tabular}
\end{center}
\end{table}

\subsection{Renewables Forecasts and Outages}\label{sec:data_forecasts}

\newcommand{\WFC}[1]{\ensuremath{\hat{W}_{#1}^{d, s}}} 
\newcommand{\SFC}[1]{\ensuremath{\hat{S}_{#1}^{d, s}}} 
\newcommand{\DFC}{\ensuremath{\hat{D}_{\text{DA}}^{d, s}}}
\newcommand{\deltaWFC}[2]{\ensuremath{\Delta\hat{W}_{#1, #2}^{d, s}}} 
\newcommand{\deltaSFC}[2]{\ensuremath{\Delta\hat{S}_{#1, #2}^{d, s}}} 
\newcommand{\deltaWFCsign}[3]{\ensuremath{\Delta\hat{W}_{#1, #2}^{d, s, #3}}} 
\newcommand{\deltaSFCsign}[3]{\ensuremath{\Delta\hat{S}_{#1, #2}^{d, s, #3}}} 
\newcommand{\sigmaWFC}{\ensuremath{\sigma^{d,s}_{\text{DA},\text{ID}}(\Delta \hat{W})}}
\newcommand{\sigmaSFC}{\ensuremath{\sigma^{d,s}_{\text{DA},\text{ID}}(\Delta \hat{S})}} 

\review{Intra-daily updated renewable production forecasts used in this paper are provided by Statkraft Markets and generated by \cite{statkraft2020a}. Day-ahead demand / system load forecasts are obtained from \cite{entsoe2020}. The forecasts have a 15-minute delivery period resolution and denote the expected produced power by all assets of the respective technology in Germany in MW. Forecasts are sampled to hourly frequency using a simple arithmetic average. A new update is available every hour. The first forecast version is issued several days before the delivery day, the latest version usually after the end of the delivery period due to ex-post updates.  Let $\WFC{v}, \SFC{v}$ denote forecasts for wind and solar production for delivery period $d, s$ available at time $v$. Forecasts for demand are not updated as regularly, hence intraday-updates are not considered in this paper. We denote demand forecasts as $\DFC$. Note that the issuance time $v$ of a new forecast does not necessarily correspond to the timing of trades on the continuous market or the 5-minute intervals used to aggregate these trades. For any forecasting study, it is important to keep in mind the information set at the point of forecasting. The start of the simulation is set to 185 minutes before the start of the delivery period. Hence, forecast versions and updates can only be considered if they are available earlier than 185 minutes before the start of the delivery period.\footnote{For example, for a product with delivery on September 1st, 12:00 to 13:00, all forecast versions available until September 1st, 8:55 can be used. For the product with delivery 13:00 to 14:00, all versions up to 9:55 can be used.} For each delivery period, two forecast versions deserve special attention: First, the latest forecast available before 12:00 on $d-1$, the deadline for submission of bids to the spot auction, is referred to as the day-ahead forecast $\WFC{\text{DA}}$ and $\SFC{\text{DA}}$. Secondly, the newest forecast available before the start of the simulation, i.e. at which $v \geq b(d,s) - 185$ holds, is denoted as the intraday forecast $\WFC{\text{ID}}$ and $\SFC{\text{ID}}$.}

\review{
An initial analysis showed that individual forecast updates immediately before the start of the simulation carries little predictive power for the whole simulation period of three hours. Therefore, the forecast updates are aggregated. We consider two aggregated measures for forecast changes: first, the aggregated change between the production forecasts at the day-ahead stage and the production forecasts at the start of the simulation. Second, we employ the volatility of all forecast changes between the day-ahead stage and the production forecasts at the start of the simulation. Let us generally define the change between two forecast versions $v_1, v_2$ as $\deltaWFC{v1}{v2} = \WFC{v_2} - \WFC{v_1}$ with $v_2$ being the newer forecast.
}

\review{
\begin{itemize}
	\item The day-ahead to simulation forecast update is defined as: $\deltaWFC{\text{DA}}{\text{ID}} = \WFC{\text{ID}} - \WFC{\text{DA}}$. The symmetry of the impact of forecast errors on $\deltaPID{t}$ is a disputed topic in the literature \citep{ziel2017, kiesel2020a, kiesel2020b} and has, so far, not been explored for the volatility of $\deltaPID{t}$. To address this issue and test for possible asymmetric effects, $\deltaWFC{\text{DA}}{\text{ID}}$ and $\deltaSFC{\text{DA}}{\text{ID}}$ are split in positive and negative updates: $\deltaWFCsign{\text{DA}}{\text{ID}}{+} = \text{max}(\deltaWFC{\text{DA}}{\text{ID}}, \; 0 )$ and $\deltaWFCsign{\text{DA}}{\text{ID}}{-} = \; \mid \text{min}(\deltaWFC{\text{DA}}{\text{ID}}, \; 0 ) \mid$.
	\item The standard deviation of the forecast updates should reflect the uncertainty about the weather situation. Highly volatile forecast updates are mirrored to starkly changing positions in renewable energy asset portfolios and should thus exercise an influence on the volatility of the price process due to quickly changing demand and supply. Let 
	\begin{equation*}
		\mathcal{V}^{d,s} = \left[d - 1, \; \text{12:00} \geq v \geq b(d, s) - 185 \right]
	\end{equation*} be the set of all forecast versions received between the day-ahead auction and the start of the simulation. The difference between two consecutive forecast versions $v, v-1$ is denoted as $\deltaWFC{v-1}{v}$. Then
$\sigmaWFC = \underset{v \in \mathcal{V}^{d,s}}{\sigma}(\deltaWFC{v-1}{v})$ 
denotes the standard deviation of all differences between two consecutive forecasts received between the day-ahead auction and the start of the simulation. It is worth noting that due to the schedule of the day-ahead and intraday markets, $\mathcal{V}$ is larger for later delivery hours, thus more forecast versions are considered for $\sigma^{d,s}_{\text{DA},0}(\Delta \hat{W})$ for later delivery hours. 
\end{itemize}
Analogously, $\deltaWFC{\text{DA}}{\text{ID}}, \deltaSFCsign{\text{DA}}{\text{ID}}{+}, \deltaSFCsign{\text{DA}}{\text{ID}}{-}$ and $\sigmaSFC$ are defined for the solar production forecasts. Panels (a) - (c) in Figure \ref{fig:summary_statistics_res} show the day-ahead versions of wind, solar and demand forecasts. For wind and solar, the change between day-ahead and intraday versions and the standard deviation are plotted as well. 
}

\newcommand{\OUT}[1]{\ensuremath{O_{#1}^{d, s}}} 
\newcommand{\deltaOUT}{\ensuremath{\Delta O_{\text{DA}, \text{ID}}^{d, s}}} 
\newcommand{\deltaOUTp}{\ensuremath{\Delta O_{\text{DA}, \text{ID}}^{d, s, \text{planned}}}} 
\newcommand{\deltaOUTu}{\ensuremath{\Delta O_{\text{DA}, \text{ID}}^{d, s, \text{unplanned}}}} 

Under the \gls{REMIT}, market participants are required to report non-availabilities of their assets and make this information available to all other market participants in order to avoid insider trading. In practice, this obligation is fulfilled by market participants by submitting non-availability messages to an inside information platform \citep{eu2011, lazarczyk2018, acer2020}. We retrieve unavailability messages from the \cite{eex2020} market transparency platform for all non-availabilities regarding the delivery periods between January 1st, 2016 and  July 31st, 2020. A non-availability message is defined by the date of publication, beginning and end of the non-availability, the type of non-availability, i.e. whether it has been planned or unplanned, the fuel type of the unavailable asset as well as the unavailable capacity in MW. The outage messages are aggregated to the total non-available generation capacity for the delivery period $d, s$ known at the time of the spot auction. Additionally, the outages are aggregated to the total non-available generation capacity known at the start of the simulation for a delivery period $d,s$. Sub-hourly outages are taken into account with the respective share of the full hour. The differences between the level of outages day-ahead and at the start of the simulation is calculated similar as the difference in the forecasts: $\deltaOUT = \OUT{\text{ID}} -  \OUT{\text{DA}}$. Afterwards, the difference $\Delta O^{d,s}_{\text{DA},0}$ is split into planned and unplanned outages denoted as $\deltaOUTp$ and $\deltaOUTu$. Figure \ref{fig:summary_statistics_res} (d) plots the aggregated outage data; Table \ref{tab:summary_statistics_wind_solar_demand} gives summary statistics. 

\begin{figure}[H]
\begin{subfigure}[c]{0.5\textwidth}
\begin{center}
\includegraphics[width=\textwidth]{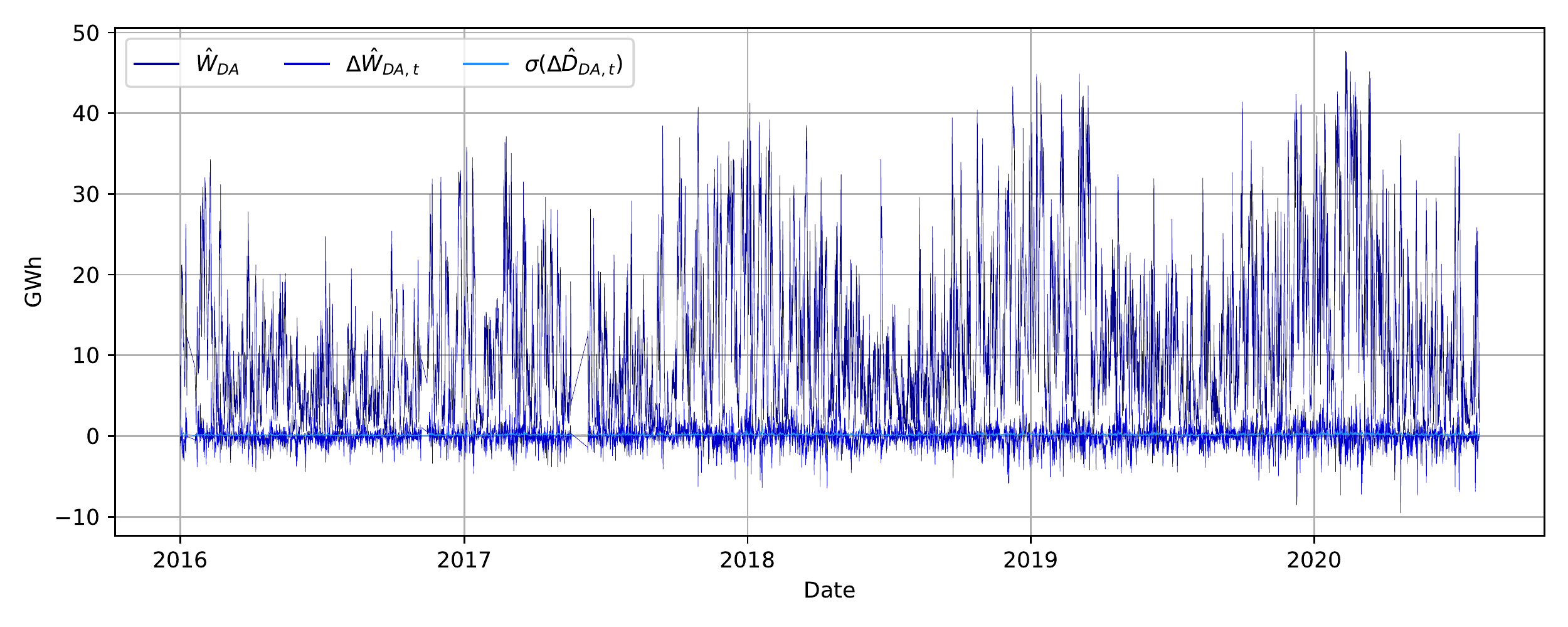}
\subcaption{\review{Day-ahead wind production forecasts $\WFC{\text{DA}}$, difference to intraday $\deltaWFC{\text{DA}}{\text{ID}}$ and standard deviation of forecast updates $\sigmaWFC$.}}
\end{center}
\end{subfigure}%
\begin{subfigure}[c]{0.5\textwidth}
\begin{center}
\includegraphics[width=\textwidth]{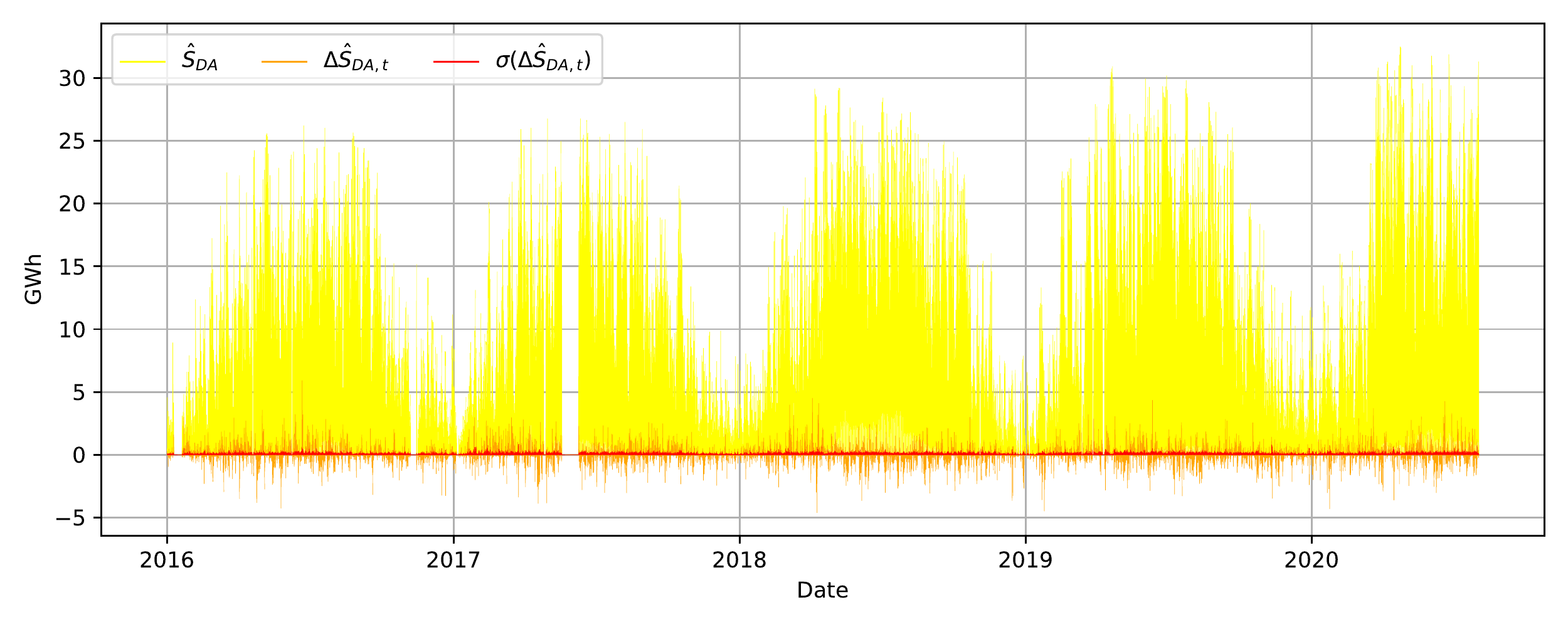}
\subcaption{\review{Day-ahead solar production forecasts $\SFC{\text{DA}}$, difference to intraday $\deltaSFC{\text{DA}}{\text{ID}}$ and standard deviation of forecast updates $\sigmaSFC$.}}
\end{center}
\end{subfigure}
\begin{subfigure}[c]{0.5\textwidth}
\begin{center}
\includegraphics[width=\textwidth]{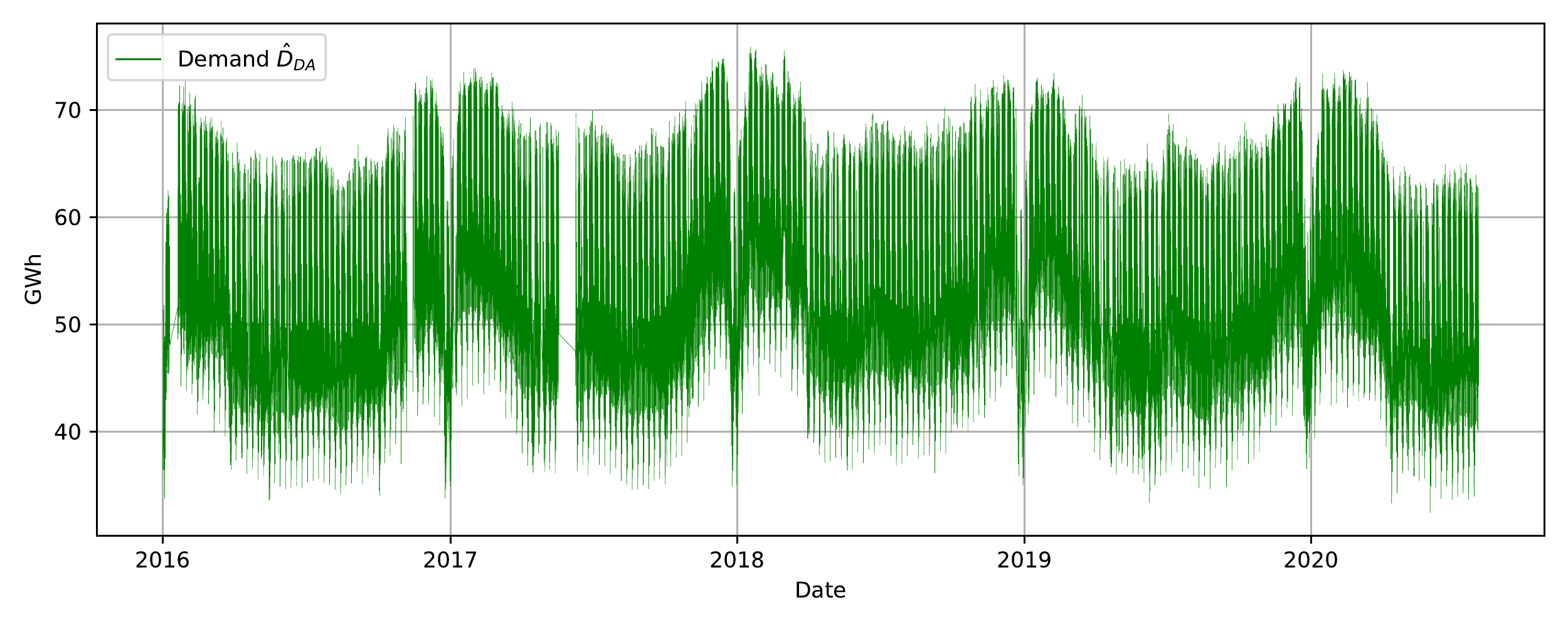}
\subcaption{\review{Day-ahead demand forecasts $\DFC$ \newline \newline.}}
\end{center}
\end{subfigure}%
\begin{subfigure}[c]{0.5\textwidth}
\begin{center}
\includegraphics[width=\textwidth]{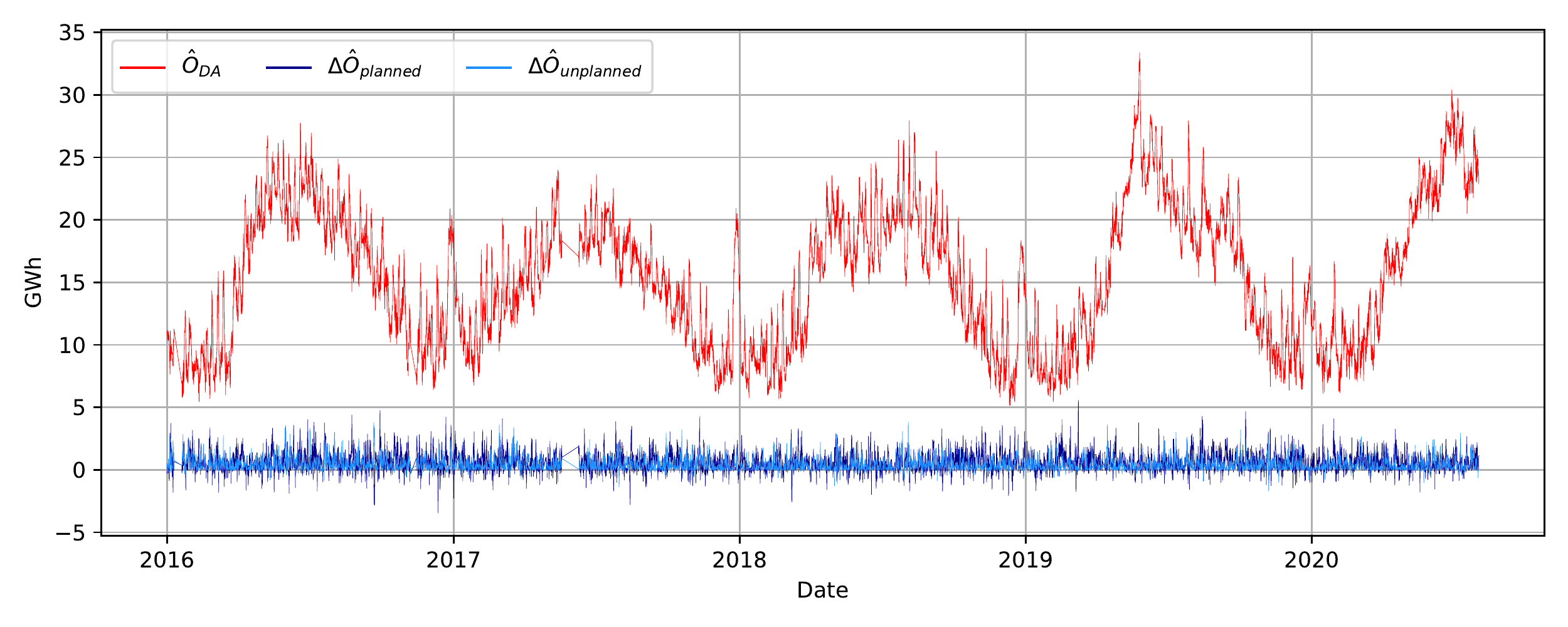}
\subcaption{\review{Day-ahead aggregated outage notifications $\OUT{\text{DA}}$ and the planned and unplanned changes $\deltaOUTp$ and $\deltaOUTu$.}}
\end{center}
\end{subfigure}
\caption[RES production forecasts and aggregated outage notifications.]{Day-ahead \gls{RES} production forecasts, forecast updates and their standard deviation and the aggregated outage notifications.} \label{fig:summary_statistics_res}
\end{figure}


\begin{table}
\caption[Summary statistics for RES forecasts and the reported outages.]{Summary Statistics for wind, solar and demand forecasts, and the reported outages. Day-ahead to intraday forecast changes and the standard deviation of versioned updates. Indices $d, s$ and $t$ omitted for better readability. Values in MW.}\label{tab:summary_statistics_wind_solar_demand}
\begin{center}
\resizebox{\textwidth}{!}{\begin{tabular}{lrrrrrrrrrr}
\toprule
{} &  \multicolumn{3}{c@{}}{Wind} & \multicolumn{3}{c@{}}{Solar} &    Demand & \multicolumn{3}{c@{}}{Outages} \\ \cmidrule(lr){2-4} \cmidrule(l){5-7} \cmidrule(l){8-8} \cmidrule(l){9-11} 
{} & $\hat{W}$ & $\Delta \hat{W}$ & $\sigma(\Delta \hat{W})$ & $\hat{S}$ & $\Delta \hat{S}$ & $\sigma(\Delta \hat{S})$ & $\hat{D}$ &       $O$ & $\Delta O_\text{planned}$ & $\Delta O_\text{unplanned}$ \\
\midrule
Count ($n$)     &  38856 &         38856 &                 38856 &  38856 &         38856 &                 38856 &  38856 &  38856 &            38856 &              38856 \\
Mean ($\mu$)    &  12372.39 &           -59.60 &                   182.78 &   4634.27 &             8.35 &                    56.22 &  59422.74 &  15506.99 &              573.24 &                410.07 \\
Std. ($\sigma$) &   9410.80 &          1291.23 &                   110.11 &   6994.23 &           521.30 &                    84.29 &  10286.86 &   5498.09 &              711.15 &                486.55 \\
Min             &    172.75 &         -9515.50 &                     6.00 &      0.00 &         -4604.50 &                     0.00 &  33927.90 &   5157.40 &            -3451.60 &              -1690.00 \\
Max             &  47708.00 &         11383.25 &                  1940.55 &  32474.50 &          5914.50 &                   842.83 &  83494.50 &  33384.40 &             5569.00 &               3847.00 \\
\bottomrule
\end{tabular}
}
\end{center}
\end{table}

\FloatBarrier
\subsection{Spot Auction Curves and Elasticity}\label{sec:data_auction_curves}

The impact of forecast errors on  $\deltaPID{t}$ depends on the steepness of the merit-order \citep{kiesel2020a, kiesel2020b, kulakov2020}. Following this thought, the volatility of the intraday price should also be influenced by the slope of the merit-order. If the market is in a steep merit-order regime, even small volume changes might have a high price impact. Thus, the expected price impact of changes in (\gls{RES}) supply is stronger. Under uncertainty of future \gls{RES} forecast updates, the expected volatility should increase with the steepness of the merit-order. If the market price corresponds to a rather flat region of the merit-order, the price impact of changes in \gls{RES} supply should be smaller and hence the volatility of $\deltaPID{t}$ should be lower. This thought will be the main intuition for the addition of a merit-order slope to the model for the volatility of the distribution of $\deltaPID{t}$.

There are different methods to model the merit-order used in practice and academia. Fundamental methods as developed by \cite{pape2016, gurtler2018, beran2019} are complex, data-intensive and rely heavily on assumptions. For this reason, \cite{kiesel2020a, kiesel2020b} develop an econometric model based on \cite{he2013} by fitting the relationship between demand forecasts and day-ahead prices to an exponential function. This yields an analytically traceable function, whose slope can easily be calculated as the first derivative. This paper develops a further method to derive the slope of the merit-order by using the day-ahead auction curves as a proxy for the supply stack. This approach is based on \cite{balardy2018} and \cite{kulakov2020} and has three advantages compared to the approach of \cite{he2013}: First, the auction curves combine all market and availability information available on $d-1$ and do not depend on a longer time frame for the estimation of the function coefficients. Second, by using the auction curves, there is no need to assume an explicit functional form for the merit-order. Lastly, the auction curves also represent negative prices, while the exponential function is only defined on the positive real line.

However, it is also important to discuss the drawbacks attached to modelling the intraday merit-order based on day-ahead information in general and attached to the auction curves especially. First, the available generation capacity can (and due to \gls{RES} will) change between $d-1$ and $d$, leading to shifts in the merit-order. Second, power plants might not be as flexible intraday as in a day-ahead planning horizon due to ramping behaviour, start-up costs or constraints due to grid service delivery. On the contrary, some power plants might be optimised predominantly intraday and not on the day-ahead auction if they are at-the-money \review{\citep[on this issue see e.g.][]{pape2016}}. Especially for the auction curves, two further problems arise: First, the demand and supply curve are both elastic curves, contrary to the common assumption of largely inflexible demand in energy markets. This problem is addressed by applying the transformation introduced by \cite{kulakov2020, kulakov2019} in the following paragraph. Thereby, all elasticity from the demand curve is shifted to the supply curve, which yields a perfectly inelastic demand and elastic supply curve. Second, the day-ahead auction curves as provided by EPEX Spot only contain standard bids. Thus, linked, block and other complex bids are excluded from the curves, which removes information about the available generation capacity. This problem cannot be addressed simply and needs to be kept in mind for the further interpretation of the results.

\begin{figure}
\begin{subfigure}[c]{0.33\textwidth}
\begin{center}
\includegraphics[width=\textwidth]{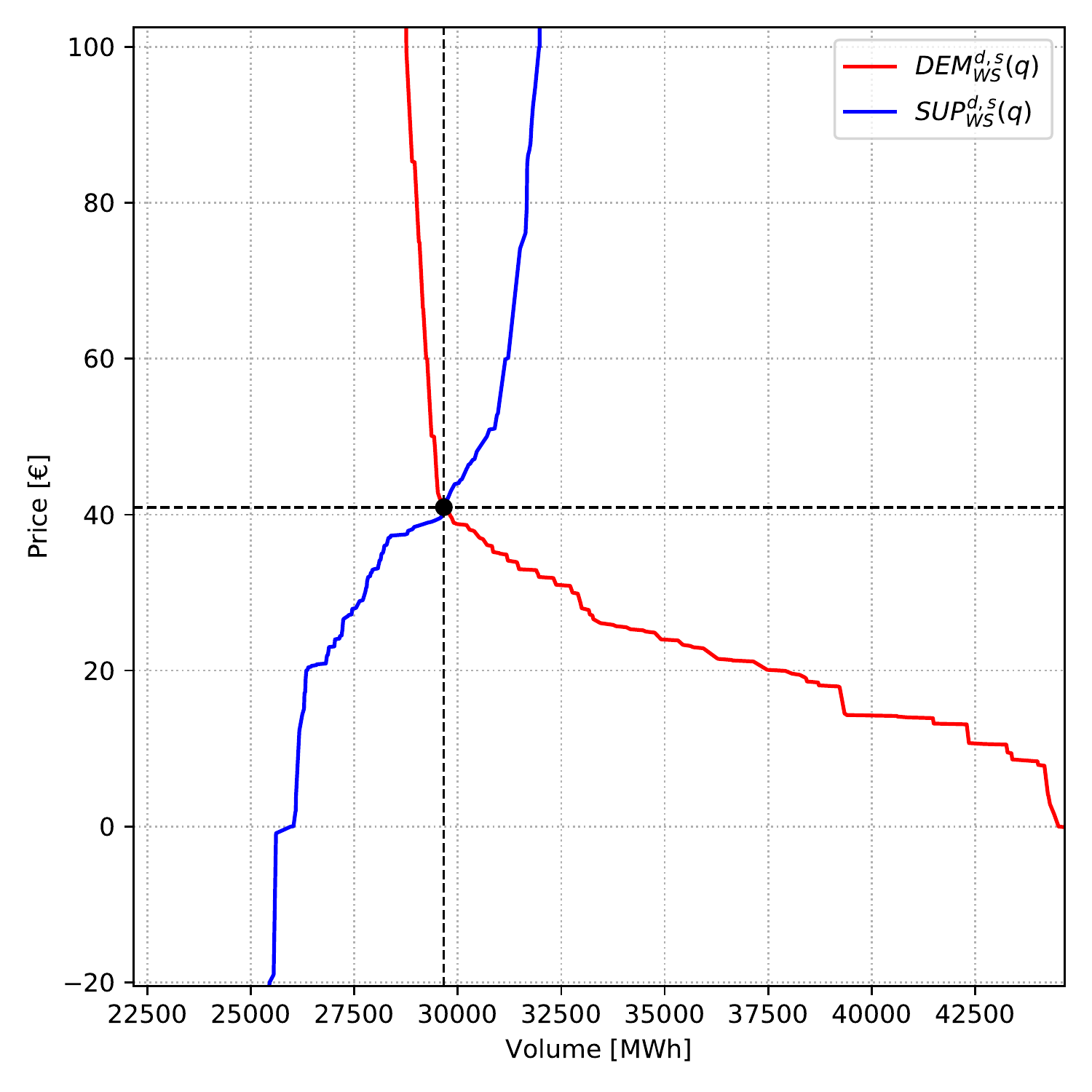}
\subcaption{Original curves.}
\label{fig_curve}
\end{center}
\end{subfigure}%
\begin{subfigure}[c]{0.33\textwidth}
\begin{center}
\includegraphics[width=\textwidth]{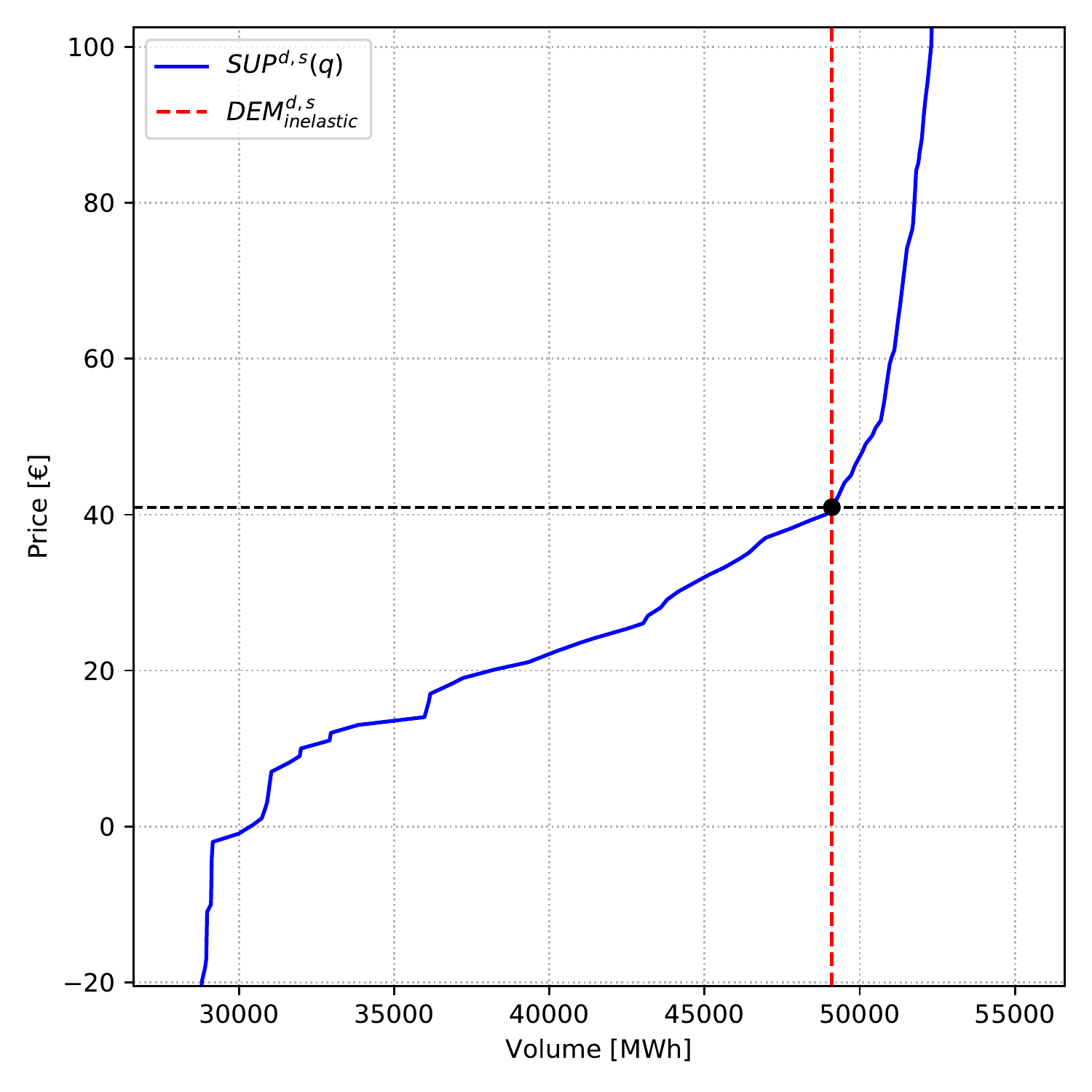}
\subcaption{Transformed curves.}
\end{center}
\end{subfigure}
\begin{subfigure}[c]{0.33\textwidth}
\begin{center}
\includegraphics[width=\textwidth]{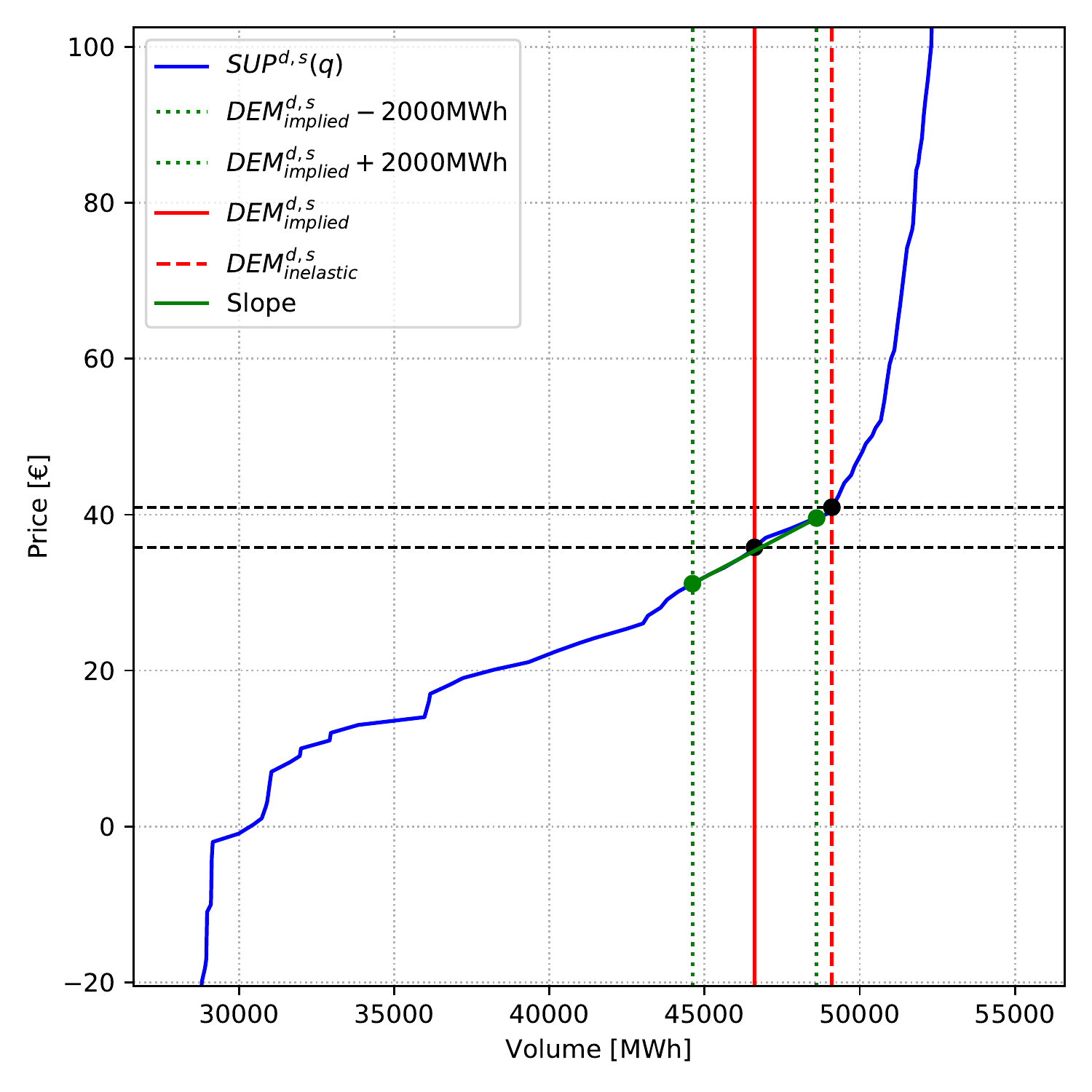}
\subcaption{Slope coefficient.}
\end{center}
\end{subfigure}
\caption[Transformed auction curves.]{Transformed auction curves for June 1, delivery hour $s = 9 $ and the calculation of the merit-order slope coefficient. Note how the intersection of $\text{\textit{SUP}}^{d,s}(\text{\textit{DEM}}^{d,s}_\text{inelastic})$ again yields $P_{Spot}^{d,s}$ for Panels (a) and (b). Figures truncated on the $y$-axis to [-20, 100].}
\label{fig:auction_curves_transformation}
\end{figure}

The intuition behind the transformation of the auction curves is outlined in detail in \cite{kulakov2019} and \cite{coulon2014}, so here only a brief introduction is given. Figure \ref{fig_curve}
 shows that the demand curve at the day-ahead auction is elastic, which is at odds with the common assumption of few price elastic consumers of electricity, especially at short notice \citep{knaut2016, coulon2014}. However, producers and consumers have the chance to sell/purchase their energy not only on the spot auction, but also in the OTC and derivative markets. In addition, there might be market participants that own both assets on the supply and demand side. Thus, arbitrage opportunities between the two markets arise that can be used by the trader. \cite{coulon2014} and \cite{kulakov2019} consider this effect by flipping the elasticity from the demand curve to the supply curve, hence obtaining a perfectly inelastic (vertical) demand curve and an elastic supply curve  to incorporate those effects. \review{The core idea here is that, at the day-ahead auction, placing a buy order for a volume $x$ for a price $y$ is the same placing a buy order for the volume $x$ at the maximum price and placing a sell order with volume $x$ for the price $y$ + the smallest tick. \cite{kulakov2019} elaborate in detail on the econometric framework, which is adopted in this paper and the implications for the different market participants.}

Figure \ref{fig:auction_curves_transformation} shows the supply and demand curves from the spot auction for the delivery day June 1st, 2017 for hour $s=9$. The intersection of supply and demand yields the spot price $P^{d,s}_\text{Spot}$. The notation follows largely \cite{kulakov2019, kulakov2020}. 
Define the supply and demand curves as a mapping of volumes to prices by $
\text{\textit{SUP}}_{WS} : \; 			(0, \infty) 					\rightarrow \left[P_\text{min}, P_\text{max}\right] $ and $
\text{\textit{DEM}}_{WS} : \; (0, \infty) 	\rightarrow \left[P_\text{min}, P_\text{max}\right].
$ Due to strict monotonicity the inverse ${\text{\textit{SUP}}_{WS}}^{-1}$ and  ${\text{\textit{DEM}}_{WS}}^{-1} $
always exist. Hence, $\text{\textit{SUP}}_{WS}^{d,s}(q) = P$ is the supply or sell curve and $\text{\textit{DEM}}_{WS}^{d,s}(q) = P$ is the demand curve at the spot auction for delivery day $d$ and hour $s$ relating the volume $q$ ought to be sold/bought to the according price $P$. The inelastic demand in the wholesale market can be calculated by
$DEM^{d,s}_\text{inelastic} = {DEM_{WS}^{d,s}}^{-1}(P_\text{min})$
where $P_\text{min} =  -500$ EUR/MWh is the minimum price at the day-ahead auction \citep[][see Chapter \ref{ch:market_structure} of this paper]{epex2018}. The transformed inverse supply curve can be written as:
\begin{equation}
{SUP^{d,s}}^{-1}(z) = 
	\underbrace{
		{SUP_{WS}^{d,s}}^{-1}(z)
	}_\text{inverted wholesale supply curve} + 
	\underbrace{
		DEM^{d,s}_\text{inelastic} - {DEM_{WS}^{d,s}}^{-1}(z)
	}_\text{flipped inverted wholesale demand curve}
\end{equation}
As the curves are monotonic, ${SUP^{d,s}}^{-1}(z)$ also defines $SUP^{d,s}(q)$. As it is clearly visible in Figure \ref{fig:auction_curves_transformation}, the original equilibrium is reached at the point
$P_\text{DA}^{d,s} = SUP^{d,s}(DEM^{d,s}_\text{inelastic})$. 
For the transformed curves it now holds that for the resulting clearing price, shifting $DEM^{d,s}_\text{inelastic}$ by some quantity $x$ equals shifting ${SUP^{d,s}}^{-1}(z)$ by $-x$, as the demand is perfectly inelastic.

\begin{figure}[htb]
\includegraphics[width=\textwidth]{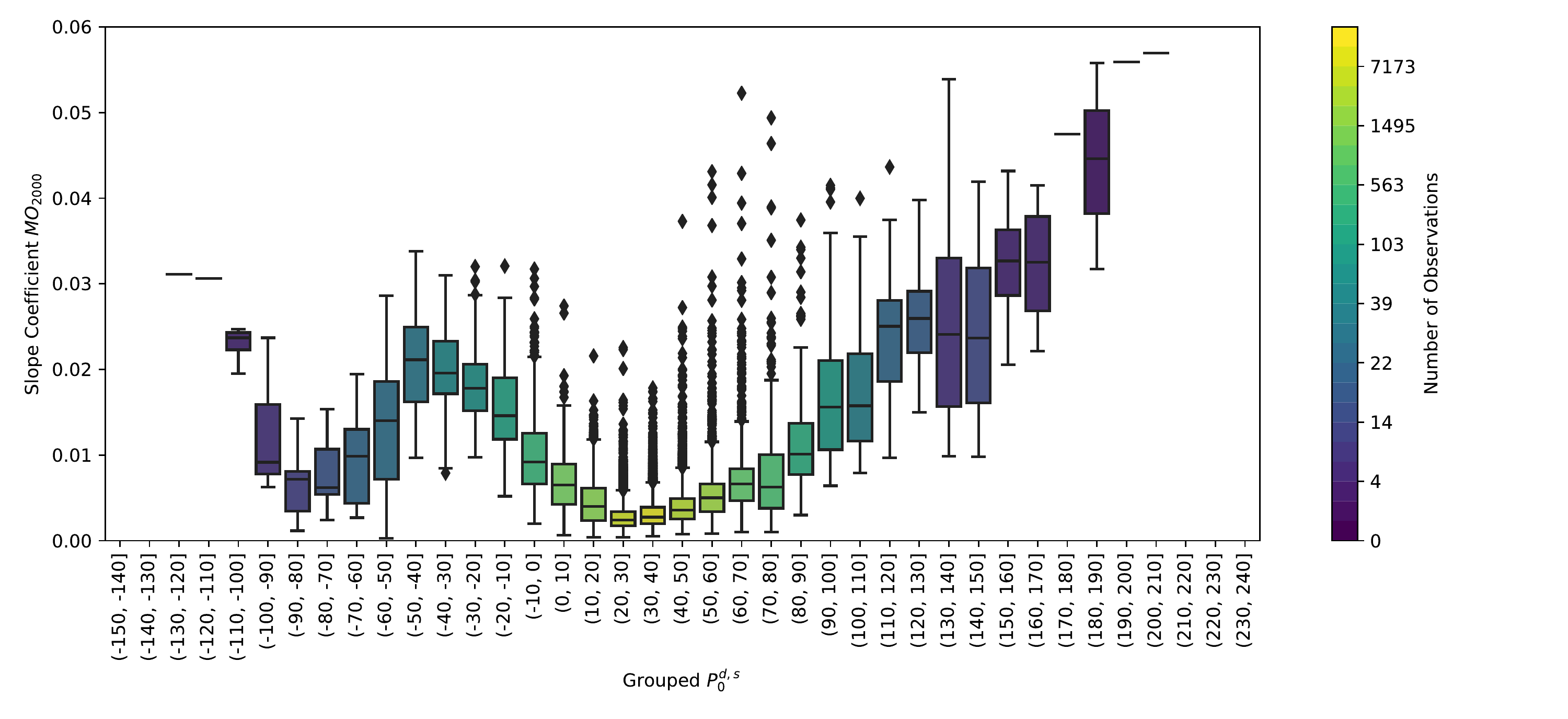}
\caption[Boxplots for the merit-order sloped grouped by the price level.]{Boxplots for the merit-order slope $\textit{MO}_{2000}$ grouped by the price level. Colouring indicates the Number of observations, note the log-scale of the colorbar. Upper and lower end of the box correspond to $Q_{0.25}$ and $Q_{0.75}$ respectively.} \label{fig:merit_order_slope}
\end{figure}

Under the assumptions that the merit-order does not change significantly between day-ahead and intraday and that the transformed supply curve is a reasonable proxy for the merit-order, the the implied intraday demand and the slope coefficient for the merit-order can be derived. The first assumption is implicitly already made by \cite{kiesel2020a, kiesel2020b}. The second assumption is discussed above. The last known 5-minute-interval-\gls{VWAP} before the start of the simulation is $\PID{0}$. Under the \gls{MEH}, this price should reflect all changes to demand and supply. As all flexibility is already included in the supply curve, the implied intraday inelastic demand at $t=0$ can be calculated as 
$\text{\textit{DEM}}^{d,s}_{\text{implied}} = {\text{\textit{SUP}}^{d,s}}^{-1} (\PID{0}).$
As already mentioned in Chapter \ref{ch:market_structure}, the lower and upper price limits at the day-ahead auction are $\left[-500, 3000\right]$ EUR/MWh, while in the intraday market these are set to $\left[-9999, 9999\right]$ EUR/MWh. Hence, it might be possible that $\PID{0}$ is outside the domain of ${\text{\textit{SUP}}^{d,s}}^{-1} (z)$. This case, however, does not occur in the dataset used in this paper. In the spirit of \cite{balardy2018} and \cite{kulakov2019}, the measure for the elasticity $\textit{MO}^{d,s}_q$ is calculated as a finite central difference quotient of the transformed supply curve around $\text{\textit{DEM}}^{d,s}_{\text{implied}}$:
\begin{align}
\text{\textit{MO}}^{d,s}_{q} &= \frac{
	\text{\textit{SUP}}^{d,s}(\text{\textit{DEM}}^{d,s}_{\text{implied}} + q) - 
	\text{\textit{SUP}}^{d,s}(\text{\textit{DEM}}^{d,s}_{\text{implied}} - q)
}{2 \cdot q}, 
\end{align}
where $q = \{500, 1000, 2000\}$ MWh. It is defined in $\text{EUR}/\text{MWh}^2$ and is the steepness of the auction curves around the price level at $t=0$. Intuitively, it can be interpreted as the expected price change for a 1 MWh change in supply. In this paper, three values for $q$ are tested as there is some arbitrariness in choosing this value. In the literature, \cite{kulakov2019} choose $q = 100$ MWh, \cite{balardy2018} chooses $q = 500$ MWh. In this paper, slightly higher $q$ are selected to accommodate the fact that the standard deviation of $\deltaWFC{\text{DA}}{\text{ID}}$ and $\deltaSFC{\text{DA}}{\text{ID}}$ is roughly between 500 MW and 1300 MW (see Table \ref{tab:summary_statistics_wind_solar_demand}). Values of $q < 500$ MWh thus might not catch the full range of volume changes occurring during the intraday trading. These volume changes in turn lead to movements along the merit-order. Figure \ref{fig:auction_curves_transformation} (c) shows the intuition of the slope coefficient for $q = 2000$. \review{Figure \ref{fig:merit_order_slope} shows boxplots of the slope coefficients for $q = 2000$ MWh by the price level. Clearly, the slope increases with increasing price level, which is in line with the classic merit order model. However, we see the slope rising as well for small and negative prices. This observation is at odds with the common assumption of \emph{zero}-marginal cost renewable production, which would imply a flat lower end of the merit order. However, many renewable assets are part of subsidy schemes, making their effective marginal costs negative. These assets are sold to the market even for negative prices, as long as the subsidy paid per produced MWh offsets negative selling prices.}


\FloatBarrier
\section{Electricity Price Models} \label{ch:models}

Recall from the introduction that the price differences $\deltaPID{t}$ follow certain distribution 
that we denote $G^{d,s}$ which is a mixture distribution
\begin{align*}
\deltaPID{t} &\sim G^{d,s} 
= (1 - \alpha_t^{d,s})\delta_0 + \alpha_t^{d,s} F^{d,s}
\end{align*}
with the Dirac distribution $\delta_0$, the continuous distribution $F^{d,s}$ and the Bernoulli variable $\alpha_t^{d,s}$ with probability $\pi^{d,s}_t$. The two-stage approach is:
\begin{enumerate}
\item Model the probability $\pi^{d,s}_t$ for a trade-event $\ALPHA{t} = 1$ by a logistic regression model. 
\item Model the distribution $F^{d,s}_t$ by skewed Student's $t$-distribution respectively the Johnson's $S_U$ and estimate $\boldsymbol{\theta}^{d,s}$ using the \review{\gls{GAMLSS}-framework  \citep{rigby2007, rigby2017, rigby2005, stasinopoulos2018gamlss}.}
\end{enumerate}
This introduces a dependence structure between the parameters of $\delta_0$ and $F$, as the probability $\pi^{d,s}$ is explained by past realisations of $\deltaPID{t}$ and $\ALPHA{t}$. In the following three sections, the logistic model, the \gls{GAMLSS} model and the used benchmarks are presented.

\subsection{Logistic regression model for $\alpha$} 

\review{The binary variable $\ALPHA{t}$ will be modelled by a regularized logistic regression model \citep{tibshirani1996, meier2008} in the implementation of \cite{rglmnet} using coordinate descent}. Generally, for a logistic model 
\begin{equation}
\text{log}\left(\frac{\pi}{1-\pi}\right) = \mathbf{X}^T\boldsymbol{\beta}
\end{equation}
for the Bernoulli variable $\alpha$ with probability $P(\alpha = 1) = \pi$, the \gls{LASSO} estimator $\hat{\mathbf{\beta}}^\text{LASSO}$ is given by
\begin{equation}
\hat{\boldsymbol{\beta}}^\text{LASSO} = 
	\text{arg} \; \text{min} \left( - l \left(\boldsymbol{\beta}, \tilde{\mathbf{X}} \right) + 
	\lambda 
	\left\lvert \left\lvert \boldsymbol{\beta}  \right\rvert \right\rvert_1 	
\right), 
\end{equation}
where $\lambda$ is a tunable shrinkage parameter. The corresponding log-likelihood $l$ is given by
\begin{equation}
l \left(\boldsymbol{\beta}, \tilde{\mathbf{X}} \right) = 
	\frac{1}{N} \sum\limits_{i=1}^N \alpha_i \tilde{\mathbf{X}}^T_i\boldsymbol{\beta} - 
	\text{log}\left(  1 +  \text{exp} \left(  \tilde{\mathbf{X}}^T_i\boldsymbol{\beta} \right) \right),
\end{equation}
where $\tilde{\mathbf{X}}$ is a standardisation of $\mathbf{X}$. The parameter $\lambda$ is optimised from an exponential grid of 100 values by choosing the minimum \gls{BIC}, i.e. $\lambda^\text{opt} = \text{arg} \; \text{min} \; \text{BIC} \left( \lambda \right)$ using the \texttt{glmnet} package by \cite{rglmnet}. 

Here, the logit function for $\ALPHA{t}$ is explained by four components: the impact of past price differences, the time to maturity and weekday effects, fundamental variables such as \gls{RES} forecasts, outages and the slope of the merit-order, and a regression on averaged past $\ALPHA{t}$. Intuitively, the probability of trades should rise with higher $\deltaPID{t}$,  closer to delivery, with increasing wind and solar forecasts and with increased recent trading activity measured by past $\ALPHA{t}$, but decrease on the weekends and the transition day Monday.

\begin{align*} 
\text{log} & \left( \frac{\pi^{d,s}_t}{1 - \pi^{d,s}_t} \right) = \beta_0 + 
	\underbrace{	
		\sum\limits_{j=1}^3 \beta_j \deltaPID{t-j} + 
		\sum\limits_{j=1}^6 \beta_{3+j} | \deltaPID{t-j} | + 
		\beta_{10} \sum\limits_{j=7}^{12} | \deltaPID{t-j}|
	}_{\text{Price differences}}  \\
	& + \underbrace{
		\beta_{11} \text{MON}(d) + \beta_{12} \text{SAT}(d) + \beta_{13} \text{SUN}(d) +  
		\sum\limits_{j=1}^{31} \beta_{13+j} \text{TTD}(t) 
	}_{\text{Time dummies}} \\
	& + \underbrace{
		\beta_{47} \DFC + \beta_{48} \WFC{\text{DA}} + \beta_{49} \SFC{\text{DA}} + \OUT{\text{DA}}
	}_{\text{Day-ahead fundamental variables}} \stepcounter{equation}\tag{\theequation}\label{eq:logistic_model_ziel}\\
	& + \underbrace{	
		\beta_{51} \deltaWFCsign{\text{DA}}{\text{ID}}{+} + \deltaWFCsign{\text{DA}}{\text{ID}}{-} + \beta_{53} \deltaSFCsign{\text{DA}}{\text{ID}}{+} + \beta_{54} \deltaSFCsign{\text{DA}}{\text{ID}}{-}
	}_{\text{Day-ahed to intraday forecast updates}} \\
	& + \underbrace{	
		\beta_{55} \sigmaWFC + \beta_{56} \sigmaSFC
	}_{\text{Standard deviation of forecast updates}} 
	+ \underbrace{	
		\beta_{57} \deltaOUTp + \beta_{58} \deltaOUTu
	}_{\text{Intraday changes in planned and unplanned outages}} \\
	& + \underbrace{
		\beta_{59} \mid \PDA - \PID{t-1} \mid
	}_{\text{Day-ahead to $t-1$ price spread}}
	+ \underbrace{	
		\sum\limits_{j=1}^3 \beta_{59+j} \textit{MO}^{d,s}_{j}
	}_{\text{Slope of the merit-order}} 
	+ \underbrace{
		\sum\limits_{j=1}^{12} \beta_{62+j} \bar{\alpha}^{d,s}_{t-j},
	}_{\text{Regression on $\bar{\alpha}^{d,s}_t$}} 
\end{align*}

where $\bar{\alpha}^{d,s}_{t-j}  = 1/j \cdot \sum_{i=1}^j \alpha_{t-i}^{d,s}$, the average of the last $j$ observed values of $\alpha^{d,s}_t$. \review{This approach to transforming lagged values is similar to HAR-type models found in the field of financial econometrics}. A thorough description of the model is omitted here and can be found in \cite{ziel2020a}. SAT($d$), SUN($d$), and MON($d$) are dummies for the weekday of $d$. TTD($t$) is a set of dummies for $t$. Accordingly, the model has more than 70 coefficients of which some tend to be highly correlated. To avoid problems with over fitting and multicollinearity, the model is estimated using the \gls{LASSO} of \cite{tibshirani1996}. \review{Note however, that for the one year training set used in this paper, we have ${365 \cdot T = 365 \cdot 31 = 11315}$ observations and are still in a setting where $n \gg{} p$. The number of observations is sufficiently larger than the number of parameters, hence identification is not an issue here.}

\subsection{GAMLSS Framework}\label{sec:models_gamlss}

\review{
This chapter briefly introduces the \gls{GAMLSS}-framework used to model \begin{equation*}
\deltaPID{t} \mid \ALPHA{t} = 1 \sim F.
\end{equation*}
The \gls{GAMLSS} is an extension of the \gls{GAM} introduced by \cite{hastie1987, hastie1990}. It allows to model not only the expected value of the a variable $Y \sim F$ but also the higher moments under a wide range of continuous and discrete distributions $F$. For an in-depth treatment we refer the reader to \cite{rigby2005, rigby2007, stasinopoulos2018gamlss} and the manual of the \texttt{R}-package \texttt{gamlss} \citep{rigby2017}. The notation in the following paragraphs follows the notation of aforementioned sources.
}

\review{
We first introduce the framework in an abstract notation. Following the mathematical formulation we will relate the abstract notation to the notation of the price differences. Let be $\boldsymbol{Y} = (Y_1, Y_2, ..., Y_n)$ be a vector of $i = 1, ..., n$ independent observations $Y_i$. The \gls{GAMLSS}-framework assumes that $Y_i$ have the probability density function 
\begin{equation*}
f(y_i \mid \mu_i, \sigma_i, \nu_i, \tau_i), 
\end{equation*}
where each of the distribution parameters can be a smooth function of the explanatory variables. We denote as $\boldsymbol{\theta}_i = (\theta_{i,1} \theta_{i,2}, \theta_{i,3}, \theta_{i,4}) = (\mu_i, \sigma_i, \nu_i, \tau_i)$ the vector of $k=1,..., 4$ distribution parameters which are usually known as the location, scale and shape parameters $\theta_{i,k}$. For the distributions used in this paper, $\nu_i$ denotes the skewness and $\tau_i$ denotes the kurtosis. $\boldsymbol{\theta}$ is a matrix whose individual components have the indices $i$ and $k$. The vectors $\boldsymbol{\theta}_i$ and $\boldsymbol{\theta}_k$ are defined along the 2 axis of $\boldsymbol{\theta}$. Formally, we have 
\begin{equation*}
Y_i \sim F(\mu_i, \sigma_i, \nu_i, \tau_i) \Leftrightarrow  Y_i \sim F(\boldsymbol{\theta}_i).
\end{equation*}
For each $k$, let $g_k(\cdot)$ be a known and monotonic link function that relates the distribution parameters $\boldsymbol{\theta}_k$ to the predictor $\boldsymbol{\eta_k}$. 
We consider the \gls{GAMLSS} model equation
\begin{equation}
g_k(\boldsymbol{\theta}_k) = \boldsymbol{\eta}_k = \boldsymbol{X}_k \boldsymbol{\beta}_k  
\end{equation}
where 
$\boldsymbol{X}_k$ is a $n \times J_k$ fixed design matrix and 
${{\boldsymbol{\beta}}'_k = (\beta_{1,k}, \beta_{2,k},...,  \beta_{J_k,k})}$ is a parameter vector of length  $J_k$. The link functions $g_k$ ensure that the estimated distribution parameters fulfil the necessary assumptions concerning their support. To improve the robustness of the estimation, the following link functions are used:
\begin{align}
g_{\text{ident}} (z ) &= z \\
g_{\text{log}}(z ) &= \log(z) \\
g_{\text{logident}}(z ) &= \text{log}( z ) \mathbbm{1}(z \leq 1) + (z - 1)  \mathbbm{1}(z  > 1) \\
g_{\text{logshift2}}(z) &= \text{log}(z - 2)
\end{align}
We use $g_{\text{ident}}$ for the location parameters $\mu$ in both distribution assumptions, and additionally for the skewness parameter $\nu$ of the skewed t-distribution which also has support $(-\infty, \infty)$. $g_{\text{logident}}$ is introduced to avoid the exponential inverse for large estimates, thus improving the robustness of the estimation \citep{ziel2021m5, ziel2020a}. We utilize it for all scale parameters $\sigma$. In addition, $g_{\text{logshift2}}$ is simply the natural logarithm shifted to 2 to preserve the condition $\nu > 2$ for the scale of the skewed $t$-distribution. For the remaining parameters we consider $g_{\text{log}}$.
}

\review{
\cite{rGamlssLasso} extend the \gls{GAMLSS} framework to allow for regularized \gls{LASSO} estimation. As with the logistic model, we employ the \gls{BIC} to select the optimal shrinkage parameter $\lambda$. The adaptive \gls{LASSO} estimator $\boldsymbol{\beta}\ast_k$ is used. It is defined as 
\begin{equation}
\boldsymbol{\beta}\ast_k = \text{argmin}_{\boldsymbol{\beta}} \mid y - \sum_{j=1}^J x_{j} \beta_{j, k} \mid^2 + \lambda_n \sum_{j=1}^J \hat{w}_{j, k} \lvert \beta_{j,k} \rvert
\end{equation}
with the weights vector $\boldsymbol{\hat{w}}_k = 1 / \mid \hat{\beta} \mid^\gamma$. $\hat{\beta}$ denotes a root-$n$ consistent estimator such as ordinary least squares  \citep{zou2006}.
}

\review{
Let us now relate the abstract notation $Y_i$ to $\deltaPID{t}$. The distribution $F(\boldsymbol{\theta}_t^{d,s})$ is fitted to all ${\deltaPID{t} \mid \ALPHA{t} = 1}$. The abstract index $i = 1, ..., N$ is replaced by the combination of the superscript index $d = 1, ..., D$ and the subscript index $t = 1, ..., T$. We fit 24 models each day, one for each delivery period $s$. The delivery periods are treated as independent. Thereby, we yield an estimated vector of four distribution parameters $\hat{\boldsymbol{\theta}}_{t}^{d,s} = (\widehat{\mu}_t^{d,s}, \widehat{\sigma}_t^{d,s}, \widehat{\tau}_t^{d,s} \widehat{\nu}_t^{d,s})$ and accordingly parameter estimates $\boldsymbol{\beta}^{d,s}_{k}$ that condition $\hat{\boldsymbol{\theta}}_t^{d,s}$ on our explanatory variables. Analogously, we can also define the the vector $\boldsymbol{\theta}_{k}^{d,s} = (\theta_{1, k}^{d,s}, ..., \theta_{T, k}^{d,s})$ along the time-axis $t$.  We explain all moments of the distribution by the same set of explanatory variables. \cite{ziel2020a} choose $F$ as Student's $t$-distribution. Here, we extend their choice to the skewed Student's $t$-distribution and Johnson's $S_U$ distribution. Both distributions have four parameters. A short description of the distributions used can be found in the Appendix \ref{app:distributions}. 
}


\review{For each distribution parameter $k$, the model for $\hat{\theta}_{t, k}^{d,s}$ reads:}
\begin{align}
\begin{split}
g_k(\theta^{d, s}_{t, k})  
	& = \beta_{k \mathbbm{1}(k \geq 2), 0} \\ \label{eq:mix_base_02}
	& + \underbrace{	
		\sum\limits_{j=1}^3 \beta_{k, j} \deltaPID{t-j}	}_{\text{Price differences}} 
	+ \underbrace{	
		\sum\limits_{j=1}^6 \beta_{k, 3+j} \mid  \deltaPID{t-j} \mid  + \beta_{10} \sum\limits_{j=7}^{12} \mid \deltaPID{t-j} \mid 
	}_{\text{Absolute price differences}} \\
	& + \underbrace{
		\beta_{k, 11} \text{MON}(d) + \beta_{k, 12} \text{SAT}(d) + \beta_{k, 13} \text{SUN}(d)
	}_{\text{Time dummies}} \\
	& + \underbrace{
		\beta_{k, 14} \hat{L}^{d,s}_\text{DA} + \beta_{k, 15} \WFC{\text{DA}} + \beta_{k, 17}  \SFC{\text{DA}} + \beta_{k, 18} \OUT{\text{DA}}
	}_{\text{Day-ahead fundamental variables}} \\
	& + \underbrace{	
		\beta_{k, 19} \deltaWFCsign{\text{DA}}{\text{ID}}{+} + \beta_{k, 20} \deltaWFCsign{\text{DA}}{\text{ID}}{-} + \beta_{k, 21} \deltaSFCsign{\text{DA}}{\text{ID}}{+} + \beta_{k, 22} \deltaSFCsign{\text{DA}}{\text{ID}}{-}
	}_{\text{Day-ahed to intraday forecast updates}} \\
	& + \underbrace{	
		\beta_{k, 23} \sigmaWFC + \beta_{k, 24} \sigmaSFC
	}_{\text{Standard deviation of forecast updates}} 
	+ \underbrace{	
		\beta_{k, 25} \deltaOUTp + \beta_{k, 26} \deltaOUTu
	}_{\text{Intraday changes in planned and unplanned outages}} \\
	& + \underbrace{
		\beta_{k, 27} \alpha^{d,s}_{t-1} + \beta_{k, 28} \alpha^{d,s}_{t-2}
	}_{\text{Lagged $\alpha^{d,s}_t$}} 
	+ \underbrace{
		\beta_{k, 29} \mid P_\text{\text{DA}}^{d,s} - P_{\text{ID}, t-1}^{d,s} \mid
	}_{\text{Day-ahead to $t-1$ price spread}} \\
	& + \underbrace{	
		\sum\limits_{j=1}^3 \beta_{k, 29+j} \textit{MO}^{d,s}_{j}
	}_{\text{Slope of the merit-order}} 
	+ \underbrace{
		\beta_{k, 32} f_\text{TTD}(t) + \beta_{k, 33}\text{SIDC}(d, t)
	}_{\text{Time to delivery and \gls{SIDC} closing}}
\end{split}
\end{align}
for $k = 1, 2, 3, 4$. The intercept is only included for $k \geq 2$ as the price differences are assumed to be centred around 0, as indicated by the summary statistics in Table \ref{tab:tab_summary_statistics_delta_p}. The volatility is expected to rise with higher absolute past price differences, on weekends, and with higher \gls{RES} generation. The strong changes between day-ahead and intraday \gls{RES} and demand forecasts should also imply higher volatility. We expect the volatility to decrease with more recent trading activity measured by lagged $\alpha^{d,s}_t$. $\text{SIDC}(d, t)$ is a dummy variable taking the value 1 for $d \geq$ June 18, 2018 and $26 \leq t \leq 31$, indicating that the cross-country order books are closed. $f_\text{TTD}(t)$ models the non-linear impact of the time to delivery and takes the form $f_\text{TTD}(t) = 1 / \sqrt{T-t+1}$. It is a deterministic transformation of the variable $t$ and can thus be calculated ex-ante. As argued already in Section \ref{sec:data_auction_curves}, we expect a steeper merit-order regime to lead to higher price volatility. Similar expectations hold for the kurtosis, i.e. we expect a steep merit-order regime to lead to heavier tails of the distribution.


\FloatBarrier
\subsection{Benchmark Models} \label{sec:models_benchmark}

Lastly, some simple benchmark models are introduced. Even though the main focus of this paper is on modelling the volatility and its influencing factor, simple benchmarks can serve as a valuable benchmark to identify potential areas for model improvement. \review{As our study shares the conceptual set-up with \cite{ziel2020a}, it is natural to employ similar benchmark models. Additionally, we compare our approach to classical time series methods such as \gls{ARIMA} models. The following section introduces these models in more detail.}

\cite{ziel2020a} introduce six simple benchmark models to evaluate the value-added by more complex models, briefly described in the following. For the exact specification we refer the reader to their work.
\begin{itemize}
\item The \textbf{Naive} benchmark randomly draws past trajectories.
\item For \textbf{MV.N} and \textbf{MV.t}, a multivariate normal respectively $t$-distribution is fitted to the vector of price differences. Forecasts are randomly drawn from the distribution. 
\item The \textbf{RW.N}, \textbf{RW.t} and \textbf{RW.t.mix.D} are random-walk type of models, where the distribution parameters are estimated from the in-sample data. \review{The \textbf{RW.t.mix.D} also includes a simple mixture term for $\alpha_t^{d,s}$ by estimating $\pi_t^{d,s}$ as the empirical mean of $\ALPHA{t}$.}
\end{itemize}

\review{The closeness of intraday electricity markets to traditional equity markets invites the use of classical time series models as benchmark. However, some attention to the unique time-structure of the intraday market is necessary: As already noted in Section \ref{ch:market_structure}, for all delivery periods $s$ on day $d$, trading starts at 15:00 on $d-1$. The same delivery hour on two following delivery days can have overlapping intraday trading sessions. Thus, we cannot simply combine all trading sessions of a product, as it is done in equity markets. We therefore can only estimate our time series models on the price differences between the start of the trading period and the start of the simulation window. The GAMLSS-based approach does not suffer from this limitation as we learn the coefficients from past data of the simulation windows directly. The following paragraphs introduce the time series benchmark models formally.}

\review{
The classic \gls{ARIMA}($p, k, q$) model is defined as follows:
\begin{equation}
\left(1 - \sum_{i=1}^p \varphi_i L^i \right) \left(1 - L\right) \deltaPID{t} = \delta + \left(1 + \sum_{i=1}^q \theta_i L^i\right)	\epsilon_t
\end{equation}
is an ARIMA($p, k, q$) process with drift $\delta / (1 - \sum{} \varphi_i$). $L$ denotes the lag operator. A full treatment of ARIMA models can be found in e.g. \cite{shumway2017}. We estimate the ARIMA($p, k, q$) models using the \texttt{auto.arima()} function in the \texttt{forecast} package \citep{rForecast}. The function uses a stepwise approach to fit the lag order for $p$ and $q$ based on the \gls{BIC} and performs the KPSS unit-root tests to evaluate the integration order $k$. For each delivery period $d,s$, we fit the model on all 5-minute intervals between the start of trading on $d-1$, 15:00 and the start of the simulation period. The models are denoted as \textbf{Auto.ARIMA}.
}
\FloatBarrier
\section{Forecasting Study and Evaluation} \label{ch:forecasting_study_evaluation}

\FloatBarrier
\subsection{Study Design and Simulation Algorithm}

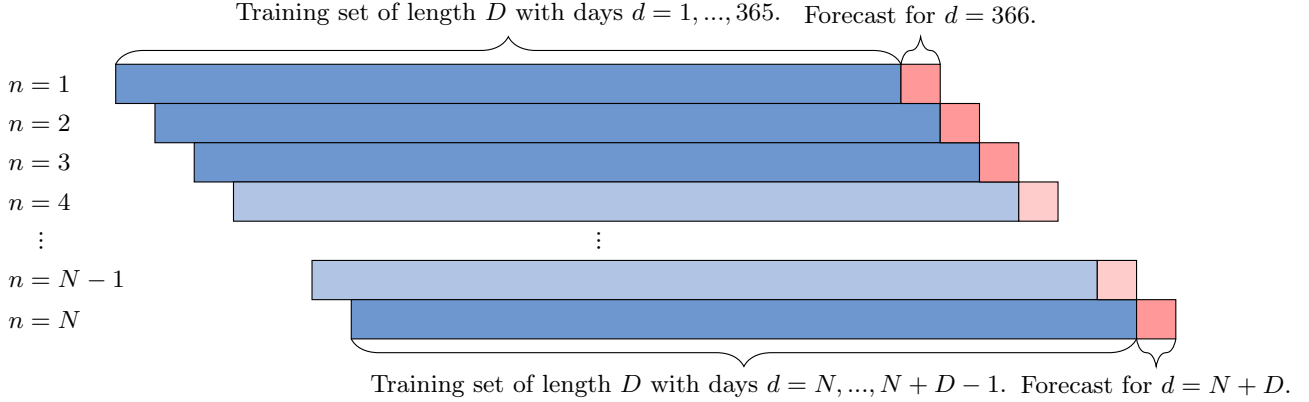
\begin{figure}[H]
\begin{center}
\resizebox{\textwidth}{!}{
\begin{tikzpicture}
\draw [decorate,decoration={brace,amplitude=10pt},xshift=0pt,yshift=0pt]
(0.0,0.0) -- (10.0,0.0) node [black,midway,yshift=20pt, text height = 10pt] {\footnotesize{Training set of length $D$ with days $d=1, ..., 365$.}};
\filldraw[fill=NavyBlue!60!white, draw=black] (0.0,-0.0) rectangle (10.0,-0.5);
\filldraw[fill=NavyBlue!60!white, draw=black] (0.5,-0.5) rectangle (10.5,-1.0);
\filldraw[fill=NavyBlue!60!white, draw=black] (1.0,-1.0) rectangle (11.0,-1.5);
\filldraw[fill=NavyBlue!30!white, draw=black] (1.5,-1.5) rectangle (11.5,-2.0);
\node [anchor=north, rotate=90](n1) at (6,-2.25) {\footnotesize{$...$}};
\filldraw[fill=NavyBlue!30!white, draw=black] (2.5,-2.5) rectangle (12.5,-3.0);
\filldraw[fill=NavyBlue!60!white, draw=black] (3.0,-3.0) rectangle (13.0,-3.5);
\draw [decorate,decoration={brace,amplitude=10pt, mirror},xshift=0pt,yshift=0pt]
(3.0,-3.5) -- (13.0,-3.5) node [black,midway,xshift=-20pt,yshift=-15pt, text height = 10pt] {\footnotesize{Training set of length $D$ with days $d = N, ..., N + D - 1$.}};

\draw [decorate,decoration={brace,amplitude=10pt},xshift=0pt,yshift=0pt]
(10.0,0.0) -- (10.5,0.0) node [black,midway,yshift=20pt, text height = 10pt] {\footnotesize{Forecast for $d = 366$.}};
\filldraw[fill=red!40!white, draw=black] (10.0,-0.0) rectangle (10.5,-0.5);
\filldraw[fill=red!40!white, draw=black] (10.5,-0.5) rectangle (11.0,-1.0);
\filldraw[fill=red!40!white, draw=black] (11.0,-1.0) rectangle (11.5,-1.5);
\filldraw[fill=red!20!white, draw=black] (11.5,-1.5) rectangle (12.0,-2.0);
\filldraw[fill=red!20!white, draw=black] (12.5,-2.5) rectangle (13.0,-3.0);
\filldraw[fill=red!40!white, draw=black] (13.0,-3.0) rectangle (13.5,-3.5);
\draw [decorate,decoration={brace,amplitude=10pt, mirror},xshift=0pt,yshift=0pt]
(13.0,-3.5) -- (13.5,-3.5) node [black,midway,yshift=-15pt, text height = 10pt] {\footnotesize{Forecast for $d = N + D$.}};

\node [anchor=west](n1) at (-1.5,-0.25) {\footnotesize{$n=1$}};
\node [anchor=west](n1) at (-1.5,-0.75) {\footnotesize{$n=2$}};
\node [anchor=west](n1) at (-1.5,-1.25) {\footnotesize{$n=3$}};
\node [anchor=west](n1) at (-1.5,-1.75) {\footnotesize{$n=4$}};
\node [anchor=north, rotate=90](n1) at (-1.1,-2.25) {\footnotesize{$...$}};
\node [anchor=west](n1) at (-1.5,-2.75) {\footnotesize{$n=N-1$}};
\node [anchor=west](n1) at (-1.5,-3.25) {\footnotesize{$n=N$}};

\end{tikzpicture}
}
\end{center}
\caption[Structure of the rolling window forecasting study.]{Structure of the rolling window forecasting study. Blue denotes in-sample data, red denotes the out-of-sample forecast.}\label{fig:study_structure}
\end{figure}

\review{We employ the well-known rolling window forecasting study design, which is common in energy price forecasting \citep[see e.g.][]{ziel2015, bunn2018, weron2018,janke2019, uniejewski2019, ziel2020a}. This setting reduces the impact of structural breaks within the data and ensures a robust setting for the comparison of predictive performance using the \gls{DM}-test \citep{diebold2002, diebold2015}. The scheme is visualized in Figure \ref{fig:study_structure}.} We train one model for each delivery hour on 365 days of in-sample data and issue forecasts for the next delivery. Subsequently, the training data set is shifted forward by one day, the models are re-trained for each delivery hour and forecasts are issued for the next day and henceforth. Keeping the length of the training set constant we thus move through the test set. Our full data set ranges from January 2016 to August 2020, holding in total $N = 1618$ days. The training set length is fixed to $D = 365$ days. The test set holds $L = 1256$ days. 

\FloatBarrier

\newcommand{\simPID}[2]{\ensuremath{{P}_{\text{ID},#1}^{d,s,[#2]}}}
\newcommand{\simDeltaPID}[2]{\ensuremath{\Delta{}P_{\text{ID},#1}^{d,s,[#2]}}}
\newcommand{\simPIDbold}[1]{\ensuremath{{\boldsymbol{P}}_{\text{ID}}^{d,s, [#1]}}}
\newcommand{\simALPHA}[2]{\ensuremath{{\alpha}_{#1}^{d,s,[#2]}}}

\newcommand{\simParameter}[3]{\ensuremath{\widehat{#1}_{#2}^{d,s,[#3]}}}

Forecasts are issued for all delivery hours $s = 0, ..., 23$. For each delivery hour, the forecast consists of \review{$j = 1, ..., M$ paths with $M = 1000$ paths of $t = 1, ..., 31$ steps.} \review{Generally, let variables with superscript $[j]$ denote simulated values on path $j$ and hence $\simPID{t}{j}$ denotes the simulation for step $t$ for delivery period $d,s$ in the path $j$.} The vector notation $\simPIDbold{j} = (\simPID{1}{j}, ..., \simPID{31}{j})$ is frequently used in the chapter on error metrics. For the simulation of the paths, an algorithm similar to the recursive Euler-Maruyama-Scheme is used \citep{asmussen2007, ziel2020a}. Each simulation starts 185 minutes and ends 30 minutes before the start of physical delivery. 
For each simulation step $t$ and path $j$, the boolean variable $\simALPHA{t}{j}$ is simulated $M$ times from the Bernoulli distribution $B(\simParameter{\pi}{t}{j})$ and the price difference $\simDeltaPID{t}{j}$ is sampled $M$ times from the distribution $F(\simParameter{\boldsymbol{\theta}}{t}{j}) = F(\simParameter{\mu}{t}{j},  \simParameter{\sigma}{t}{j}, \simParameter{\nu}{t}{j}$ and $\simParameter{\tau}{t}{j})$. The price $\simPID{t}{j}$ is then calculated as:
\begin{equation}
\simPID{t}{j} = \simPID{t-1}{j} + \simDeltaPID{t}{j} \cdot \simALPHA{t}{j}.
\end{equation}
The algorithm is visualized in Figure \ref{fig:simulation_algorithm}. The estimates for $\simParameter{\pi}{t}{j}, \simParameter{\mu}{t}{j},  \simParameter{\sigma}{t}{j}, \simParameter{\nu}{t}{j}$ and $\simParameter{\tau}{t}{j}$ for the first step $t = 1$ are all equal, but begin to differ from $t \geq 2$ onwards as the paths develop individually. Therefore, the prediction matrix needs to be updated dynamically for each path and each step.

\begin{figure}[htb]
\begin{center}
\resizebox{\textwidth}{!}{
\begin{tikzpicture}

\node [anchor=south](n1) at (03.5, 0.125) {Predict based on $t-1$.};
\node [anchor=south](n1) at (11.5, 0.125) {Simulate $j= 1, ..., M$.};
\node [anchor=south](n1) at (19.5, 0.125) {Calculate $\simPID{t}{j}$ for each path.};

\filldraw[fill=NavyBlue!40!white, draw=black] (0.0,-0.0) rectangle (7,-1.5) 
	node[pos=.5, text width=6.8cm, align=center] (predict1) {Predict $\simParameter{\pi}{1}{j}$ \\ and $\simParameter{\boldsymbol{\theta}}{1}{j}$};
\filldraw[fill=Red!40!white, draw=black] (8.0,-0.0) rectangle (15,-1.5) 
	node[pos=.5, text width=6.8cm, align=center] (sim1) {$\simALPHA{1}{j} \sim B(\simParameter{\pi}{1}{j})$ \\ $\simDeltaPID{1}{j} \sim F(\simParameter{\boldsymbol{\theta}}{1}{j})$};
\filldraw[fill=SeaGreen!60!white, draw=black] (16.0,-0.0) rectangle (23,-1.5) 
	node[pos=.5, text width=6.8cm,align=center] (calc1){$\simPID{1}{j} = P_{\text{ID}, 0}^{d,s} +  \simDeltaPID{1}{j} \cdot \simALPHA{1}{j}$};
		
\draw[-{Latex[length=5]}] (07.1,-0.75)--(07.9,-0.75) ;
\draw[-{Latex[length=5]}] (15.1,-0.75)--(15.9,-0.75) ;
\draw[-{Latex[length=5]}, color=black, rounded corners=5] (19.5, -1.5) .. controls (19.5, -2.5) and (3.5, -1.5) .. (3.5, -2.5);

\filldraw[fill=NavyBlue!30!white, draw=black] (0.0,-2.5) rectangle (7,-4.0) 
	node[pos=.5, text width=7.8cm, align=center] (predict1) {Predict $\simParameter{\pi}{2}{j}$ \\ and $\simParameter{\boldsymbol{\theta}}{2}{j}$};
\filldraw[fill=Red!30!white, draw=black] (8.0,-2.5) rectangle (15,-4.0) 
	node[pos=.5, text width=7.8cm, align=center] (sim1) {$\simALPHA{2}{j} \sim B(\simParameter{\pi}{2}{j})$ \\ $\simDeltaPID{2}{j} \sim F(\simParameter{\boldsymbol{\theta}}{2}{j})$};
\filldraw[fill=SeaGreen!50!white, draw=black] (16.0,-2.5) rectangle (23,-4.0) 
	node[pos=.5, text width=7.8cm,align=center] (calc1){$\simPID{2}{j} = \simPID{1}{j} + \simDeltaPID{2}{j} \cdot \simALPHA{2}{j}$};

\draw[-{Latex[length=5]}] (7.1,-3.25)--(7.9,-3.25) ;
\draw[-{Latex[length=5]}] (15.1,-3.25)--(15.9,-3.25) ;
\draw[-{Latex[length=5]}, color=black, rounded corners=5] (19.5, -4.0) .. controls (19.5, -5.0) and (3.5, -4.0) .. (3.5, -5);

\filldraw[fill=NavyBlue!20!white, draw=black] (0.0,-5) rectangle (7,-6.5) 
	node[pos=.5, text width=7.8cm, align=center] (predict1) {Predict $\simParameter{\pi}{3}{j}$ \\ and $\simParameter{\boldsymbol{\theta}}{3}{j}$};
\draw[-{Latex[length=5]}] (7.1,-5.75)--(7.9,-5.75) ;

\node [anchor=south](n1) at (11.5, -6.5) {\LARGE{...}};
\node [anchor=south](n1) at (19.5, -6.5) {\LARGE{...}};

\draw[-{Latex[length=5]},dotted, line width=1.5pt, color=black, rounded corners=5] (19.5, -6.5) .. controls (19.5, -7.5) and (3.5, -6.5) .. (3.5, -7.5);

\filldraw[fill=NavyBlue!40!white, draw=black] (0.0,-7.5) rectangle (7,-9) 
	node[pos=.5, text width=7.8cm, align=center] (predict1) {Predict $\simParameter{\pi}{T}{j}$ \\ and $\simParameter{\boldsymbol{\theta}}{T}{j}$};
\filldraw[fill=Red!40!white, draw=black] (8,-7.5) rectangle (15,-9) 
	node[pos=.5, text width=7.8cm, align=center] (sim1) {$\simALPHA{T}{j} \sim B(\simParameter{\pi}{T}{j})$ \\ $\simDeltaPID{T}{j} \sim F(\simParameter{\boldsymbol{\theta}}{T}{j})$};
\filldraw[fill=SeaGreen!60!white, draw=black] (16.0,-7.5) rectangle (23,-9) 
	node[pos=.5, text width=7.8cm, align=center] (calc1){$\simPID{T}{j} = \simPID{T-1}{j} + \simDeltaPID{T}{j} \cdot \simALPHA{T}{j}$};

\draw[-{Latex[length=5]}] (7.1,-8.25)--(7.9,-8.25) ;
\draw[-{Latex[length=5]}] (15.1,-8.25)--(15.9,-8.25) ;

\node [anchor=west](t1) at (-1.5, -0.75) {$t = 1$};
\node [anchor=west](t2) at (-1.5, -3.25) {$t = 2$};
\node [anchor=west](t3) at (-1.5, -5.75) {$t = 3$};
\node [anchor=west](t3) at (-1.5, -8.25) {$t = T$};
\end{tikzpicture}}
\end{center}
\caption[Simulation Algorithm.]{Simulation Algorithm. After $t=3$, the steps until $t=T$ are omitted. For each path $j$, an individual regression matrix is created with the information of the path's past development. Together, the results of the right, green column yield the path vector $\simPIDbold{j}$.} \label{fig:simulation_algorithm}
\end{figure}

\FloatBarrier
\subsection{Forecast Evaluation}

The mean and median trajectory are evaluated using the \gls{RMSE} and \gls{MAE} respectively. For the probabilistic evaluation of the generated scenarios the \gls{ES}, \gls{CRPS} and the empirical coverage ratio are used. Additionally, the \gls{WS} is used to evaluate the coverage of an $(1-\alpha) \cdot 100 \%$-\gls{PI}. The \gls{ES}, \gls{CRPS} and the \gls{WS} are strictly proper scoring rules \citep{weron2018, gneiting2007, ziel2019b}. To draw conclusions about the statistical significance of the difference in forecasting performance for each model, the \gls{DM}-test is used. All measures are widely employed in academia and practice. 

Formally, the \gls{RMSE} and \gls{MAE} are defined as:
\begin{equation}
\text{RMSE} = \sqrt{
	\frac{1}{NST}
	\sum\limits^N_{d=1}
	\sum\limits^S_{s=1}
	\sum\limits^T_{t=1}
	\left(\PID{t} - \frac{1}{M} \sum\limits^M_{j=1} \simPID{t}{j} \right) ^2 }, 
\end{equation}
and:
\begin{equation}
\text{MAE} =
	\frac{1}{NST}
	\sum\limits^N_{d=1}
	\sum\limits^S_{s=1}
	\sum\limits^T_{t=1}
	\left\lvert \PID{t} - \text{med}(\simPID{t}{j}) \right\rvert , 
\end{equation}
For an $(1-\alpha) \cdot 100 \%$-\gls{PI} with the lower and upper bounds $L_t, U_t$ and prediction interval width $\delta_t = \hat{U}_t - \hat{L}_t$, the empirical \gls{CR} is defined as:
\begin{equation}
\text{CR} = 
	\frac{1}{NST}
	\sum\limits^N_{d=1}
	\sum\limits^S_{s=1}
	\sum\limits^T_{t=1}
	\begin{cases}
		1 & \text{for} \; \PID{t} \in [ \hat{L}_t^{d,s}, \hat{U}_t^{d,s} ] \\
		0 & \text{else.} \\
	\end{cases}
\end{equation}
The $\text{WS}_t^{d,s}$ is defined as:
\begin{equation}
\text{WS}_t^{d,s} = 
	\begin{cases}
		\delta_t, & \text{for} \; \PID{t}  \in [ \hat{L}_t^{d,s}, \hat{U}_t^{d,s} ] \\
		\delta_t + \frac{2}{\alpha}(\hat{L}_t^{d,s} - \PID{t}), & \text{for} \; \PID{t}  < \hat{L}_t^{d,s} \\
		\delta_t + \frac{2}{\alpha}(\PID{t} - \hat{U}_t^{d,s}), & \text{for} \; \PID{t}  > \hat{U}_t^{d,s} \\
	\end{cases}
\end{equation}
and aggregated as:
\begin{equation}
\text{WS} = 
	\frac{1}{NST}
	\sum\limits^N_{d=1}
	\sum\limits^S_{s=1}
	\sum\limits^T_{t=1}
	\text{WS}_t^{d,s}.
\end{equation}
For both, \gls{CR} and \gls{WS}, the upper and lower bounds of the $(1- \alpha) \cdot 100 \%$ \gls{PI} are defined by the respective quantiles $\hat{L}_t^{d,s} = Q^{\alpha/2}_{j=1,...,M}(\simPID{t}{j})$ and $\hat{U}_t^{d,s} = Q^{1-\alpha/2}_{j=1,...,M}(\simPID{t}{j})$, where $Q^\tau_{j=1,...,M}(\simPID{t}{j})$ denotes the $\tau$-th quantile of $M$ simulated $\simPID{t}{j}$ prices. Comparing both, \gls{WS} and \gls{CR}, one can see how the \gls{WS} penalizes for an observation outside the interval and rewards the forecaster at the same time for a more narrow \gls{PI}. Contrary to the \gls{CR}, the \gls{WS} is a strictly proper evaluation measure \citep{weron2018}. 

The \gls{CRPS} \cite[see e.g.][]{gneiting2007, weron2018} is approximated by the \gls{PB}
\begin{equation}
\text{CRPS}^{d,s}_t = \frac{1}{R} \sum\limits_{\tau \in \mathcal{T}} \text{PB}^{d,s}_{t, \tau}
\end{equation}
for a dense equidistant grid of probabilities $\mathcal{T} = \{0.01, ... 0.99\}$ of size $R = 99$. $\text{PB}^{d,s}_{t,\tau}$ denotes the pinball loss for probability $\tau$. The formula is given by:
\begin{equation}
\text{PB}^{d,s}_{t, \tau} = 
	\begin{cases}
      (1 - \tau) \cdot (Q^\tau_{j=1,...,M}(\simPID{t}{j}) - P_{t,j}^{d,s}) & \text{for} \; \PID{t} \leq  Q^\tau_{j=1,...,M}(\simPID{t}{j})\\
      \tau \cdot (\PID{t} - Q^\tau_{j=1,...,M}(\simPID{t}{j}))  & \text{else. }\\
    \end{cases}  
\end{equation}
The overall \gls{CRPS} is calculated by the average:
\begin{equation}
\text{CRPS}  =
	\frac{1}{NST}
	\sum\limits^N_{d=1}
	\sum\limits^S_{s=1}
	\sum\limits^T_{t=1}
	\text{CRPS}^{d,s}_t.
\end{equation}
The pinball loss is also used to evaluate the performance of different models in specific quantile levels. For this reason, the \gls{PB} is aggregated as follows:
\begin{equation}
\text{PB}_\tau  =
	\frac{1}{NST}
	\sum\limits^N_{d=1}
	\sum\limits^S_{s=1}
	\sum\limits^T_{t=1}
	\text{PB}^{d,s}_{t, \tau}.
\end{equation}
To measure the quality of the generated paths, \cite{ziel2020a} propose the \gls{ES}. It is a generalisation of the \gls{CRPS} for two dimensions. Thereby, not only the approximation of the marginal distribution is evaluated, but the generated multivariate distribution \citep{gneiting2007, ziel2019b}:
\begin{align}
\begin{split}
\text{ES}^{d,s} 
	=  \frac{1}{M} \sum\limits_{j=1}^M 
\left\lvert \left\lvert \boldsymbol{P}^{d,s}_\text{ID} - \simPIDbold{j} \right\rvert \right\rvert_2 
	- \frac{1}{\cdot M \cdot (M-1)} \sum\limits_{j=1}^M \sum\limits_{i=j+1}^M
\left\lvert \left\lvert \simPIDbold{j} - \simPIDbold{i} \right\rvert \right\rvert_2.
\end{split}
\end{align}
The average yields the overall energy score for each model: 
\begin{equation}
\text{ES}  =
	\frac{1}{NS}
	\sum\limits^N_{d=1}
	\sum\limits^S_{s=1}
	\text{ES}^{d,s}.
\end{equation}

\newcommand{\loss}[3]{\ensuremath{L_{#1}^{#2, #3}}} 
\newcommand{\boldloss}[2]{\ensuremath{\boldsymbol{L}_{#1}^{#2}}} 

\review{The aforementioned measures provide insight in the accuracy of different forecasting models. To evaluate the statistical significance of the difference in forecast accuracy of two models $A$ and $B$, the \gls{DM}-test \citep{diebold2002, diebold2015} is routinely employed in the field of energy price forecasting \citep{weron2018, ziel2018day, janke2019}. It originally stems from the field of point forecasting, however \cite{diebold2015} notes that the test is agnostic to the scoring rule used to evaluate forecasts. Hence, using strictly proper probabilistic scoring rules, such as the \gls{CRPS} and \gls{ES} loss, the \gls{DM} test can be applied to probabilistic forecasts as well \citep[see e.g.][]{diebold2015, weron2018}. Following \cite{ziel2020a} and \cite{ziel2018day}, the \gls{DM}-test is employed in a multivariate fashion. Hence, let $\boldloss{A}{d} = (\loss{A}{d}{1}, ..., \loss{A}{d}{S})$ and $\boldloss{A}{d}  = (\loss{B}{d}{1}, ..., \loss{B}{d}{S})$ denote the out-of-sample loss vectors for model $A$ and $B$ for day $d$ and delivery period $s$ of length $N$. For models $A, B$, the $N \times S$ vector of losses are reduced to an $N \times 1$ vector by taking the 1-norm. The difference between both is the loss differential used in the DM-test.
\begin{equation}\label{eq:loss_series}
\Delta \boldloss{A, B}{d}  = \left\lvert \left\lvert  \boldloss{A}{d} \right\rvert \right\rvert_1 - 
	\left\lvert \left\lvert  \boldloss{B}{d} \right\rvert \right\rvert_1.
\end{equation} 
For example, for the \gls{ES} and the \textbf{Naive} model, the loss vector $\boldsymbol{L}_{\text{Naive}}^d =  (\text{ES}_{\text{Naive}}^{d, 1}, ..., \text{ES}_{\text{Naive}}^{d, S})$.}

\review{We test the loss differential series for stationarity using the augmented Dickey-Fuller (ADF) test \citep{dickey1979, dickey1981} and reject the $H_0$ of unit root at the 5\%{} significance level for all loss differential series.} \cite{harvey1997} propose the usage of the $t$-distribution with $\nu = N-1$ degrees of freedom rather than the normal distribution, as well as the introduction of a bias correction. Formally, the corrected test statistic is defined as 
$t_\text{DM}^{\text{HLN}, h=1} = \sqrt{\frac{N+3}{N}} \cdot t_\text{DM} \sim t(0, 1, N-1) $
under the $H_0$, where $N$ is the length of the loss differential series $\Delta L^d_{A,B}$ and $h$ denotes the forecast horizon. \review{The standard deviation is computed using an autocorrelation-consistent estimator.} For each model pair, two one-sided \gls{DM}-tests are computed. The first test has the $H_0$ that the forecasts of model $A$ are significantly better than the forecasts of model $B$. For the second test, the $H_0$ is that the forecasts of model $B$ are significantly better than the forecasts of model $A$. These tests are complimentary. We use the implementation in the \texttt{R}-package \texttt{forecast} \citep{hyndman2008, rForecast}. 

\FloatBarrier
\section{Results} \label{ch:results}
The following chapter presents the results of the forecasting study. It is split into two parts: First, we show the error metrics for the out-of-sample analysis. Additionally, we show the in-sample coefficients for the model using Johnson's $S_U$ distribution.  

\FloatBarrier
\subsection{Out-of-sample Analysis: Forecasting Performance on Test Data} \label{sec:results_out_of_sample}

First, the aggregate error statistics will be presented, followed by the scoring rules considering the marginal fit relative to the time to delivery and the quantile range $\mathcal{T}$. Statistical significance is evaluated using the Diebold-Mariano test. 

The \textbf{Naive} performs best in terms of RMSE and MAE, while \textbf{Mix.JSU} performs best across the probabilistic evaluation using the CRPS and ES scoring rules. Its superior performance in terms of ES is statistically significant according to the \gls{DM}-test. \review{The GAMLSS-based model assuming the skew-$t$ distribution however shows a very poor performance for hour 6, which yields an overall poor performance. For this delivery period, we can trace the high error back to outliers and extreme $\deltaPID{t}$ larger than 2000 EUR/MWh on March 11th, 2020. This indicates that Johnson's $S_U$ is more robust towards outliers. An investigation of the loss time series for \textbf{Mix.JSU} and \textbf{Mix.SST} shows the deteriorating forecasting performance of the \textbf{Mix.SST} after March 11th, 2020 clearly (see Figure \ref{fig:error_outliers} in Appendix \ref{app:plots}). The \textbf{Auto.ARIMA} performs surprisingly bad in terms of the RMSE and MAE and somewhat better in terms of the CRPS and ES. With respect to the other benchmark models, we see an overall mixed performance. We note a worse performance for the benchmark models for the probabilistic measures CRPS and ES compared to the \textbf{Naive} and \textbf{Mix.JSU}.
}

\begin{table}[H]
\caption[Aggregate error statistics for all scoring rules.]{
	Aggregate error statistics for the \gls{MAE}, \gls{RMSE}, \gls{CRPS}, \gls{ES} and the \gls{CR} and \gls{WS} for the 50\%, 90\% and 99\% prediction interval. Colour indicates performance. The best value for each scoring rule is \textbf{\underline{highlighted}}.
}
\label{tab:error_statistics}
\begin{center}
\resizebox{\textwidth}{!}{\begin{tabular}{lrrrrrrrrrr}
\toprule
\multicolumn{1}{c}{ } & \multicolumn{4}{c}{Aggregate Statistics} & \multicolumn{3}{c}{Coverage Ratio} & \multicolumn{3}{c}{Winkler Score} \\
\cmidrule(l{3pt}r{3pt}){2-5} \cmidrule(l{3pt}r{3pt}){6-8} \cmidrule(l{3pt}r{3pt}){9-11}
 & MAE & RMSE & CRPS & ES & $	\text{CR}_{0.5}$ & $	\text{CR}_{0.9}$ & $	\text{CR}_{0.99}$ & $	\text{WS}_{0.5}$ & $	\text{WS}_{0.9}$ & $	\text{WS}_{0.99}$\\ 
 \midrule
Naive & {\cellcolor[HTML]{FDE725}} \color[HTML]{000000} \underline{\bfseries 3.178} & {\cellcolor[HTML]{FDE725}} \color[HTML]{000000} \underline{\bfseries 6.564} & {\cellcolor[HTML]{D5E21A}} \color[HTML]{000000} 1.222 & {\cellcolor[HTML]{8ED645}} \color[HTML]{000000} 17.271 & {\cellcolor[HTML]{FDE725}} \color[HTML]{000000} \underline{\bfseries 0.491} & {\cellcolor[HTML]{FDE725}} \color[HTML]{000000} \underline{\bfseries 0.892} & {\cellcolor[HTML]{E7E419}} \color[HTML]{000000} 0.984 & {\cellcolor[HTML]{29AF7F}} \color[HTML]{F1F1F1} 8.676 & {\cellcolor[HTML]{440154}} \color[HTML]{F1F1F1} 12.922 & {\cellcolor[HTML]{440154}} \color[HTML]{F1F1F1} 36.218 \\
Auto.ARIMA & {\cellcolor[HTML]{39558C}} \color[HTML]{F1F1F1} 3.295 & {\cellcolor[HTML]{440154}} \color[HTML]{F1F1F1} 7.240 & {\cellcolor[HTML]{440154}} \color[HTML]{F1F1F1} 1.313 & {\cellcolor[HTML]{440154}} \color[HTML]{F1F1F1} 18.623 & {\cellcolor[HTML]{56C667}} \color[HTML]{000000} 0.407 & {\cellcolor[HTML]{440154}} \color[HTML]{F1F1F1} 0.713 & {\cellcolor[HTML]{440154}} \color[HTML]{F1F1F1} 0.845 & {\cellcolor[HTML]{440154}} \color[HTML]{F1F1F1} 9.544 & {\cellcolor[HTML]{FDE725}} \color[HTML]{000000} \underline{\bfseries 9.269} & {\cellcolor[HTML]{FDE725}} \color[HTML]{000000} \underline{\bfseries 13.656} \\
MV.N & {\cellcolor[HTML]{C0DF25}} \color[HTML]{000000} 3.193 & {\cellcolor[HTML]{F4E61E}} \color[HTML]{000000} 6.570 & {\cellcolor[HTML]{48186A}} \color[HTML]{F1F1F1} 1.275 & {\cellcolor[HTML]{440154}} \color[HTML]{F1F1F1} 18.009 & {\cellcolor[HTML]{25858E}} \color[HTML]{F1F1F1} 0.680 & {\cellcolor[HTML]{B5DE2B}} \color[HTML]{000000} 0.927 & {\cellcolor[HTML]{B0DD2F}} \color[HTML]{000000} 0.972 & {\cellcolor[HTML]{440154}} \color[HTML]{F1F1F1} 9.364 & {\cellcolor[HTML]{440154}} \color[HTML]{F1F1F1} 16.392 & {\cellcolor[HTML]{440154}} \color[HTML]{F1F1F1} 25.500 \\
MV.t & {\cellcolor[HTML]{CAE11F}} \color[HTML]{000000} 3.191 & {\cellcolor[HTML]{F4E61E}} \color[HTML]{000000} 6.570 & {\cellcolor[HTML]{2CB17E}} \color[HTML]{F1F1F1} 1.240 & {\cellcolor[HTML]{1FA188}} \color[HTML]{F1F1F1} 17.495 & {\cellcolor[HTML]{2DB27D}} \color[HTML]{F1F1F1} 0.622 & {\cellcolor[HTML]{AADC32}} \color[HTML]{000000} 0.930 & {\cellcolor[HTML]{EFE51C}} \color[HTML]{000000} 0.986 & {\cellcolor[HTML]{31678E}} \color[HTML]{F1F1F1} 8.805 & {\cellcolor[HTML]{440154}} \color[HTML]{F1F1F1} 15.972 & {\cellcolor[HTML]{440154}} \color[HTML]{F1F1F1} 33.279 \\
RW.N & {\cellcolor[HTML]{7FD34E}} \color[HTML]{000000} 3.209 & {\cellcolor[HTML]{E5E419}} \color[HTML]{000000} 6.577 & {\cellcolor[HTML]{440154}} \color[HTML]{F1F1F1} 1.472 & {\cellcolor[HTML]{440154}} \color[HTML]{F1F1F1} 20.030 & {\cellcolor[HTML]{440154}} \color[HTML]{F1F1F1} 0.824 & {\cellcolor[HTML]{2FB47C}} \color[HTML]{F1F1F1} 0.970 & {\cellcolor[HTML]{FDE725}} \color[HTML]{000000} \underline{\bfseries 0.989} & {\cellcolor[HTML]{440154}} \color[HTML]{F1F1F1} 12.098 & {\cellcolor[HTML]{440154}} \color[HTML]{F1F1F1} 25.726 & {\cellcolor[HTML]{440154}} \color[HTML]{F1F1F1} 39.947 \\
RW.t & {\cellcolor[HTML]{B8DE29}} \color[HTML]{000000} 3.195 & {\cellcolor[HTML]{28AE80}} \color[HTML]{F1F1F1} 6.686 & {\cellcolor[HTML]{440154}} \color[HTML]{F1F1F1} 1.323 & {\cellcolor[HTML]{440154}} \color[HTML]{F1F1F1} 18.393 & {\cellcolor[HTML]{3C508B}} \color[HTML]{F1F1F1} 0.748 & {\cellcolor[HTML]{34B679}} \color[HTML]{F1F1F1} 0.968 & {\cellcolor[HTML]{E7E419}} \color[HTML]{000000} 0.996 & {\cellcolor[HTML]{440154}} \color[HTML]{F1F1F1} 9.888 & {\cellcolor[HTML]{440154}} \color[HTML]{F1F1F1} 23.068 & {\cellcolor[HTML]{440154}} \color[HTML]{F1F1F1} 63.616 \\
RW.t.Mix.D & {\cellcolor[HTML]{C5E021}} \color[HTML]{000000} 3.192 & {\cellcolor[HTML]{95D840}} \color[HTML]{000000} 6.616 & {\cellcolor[HTML]{440154}} \color[HTML]{F1F1F1} 1.304 & {\cellcolor[HTML]{440154}} \color[HTML]{F1F1F1} 18.171 & {\cellcolor[HTML]{33628D}} \color[HTML]{F1F1F1} 0.726 & {\cellcolor[HTML]{3BBB75}} \color[HTML]{F1F1F1} 0.964 & {\cellcolor[HTML]{E7E419}} \color[HTML]{000000} 0.996 & {\cellcolor[HTML]{440154}} \color[HTML]{F1F1F1} 9.624 & {\cellcolor[HTML]{440154}} \color[HTML]{F1F1F1} 21.941 & {\cellcolor[HTML]{440154}} \color[HTML]{F1F1F1} 61.102 \\
Mix.JSU & {\cellcolor[HTML]{EFE51C}} \color[HTML]{000000} 3.182 & {\cellcolor[HTML]{F1E51D}} \color[HTML]{000000} 6.571 & {\cellcolor[HTML]{FDE725}} \color[HTML]{000000} \underline{\bfseries 1.218} & {\cellcolor[HTML]{FDE725}} \color[HTML]{000000} \underline{\bfseries 17.127} & {\cellcolor[HTML]{27AD81}} \color[HTML]{F1F1F1} 0.628 & {\cellcolor[HTML]{6ECE58}} \color[HTML]{000000} 0.947 & {\cellcolor[HTML]{FDE725}} \color[HTML]{000000} 0.991 & {\cellcolor[HTML]{FDE725}} \color[HTML]{000000} \underline{\bfseries 8.519} & {\cellcolor[HTML]{440154}} \color[HTML]{F1F1F1} 16.763 & {\cellcolor[HTML]{440154}} \color[HTML]{F1F1F1} 31.549 \\
Mix.SST & {\cellcolor[HTML]{440154}} \color[HTML]{F1F1F1} 4.509 & {\cellcolor[HTML]{440154}} \color[HTML]{F1F1F1} 109.661 & {\cellcolor[HTML]{440154}} \color[HTML]{F1F1F1} 1.748 & {\cellcolor[HTML]{440154}} \color[HTML]{F1F1F1} 28.247 & {\cellcolor[HTML]{42BE71}} \color[HTML]{F1F1F1} 0.604 & {\cellcolor[HTML]{93D741}} \color[HTML]{000000} 0.936 & {\cellcolor[HTML]{FDE725}} \color[HTML]{000000} 0.989 & {\cellcolor[HTML]{440154}} \color[HTML]{F1F1F1} 12.427 & {\cellcolor[HTML]{440154}} \color[HTML]{F1F1F1} 17.356 & {\cellcolor[HTML]{440154}} \color[HTML]{F1F1F1} 33.674 \\
\bottomrule
\end{tabular}
}
\end{center}
\end{table}

\begin{figure}[htb]
\begin{subfigure}[c]{0.5\textwidth}
\includegraphics[width=\textwidth]{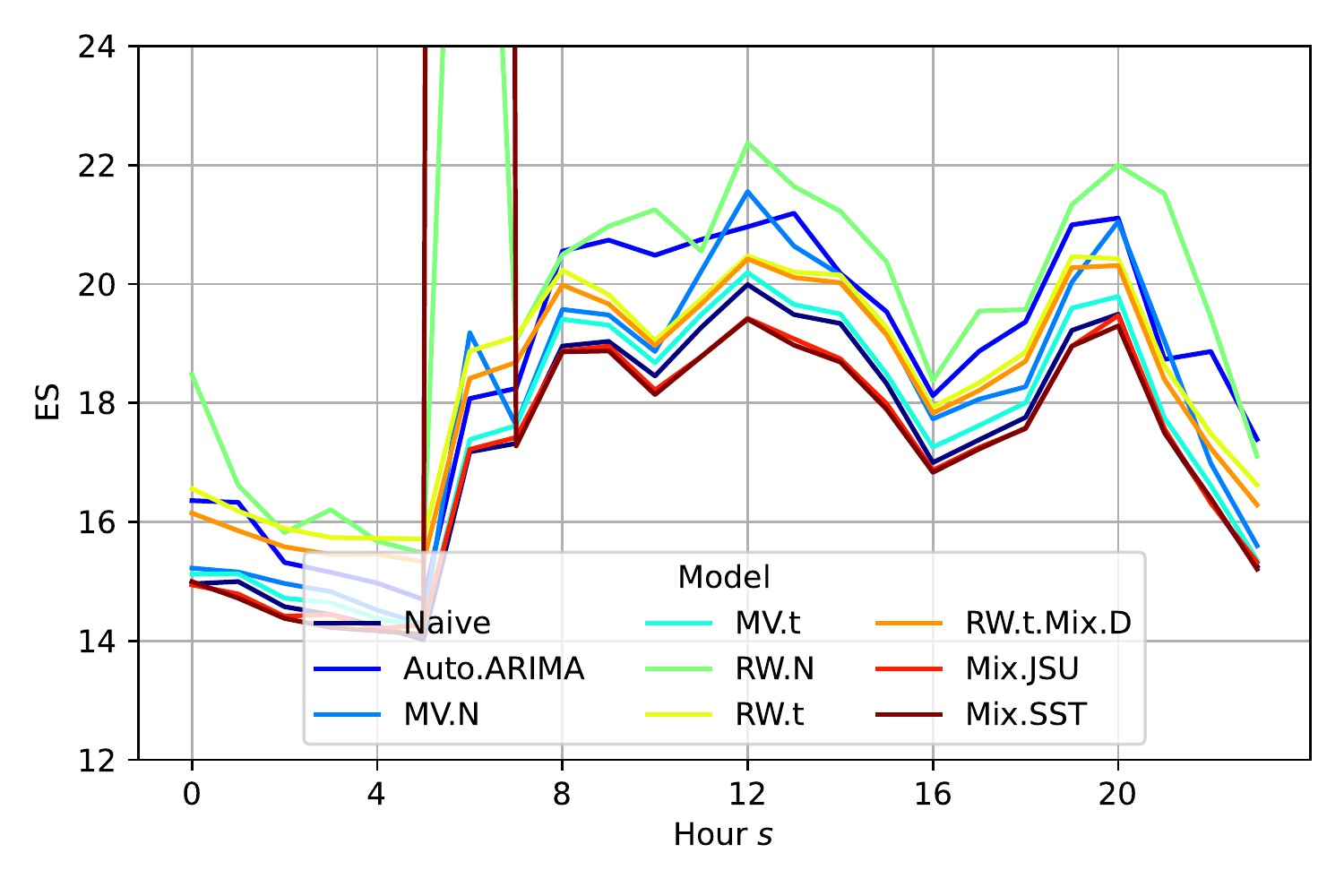}
\subcaption{\gls{ES} over hours $s=0,\ldots, 23$.}
\end{subfigure}%
\begin{subfigure}[c]{0.5\textwidth}
\includegraphics[width=\textwidth]{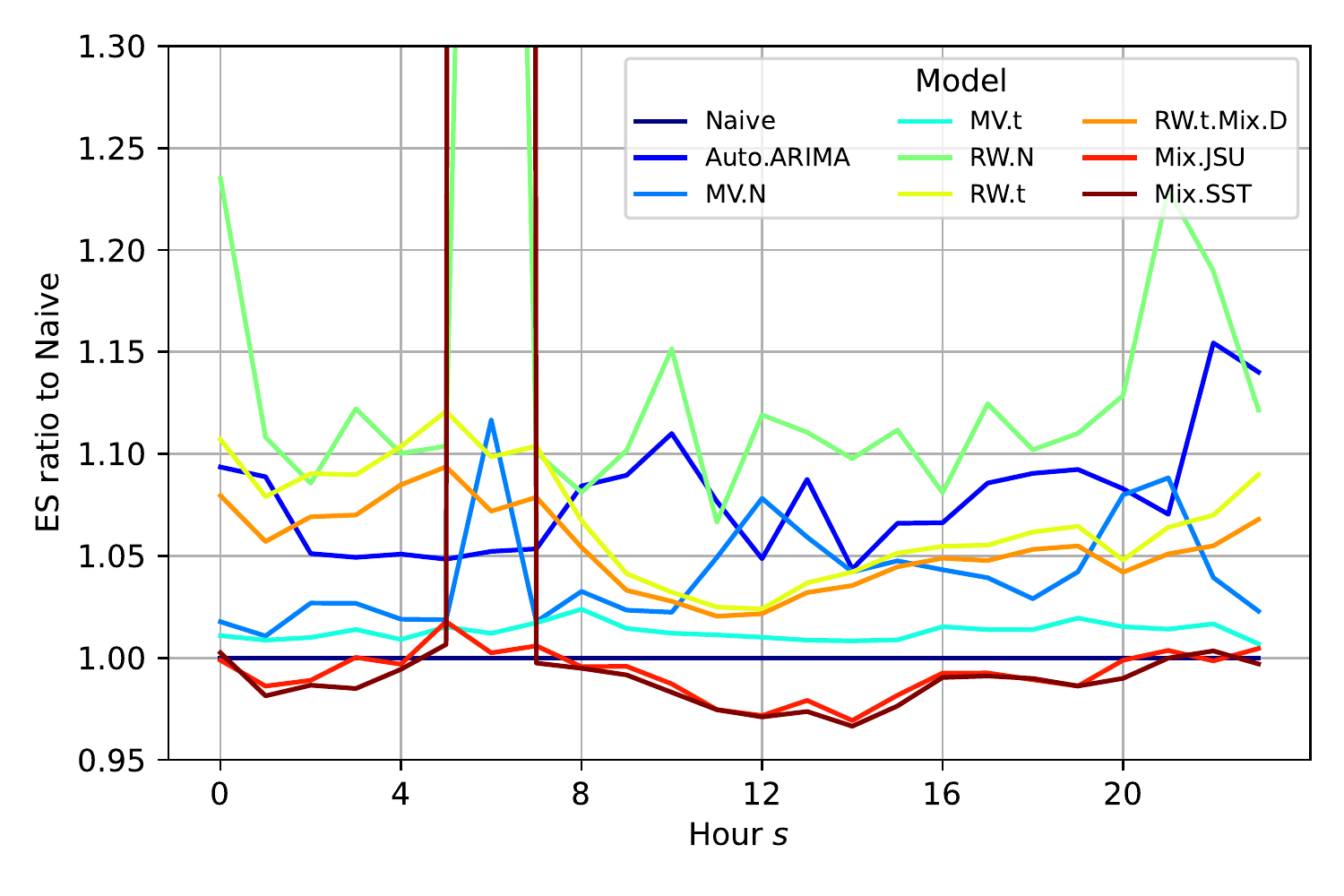}
\subcaption{Relative to \textbf{Naive} over hours $s=0,\ldots, 23$.}
\end{subfigure}%
\caption[Energy Score (ES) and its ratio to \textbf{Naive} over the delivery hours $s$.]{Plot (a) shows the \gls{ES} and (b) its ratio to \textbf{Naive} over the delivery hours $s$.} \label{fig:errors_es_s}
\end{figure}

Figure \ref{fig:errors_es_s} shows the ES relative to the delivery hours $s$. Relative to the \textbf{Naive}, the \gls{GAMLSS}-based models show an improved forecasting performance in the peak hours (Plot \ref{fig:errors_es_s} b). The error of the \textbf{RW.N} and \textbf{Mix.SST} models explodes for hour $s=6$.

\begin{figure}[htb]
\begin{subfigure}[c]{0.5\textwidth}
\includegraphics[width=\textwidth]{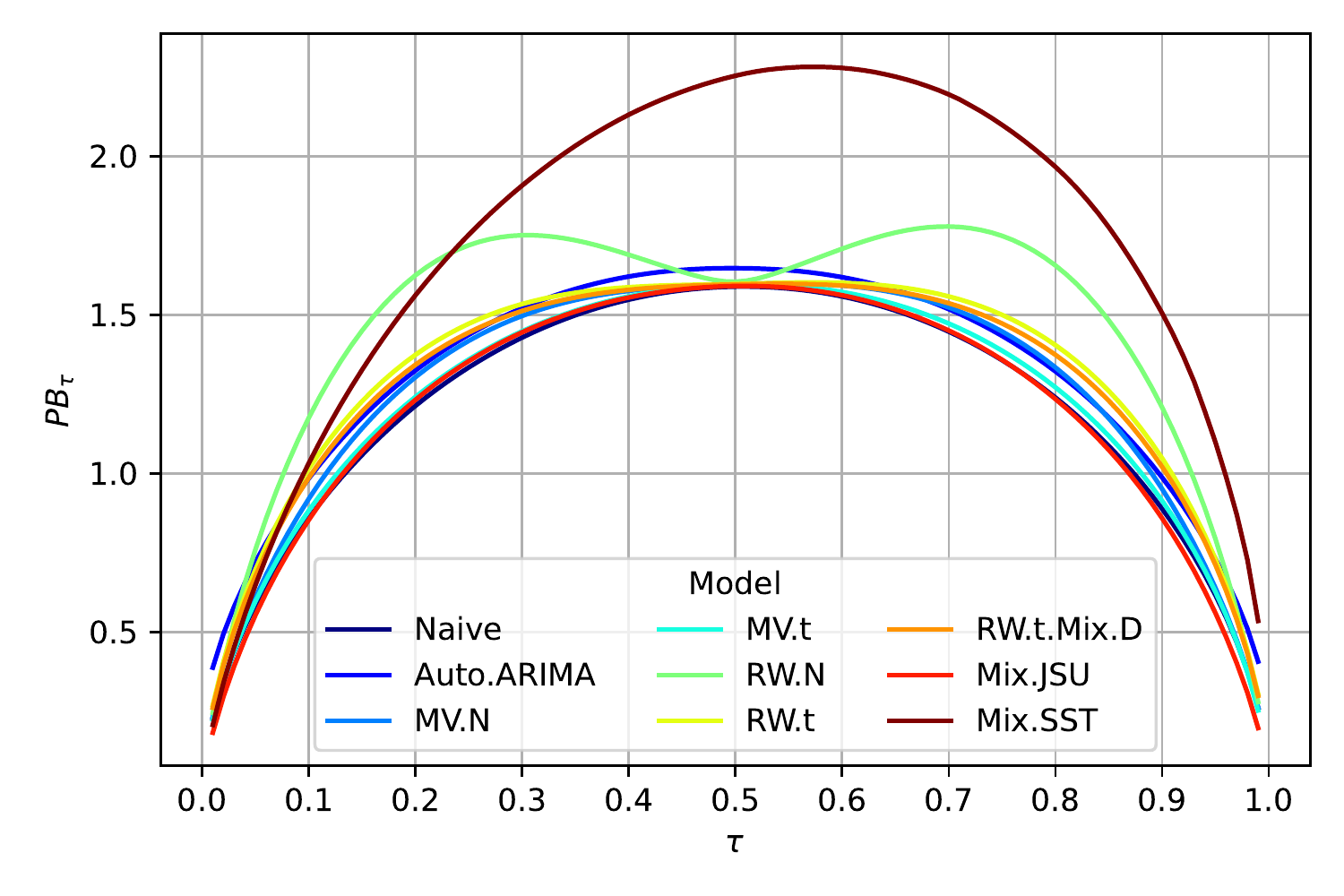}
\subcaption{$\text{PB}_\tau$.}
\end{subfigure}%
\begin{subfigure}[c]{0.5\textwidth}
\includegraphics[width=\textwidth]{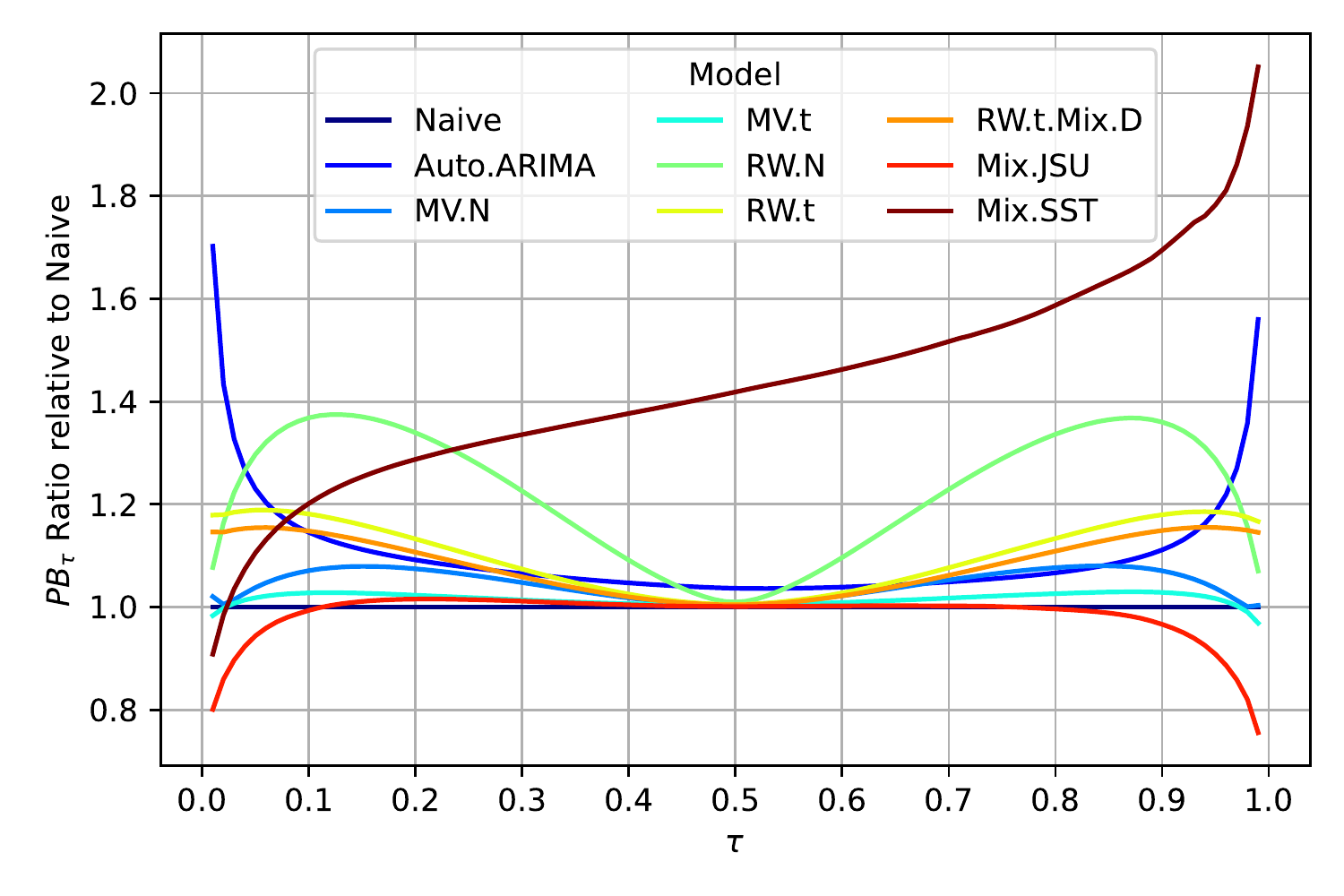}
\subcaption{Relative to \textbf{Naive}.}
\end{subfigure}
\caption[PB and its ratio to \textbf{Naive} over the quantile range $\mathcal{T}$.]{Plot (a) shows the \gls{PB} and (b) its ratio to \textbf{Naive} over the quantile range $\mathcal{T}$.} \label{fig:errors_pinball_tau}
\end{figure}

\begin{figure}[htb]
\begin{subfigure}[c]{0.5\textwidth}
\includegraphics[width=\textwidth]{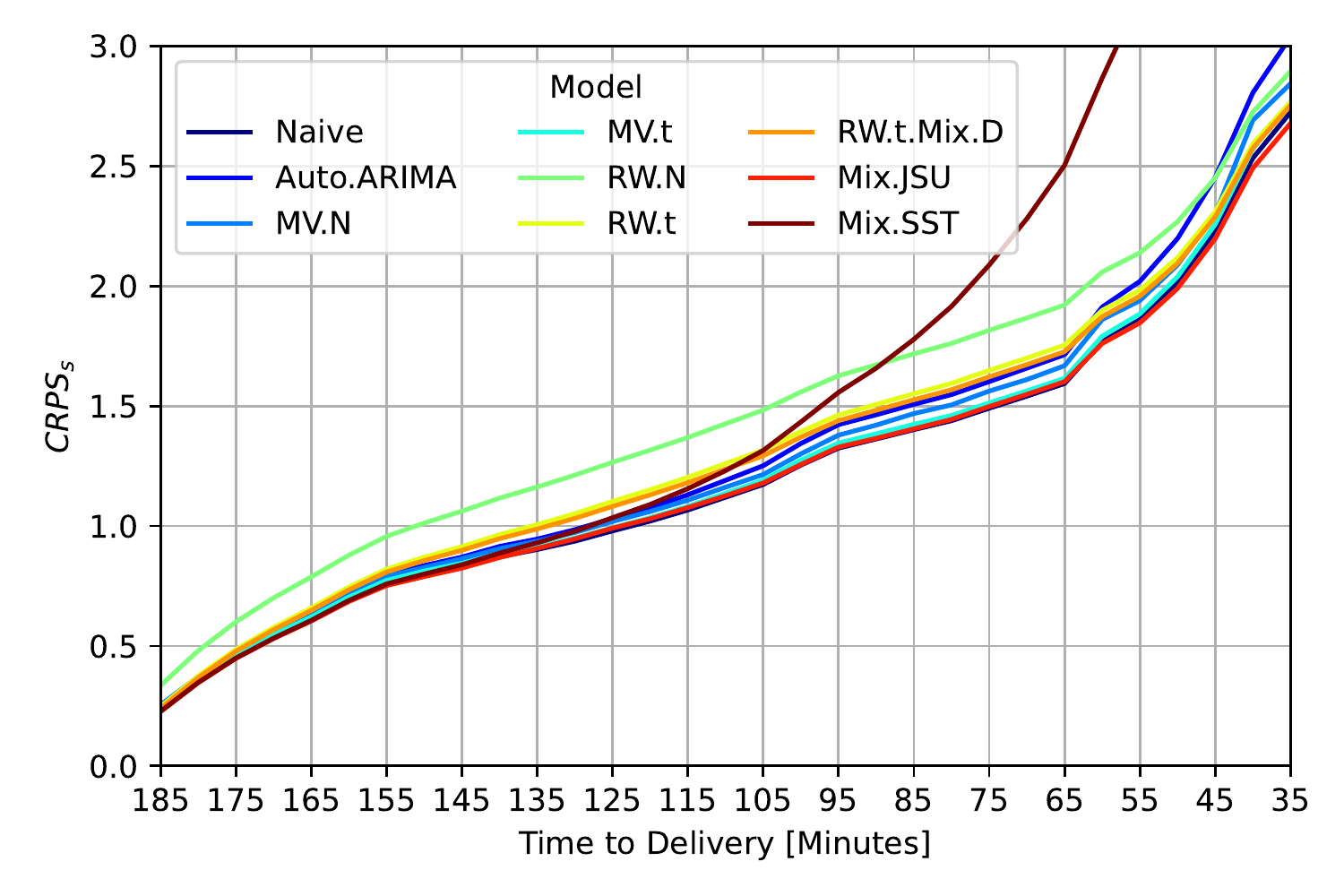}
\subcaption{$\text{\gls{CRPS}}_t$ over $t$.}
\end{subfigure}%
\begin{subfigure}[c]{0.5\textwidth}
\includegraphics[width=\textwidth]{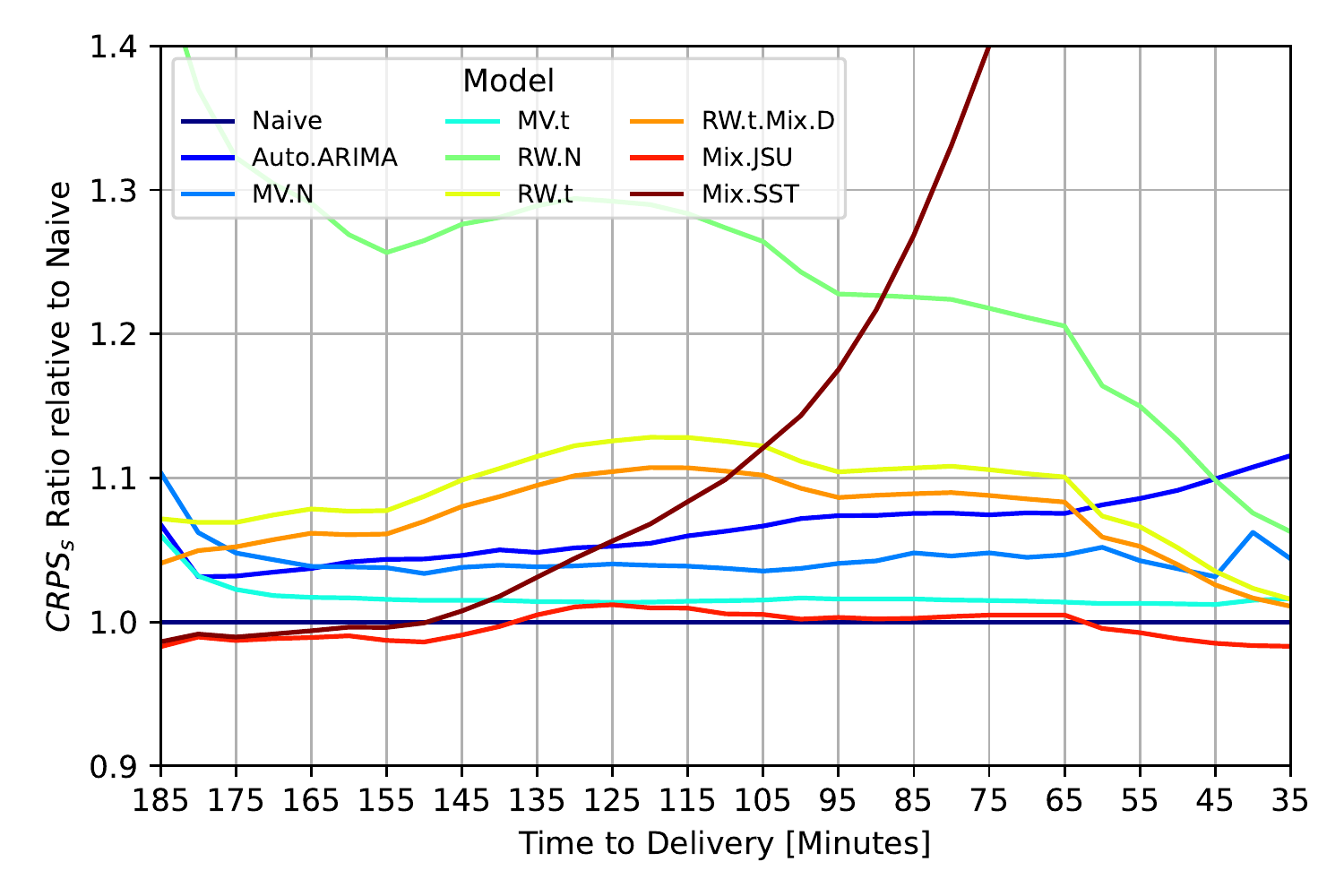}
\subcaption{Relative to \textbf{Naive}.}
\end{subfigure}%
\caption[CRPS and its ratio to \textbf{Naive} over the time to delivery.]{Plot (a) shows the CRPS and (b) its ratio to \textbf{Naive} over the time to delivery.} \label{fig:errors_crps_t}
\end{figure}

The \gls{PB} over $\mathcal{T} = 0.01, ..., 0.99$ is shown in Figure \ref{fig:errors_pinball_tau}. Again, subfigure (a) represents absolute values and (b) depicts all models relative to the \textbf{Naive}. All models show similar performance in the central quantiles, as already indicated by the very close values for the \gls{MAE}. Relative to the \textbf{Naive}, most other benchmark models show worse performance in the tails of the distribution. The \textbf{Mix.JSU} shows an improved modelling of the tails compared to the \textbf{Naive}. The \textbf{Mix.SST} again shows a weak performance given by its sensitivity to outliers. The development of the \gls{CRPS} throughout the simulation window is shown in Figure \ref{fig:errors_crps_t}. Again, (a) shows absolute values while (b) shows the error relative to \textbf{Naive}. The CRPS is rising through the simulation window, especially for the last 60 to 30 minutes of trading. \review{The relative error of most models towards the \textbf{Naive} decreases throughout the simulation window, however, it increases for the \textbf{Auto.ARIMA}. This might indicate that learning the model parameters of past trading sessions can be beneficial compared to learning the parameters only from the trading session of interest, before the start of the simulation, as market behaviour changes throughout the session.}

\begin{figure}[htb]
\begin{subfigure}[c]{0.5\textwidth}
\includegraphics[width=\textwidth]{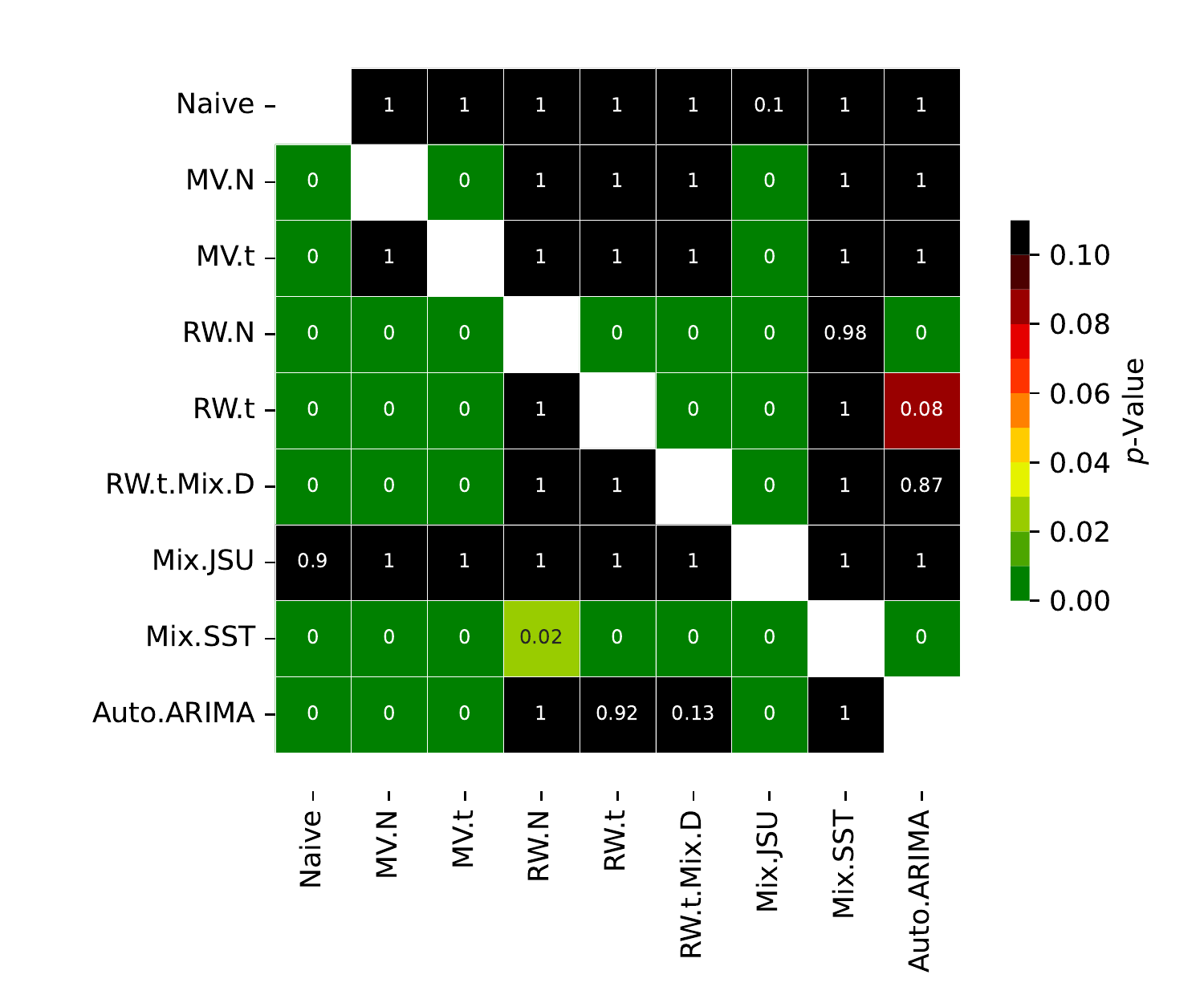}
\subcaption{Continuous Ranked Probability Score.}
\end{subfigure}%
\begin{subfigure}[c]{0.5\textwidth}
\includegraphics[width=\textwidth]{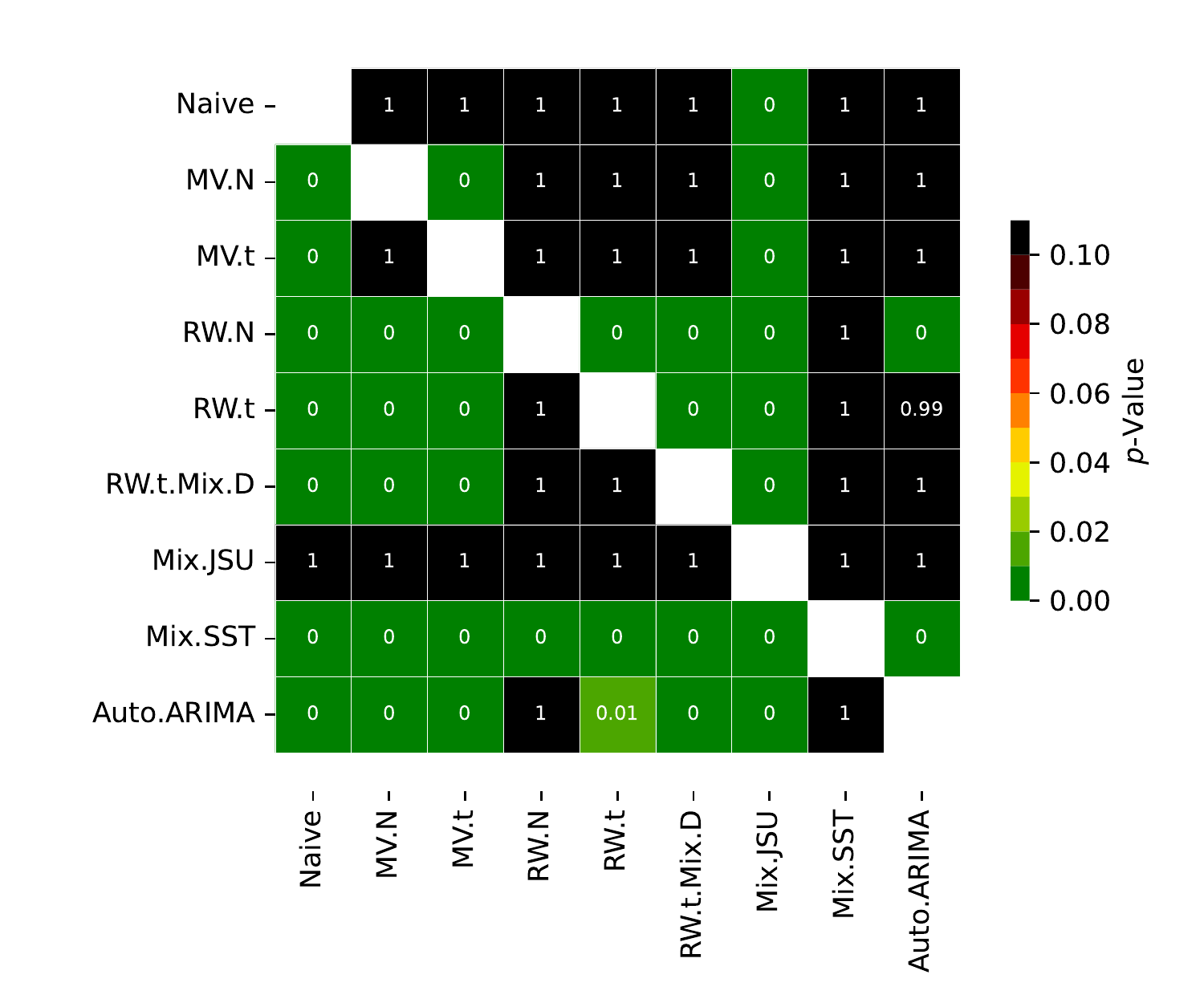}
\subcaption{Energy Score.}
\end{subfigure}
\caption[Pairwise $p$-values for the DM-test for the CRPS and ES loss.]{Pairwise $p$-values for the \gls{DM}-test for the \gls{CRPS} and \gls{ES} loss. The closer the $p$-value to 0, the more significant is the difference between the model on the column (better) and the model on the row (worse).} \label{fig:errors_diebold_mariano}
\end{figure}

The results are largely confirmed as statistically significant by the Diebold-Mariano-Test. Figure \ref{fig:errors_diebold_mariano} shows the $p$-values for the pairwise \gls{DM}-tests for the \gls{ES} and \gls{CRPS}. The lower the $p$-value, the more significant is the difference in the forecasting performance, which implies that the model on the column (or $x$-axis) outputs superior forecasts than the model on the row (or $y$-axis). Generally, the $p$-values for the \gls{CRPS} and \gls{ES} are rather close. This makes sense, as a good coupling to the path's distribution should be closely related to a good fit on the marginal distribution. The other way, however, is not necessarily true. Inside the group of the benchmark models, the \textbf{Naive} model is confirmed as the superior model as it yields significantly better forecasting performance than all other benchmark models. \review{The \textbf{Auto.ARIMA} is significantly better as the \textbf{RW.N} only.} The \textbf{Mix.JSU} yields significantly better forecasting performance than all other models in terms of the ES. The \textbf{Mix.SST} yields significantly worse forecasting accuracy in terms of both CRPS and ES than all other models, which is expected given the results shown in Figures \ref{fig:errors_es_s} to \ref{fig:errors_crps_t}.

\FloatBarrier
\subsection{In-sample Analysis: Estimated Coefficients and their Development}

Given the strong probabilistic forecasting performance of the \textbf{Mix.JSU} we turn to an in-sample analysis of the estimated coefficients. \review{Compared to black-box deep learning algorithms, the parametric \gls{GAMLSS} framework used in this paper allows for explainable machine learning by quantitatively and qualitatively analysing the estimated coefficients. Hence, we can gain further insight in the driving variables for all distribution parameters.} Tables \ref{tab:coefficients_mu} to \ref{tab:coefficients_tau} present the estimated \review{scaled} coefficients for $d$ = January 22nd, 2017, the first out-of-sample day. \review{Scaled coefficient correspond to mean-variance scaled inputs. Hence, the coefficients are hence unit-free and can be compared in the magnitude.} The background colouring indicates the share of non-zero estimates for the whole out-of-sample data set. Green indicates that few estimates are set to zero by the sparsity property of the LASSO, the darker the red, the more estimates are set to zero. 

For $\mu$, only the first lag of $\deltaPID{t}$ shows more than a couple non-zero values for the first day of the test set. This variable yields non-zero estimates as well across the test set for the late morning to afternoon peak hours. This result is similar to the findings of \cite{ziel2020b}, who find the most recent price to be among the most important features for forecasting the $\text{ID}_3$ as well as with the results of \cite{kiesel2020a, kiesel2020b}, who find that lagged prices are an important predictor. The fact that other fundamental and trading related information, especially intraday forecast changes, do not yield additional predictive power suggests that this information is contained in the price already. These results support the notion of weak-form market efficiency already indicated by \cite{ziel2020b} and \cite{kuppelwieser2021}.

\begin{figure}[htb]
\includegraphics[width=\textwidth]{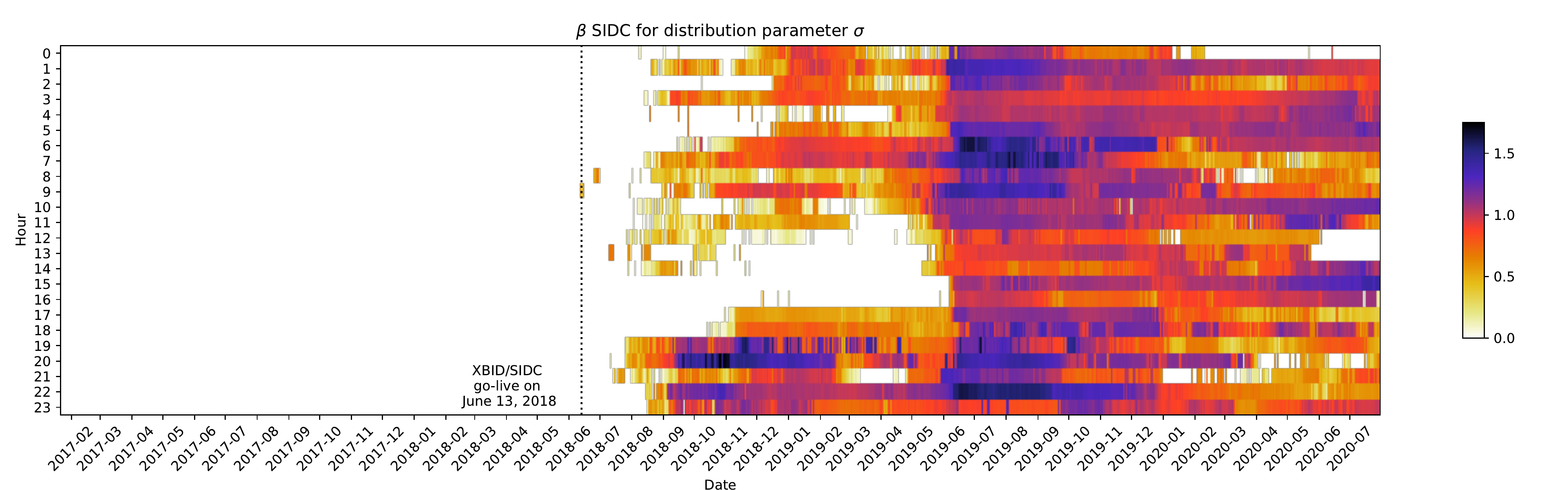}
\caption{Estimated coefficients for the SIDC dummy across the test set.} \label{fig:sidc_dummy_fig}
\end{figure}

For the volatility $\sigma$, we present coefficients in similar fashion in Table \ref{tab:coefficients_sigma}. For the first day of the test set, we yield non-zero estimates for the coefficients for the merit-order slope, for the intercept and for the transformed time to delivery. For a few hours, the coefficient for lagged values of $\ALPHA{t}$ has a negative non-zero estimate as well. The large and positive coefficients for the merit-order slope confirm our initial assumption that the shape of the merit-order is a driving factor for the volatility in intraday markets. Intuitively, this is derived from the observation that on a steep merit-order, a slight change in supply or demand has a higher impact on the price than in a flat regime. Moving this effect from a threshold variable for the size of $\deltaPID{t}$ to the volatility parameter of the distribution of $\deltaPID{t}$ thus generalizes the results of \cite{kiesel2020a, kiesel2020b}. Contrary to \cite{baule2021}, we find little predictive power for the spread between spot and intraday price as well as for the fundamental forecasts and their intraday changes for the volatility. Remember that the coefficients in Table \ref{tab:coefficients_sigma} correspond to January 22nd, 2017, well before the introduction of SIDC. Thus, the SIDC variable is zero for this training period. We show the evolution of the estimated coefficient across the rolling training set in Figure \ref{fig:sidc_dummy_fig}. After the launch of SIDC on June 13, 2018, the dummy is first included in the rolling training set. A sizeable positive estimate is visible, i.e. the volatility rises after gate closure of the cross-border shared order books 60 minutes before delivery. The effect is the strongest in 2019 and 2020 for the morning and afternoon peak hours and less clear for the solar peak hours around noon. Our findings are consistent with \cite{ziel2020a} and contradict \cite{kath2019}, who finds no evidence of rising volatility due to SIDC. 

\begin{table}[H]
\begin{center}
\caption{Estimated \review{scaled} coefficients for $\mu$ (expected value) on the first day of the test set.}
\label{tab:coefficients_mu}
\setlength{\tabcolsep}{2pt}
\resizebox{\textwidth}{!}{
\begin{tabular}{l>{\raggedright\arraybackslash}p{1.25cm}>{\raggedright\arraybackslash}p{1.25cm}>{\raggedright\arraybackslash}p{1.25cm}>{\raggedright\arraybackslash}p{1.25cm}>{\raggedright\arraybackslash}p{1.25cm}>{\raggedright\arraybackslash}p{1.25cm}>{\raggedright\arraybackslash}p{1.25cm}>{\raggedright\arraybackslash}p{1.25cm}>{\raggedright\arraybackslash}p{1.25cm}>{\raggedright\arraybackslash}p{1.25cm}>{\raggedright\arraybackslash}p{1.25cm}>{\raggedright\arraybackslash}p{1.25cm}>{\raggedright\arraybackslash}p{1.25cm}>{\raggedright\arraybackslash}p{1.25cm}>{\raggedright\arraybackslash}p{1.25cm}>{\raggedright\arraybackslash}p{1.25cm}>{\raggedright\arraybackslash}p{1.25cm}>{\raggedright\arraybackslash}p{1.25cm}>{\raggedright\arraybackslash}p{1.25cm}>{\raggedright\arraybackslash}p{1.25cm}>{\raggedright\arraybackslash}p{1.25cm}>{\raggedright\arraybackslash}p{1.25cm}>{\raggedright\arraybackslash}p{1.25cm}>{\raggedright\arraybackslash}p{1.25cm}}
\toprule
  & 0 & 1 & 2 & 3 & 4 & 5 & 6 & 7 & 8 & 9 & 10 & 11 & 12 & 13 & 14 & 15 & 16 & 17 & 18 & 19 & 20 & 21 & 22 & 23\\
\bottomrule
\end{tabular}
}
\end{center}
\end{table}

\begin{table}[H]
\begin{center}
\caption{Estimated \review{scaled} coefficients for $\sigma$ (scale / volatility) on the first day of the test set.}\label{tab:coefficients_sigma}
\resizebox{\textwidth}{!}{
\begin{tabular}{l>{\raggedright\arraybackslash}p{1.25cm}>{\raggedright\arraybackslash}p{1.25cm}>{\raggedright\arraybackslash}p{1.25cm}>{\raggedright\arraybackslash}p{1.25cm}>{\raggedright\arraybackslash}p{1.25cm}>{\raggedright\arraybackslash}p{1.25cm}>{\raggedright\arraybackslash}p{1.25cm}>{\raggedright\arraybackslash}p{1.25cm}>{\raggedright\arraybackslash}p{1.25cm}>{\raggedright\arraybackslash}p{1.25cm}>{\raggedright\arraybackslash}p{1.25cm}>{\raggedright\arraybackslash}p{1.25cm}>{\raggedright\arraybackslash}p{1.25cm}>{\raggedright\arraybackslash}p{1.25cm}>{\raggedright\arraybackslash}p{1.25cm}>{\raggedright\arraybackslash}p{1.25cm}>{\raggedright\arraybackslash}p{1.25cm}>{\raggedright\arraybackslash}p{1.25cm}>{\raggedright\arraybackslash}p{1.25cm}>{\raggedright\arraybackslash}p{1.25cm}>{\raggedright\arraybackslash}p{1.25cm}>{\raggedright\arraybackslash}p{1.25cm}>{\raggedright\arraybackslash}p{1.25cm}>{\raggedright\arraybackslash}p{1.25cm}}
\toprule
  & 0 & 1 & 2 & 3 & 4 & 5 & 6 & 7 & 8 & 9 & 10 & 11 & 12 & 13 & 14 & 15 & 16 & 17 & 18 & 19 & 20 & 21 & 22 & 23\\
\bottomrule
\end{tabular}
}
\end{center}
\end{table}

\begin{table}[H]
\begin{center}
\caption{Estimated \review{scaled} coefficients for $\nu$ (skewness) on the first day of the test set.}\label{tab:coefficients_nu}
\resizebox{\textwidth}{!}{
\begin{tabular}{l>{\raggedright\arraybackslash}p{1.25cm}>{\raggedright\arraybackslash}p{1.25cm}>{\raggedright\arraybackslash}p{1.25cm}>{\raggedright\arraybackslash}p{1.25cm}>{\raggedright\arraybackslash}p{1.25cm}>{\raggedright\arraybackslash}p{1.25cm}>{\raggedright\arraybackslash}p{1.25cm}>{\raggedright\arraybackslash}p{1.25cm}>{\raggedright\arraybackslash}p{1.25cm}>{\raggedright\arraybackslash}p{1.25cm}>{\raggedright\arraybackslash}p{1.25cm}>{\raggedright\arraybackslash}p{1.25cm}>{\raggedright\arraybackslash}p{1.25cm}>{\raggedright\arraybackslash}p{1.25cm}>{\raggedright\arraybackslash}p{1.25cm}>{\raggedright\arraybackslash}p{1.25cm}>{\raggedright\arraybackslash}p{1.25cm}>{\raggedright\arraybackslash}p{1.25cm}>{\raggedright\arraybackslash}p{1.25cm}>{\raggedright\arraybackslash}p{1.25cm}>{\raggedright\arraybackslash}p{1.25cm}>{\raggedright\arraybackslash}p{1.25cm}>{\raggedright\arraybackslash}p{1.25cm}>{\raggedright\arraybackslash}p{1.25cm}}
\toprule
  & 0 & 1 & 2 & 3 & 4 & 5 & 6 & 7 & 8 & 9 & 10 & 11 & 12 & 13 & 14 & 15 & 16 & 17 & 18 & 19 & 20 & 21 & 22 & 23\\
\bottomrule
\end{tabular}
}
\end{center}
\end{table}

\begin{table}[H]
\begin{center}
\caption{Estimated \review{scaled} coefficients for $\tau$ (kurtosis) on the first day of the test set.}\label{tab:coefficients_tau}
\resizebox{\textwidth}{!}{
\begin{tabular}{l>{\raggedright\arraybackslash}p{1.25cm}>{\raggedright\arraybackslash}p{1.25cm}>{\raggedright\arraybackslash}p{1.25cm}>{\raggedright\arraybackslash}p{1.25cm}>{\raggedright\arraybackslash}p{1.25cm}>{\raggedright\arraybackslash}p{1.25cm}>{\raggedright\arraybackslash}p{1.25cm}>{\raggedright\arraybackslash}p{1.25cm}>{\raggedright\arraybackslash}p{1.25cm}>{\raggedright\arraybackslash}p{1.25cm}>{\raggedright\arraybackslash}p{1.25cm}>{\raggedright\arraybackslash}p{1.25cm}>{\raggedright\arraybackslash}p{1.25cm}>{\raggedright\arraybackslash}p{1.25cm}>{\raggedright\arraybackslash}p{1.25cm}>{\raggedright\arraybackslash}p{1.25cm}>{\raggedright\arraybackslash}p{1.25cm}>{\raggedright\arraybackslash}p{1.25cm}>{\raggedright\arraybackslash}p{1.25cm}>{\raggedright\arraybackslash}p{1.25cm}>{\raggedright\arraybackslash}p{1.25cm}>{\raggedright\arraybackslash}p{1.25cm}>{\raggedright\arraybackslash}p{1.25cm}>{\raggedright\arraybackslash}p{1.25cm}}
\toprule
  & 0 & 1 & 2 & 3 & 4 & 5 & 6 & 7 & 8 & 9 & 10 & 11 & 12 & 13 & 14 & 15 & 16 & 17 & 18 & 19 & 20 & 21 & 22 & 23\\
\bottomrule
\end{tabular}
}
\end{center}
\end{table}

For the skewness parameter $\nu$ none of the variables apart from the intercept yield non-zero coefficient estimates. The intercept is slightly negative in the night hours and positive in the morning and afternoon hours. We conclude that the intraday price returns do not exhibit any strong skewness within the individual trading sessions.

Lastly, we turn to Table \ref{tab:coefficients_tau} giving the estimated coefficients for the kurtosis parameter $\tau$. We find a negative impact of lagged $\ALPHA{t}$ and a positive impact of lagged $\deltaPID{t}$. Thus, we expect the distribution of $\deltaPID{t}$ to be lighter-tailed if there has been no trade in the preceding 15 minutes of trading. On the other hand, large absolute price changes in the previous 15 minutes of trading increase $\tau$ and thus the heaviness of the tails. The impact of lagged $\ALPHA{t}$ and $\deltaPID{t}$ is more pronounced during the night hours. During the day hours, there are some none-zero estimates for wind and solar forecasts. This is consistent with \cite{kiesel2020b}'s finding that the behaviour of night contracts is more driven by trading-related variables than fundamentals. For $\tau$, we find a negative impact of the merit-order slope parameter. This implies that a steep-merit leads to heavier tails for the distribution of $\deltaPID{t}$. Thus, if the merit-order is steep, not only the volatility level is elevated, but also the likelihood of spikes is higher. Lastly, we find that with decreasing time to delivery the heaviness of the distribution's tails decreases.  

\FloatBarrier
\section{Discussion and Conclusion} \label{ch:discussion_conclusion}

This paper develops a simulation-based forecasting model for the intraday price process in the last three hours of each product's trading window. We expand the key work of \cite{ziel2020a} in four dimensions by (\textit{i}) investigating distributions with potential skewness and modelling all moments explicitly, (\textit{ii}) adding intra-daily forecast updates and (\textit{iii}) a novel measure for the merit-order slope, derived from day-ahead auction curves, and (\textit{iv}) employing a regularized estimation using the \gls{GAMLSS}-\gls{LASSO} for all distribution moments.  

Our results are two-fold: \review{First, we show that the proposed method is able to generate high quality ensembles for the intraday markets, whose predictive performance is significantly better than benchmark models such as random walk or \gls{ARIMA}-type processes on a wide range of probabilistic scoring rules. The improvement in accuracy is especially distinct in the tails of the predictive distribution. Thus, our results can be applied directly to trading problems as proposed by \cite{serafin2022} or plugged into any optimization method relying on accurate sampling methods.} Second, the GAMLSS framework's explicit traceability and the regularized estimation allows to draw conclusions on the impact of explanatory variables. Qualitatively, our results for the expected value of the intraday return distribution imply weak-form efficient markets, as the inclusion of additional variables does not improve the prediction of the expected value significantly. Additionally, we find evidence for a merit-order effect in the volatility and kurtosis of the return distribution. A steep merit-order regime leads to higher volatility and heavier tails. What is more, we find that the volatility rises with decreasing time to delivery and rises with the closure of the pan-European order book sharing (SIDC). On the other hand, the kurtosis is driven by trading-related variables such as trade events and lagged prices. We find however, that the skewness is close to zero for all hours, and that none of the analysed variables show predictive power.  

This paper's result opens several new research strings: the models used can be improved by the inclusion of cross-product effects and neighbouring products as additional input variables. However, due to the structure of intraday markets with parallel and overlapping trading sessions, this task is non-trivial. A second interesting research avenue is the relationship between trading volume, liquidity and  volatility in intraday markets. Further research is also needed to better understand the impact of fundamental variables for modelling the volatility, kurtosis and skewness of the distribution of intraday price returns. The influence of the merit-order shape as explanatory variable for the volatility warrants further research into its modelling for short-term markets. 

\FloatBarrier
\section*{Acknowledgements}

This paper is based on research conducted during a joint project of Simon Hirsch and Statkraft Trading GmbH. Simon Hirsch is grateful to Statkraft, especially Patrick Otto, Dr.\ Konstantin Wiegandt and Dr.\ Daniel Gruhlke for the support received while writing his thesis. The authors are grateful to energy \& meteo systems GmbH for providing the forecasts used in the paper. The views and opinions expressed in this paper are the author's own and do not reflect the views of Statkraft Trading GmbH or energy \& meteo systems GmbH. The authors are grateful to helpful discussions at the 30. GEE Doctoral Workshop, Essen, 2022. 

\section*{Data Statement}

Due to the commercial nature of production forecasts the dataset remains confidential and cannot be shared.

\section*{Declaration of Interest}

Simon Hirsch is employed by Statkraft Trading GmbH. The authors declare no conflict of interest. 

\bibliographystyle{plainnat}
\bibliography{references}

\appendix

\review{\section{Aggregation of Intraday Trades}\label{app:trade_aggregation}}

\review{The following section gives a detailed definition of the volume-weighted prices $\PID{t}$ and the price changes $\deltaPID{t}$. We start with the definition of a single trade and describe how we aggregate trades.}

\review{A trade on the intraday market for the delivery period $d,s$ is identified by the unique time stamp $i$. For each trade, we have the transacted price $P_{\text{trade}, i}^{d,s}$ in EUR/MWh and the volume $V_{\text{trade}, i}^{d,s}$ in MW. We now define two sets to aggregate the trades:
\begin{enumerate}
	\item Let $\mathbb{I}^{d,s}$ denote the set of all trade timestamps for the delivery period $d,s$.
	\item Let $b(d,s)$ denote the start of the delivery period $d, s$ and let $$\mathbb{T}^{d,s}_{x,y} =[b(d,s) - x - y, \; b(d,s) - x), \; \text{$x \geq 0$ and $y > 0$}$$ denote the left closed time interval between $x + y$ and $x$ minutes before the delivery of product $d,s$.
\end{enumerate}
We can then define the the 5-minute volume-weighted average price of the interval $x,y$ for the delivery period $d,s$ as:
\begin{align}
_{x}\text{ID}_y^{d,s} = & \frac{1}{\sum{}_{i^{d,s} \in \mathbb{I}^{d,s} \cap \mathbb{T}^{d,s}_{x,y}} V_{\text{trade}, i}^{d,s}} \cdot \sum{}_{i^{d,s} \in \mathbb{I}^{d,s} \cap \mathbb{T}^{d,s}_{x,y}} V_{\text{trade}, i}^{d,s} P_{\text{trade}, i}^{d,s}  \label{eqn:idxy} \\
\PID{t} = & _{(T-t)\cdot{}y + 30}\text{ID}_y^{d,s} \label{eqn:delta_p} \\
\deltaPID{t} = & \PID{t} - \PID{t-1}
\end{align}
with $y=5$ minutes. The shift of 30 minutes between Equations \ref{eqn:delta_p} and \ref{eqn:idxy} is due to the fact that we are not taking the control zone trading between between 30 and 5 minutes before start of physical delivery in the German market into account. If there are no trades observed in the interval of length $y$ ending $x$ minutes before delivery, the value is set to the previous value. If no trade happened since the start of the trading session, the corresponding day-ahead spot price $\PDA$ is used. The boolean variable $\alpha_t^{d,s}$ denotes no-trade periods and is set to 1 if there is at least one trade within the 5-minute interval, and set to 0 if there is none.}

\section{Distributions} \label{app:distributions}

\FloatBarrier
\paragraph{Johnson's $S_U$ Distribution:} Proposed by \cite{johnson1949}, the $S_U$ family of distributions is a transformation of the normal distribution. It is a four parameter distribution with the parameter vector $\Theta^{S_U} = (\mu, \sigma, \nu, \tau)$ for the location and scale of the distribution and two shape parameters $\nu$ and $\tau$ for the kurtosis and skewness of the distribution.

\begin{figure}
\begin{subfigure}[c]{0.5\textwidth}
\includegraphics[width=\textwidth]{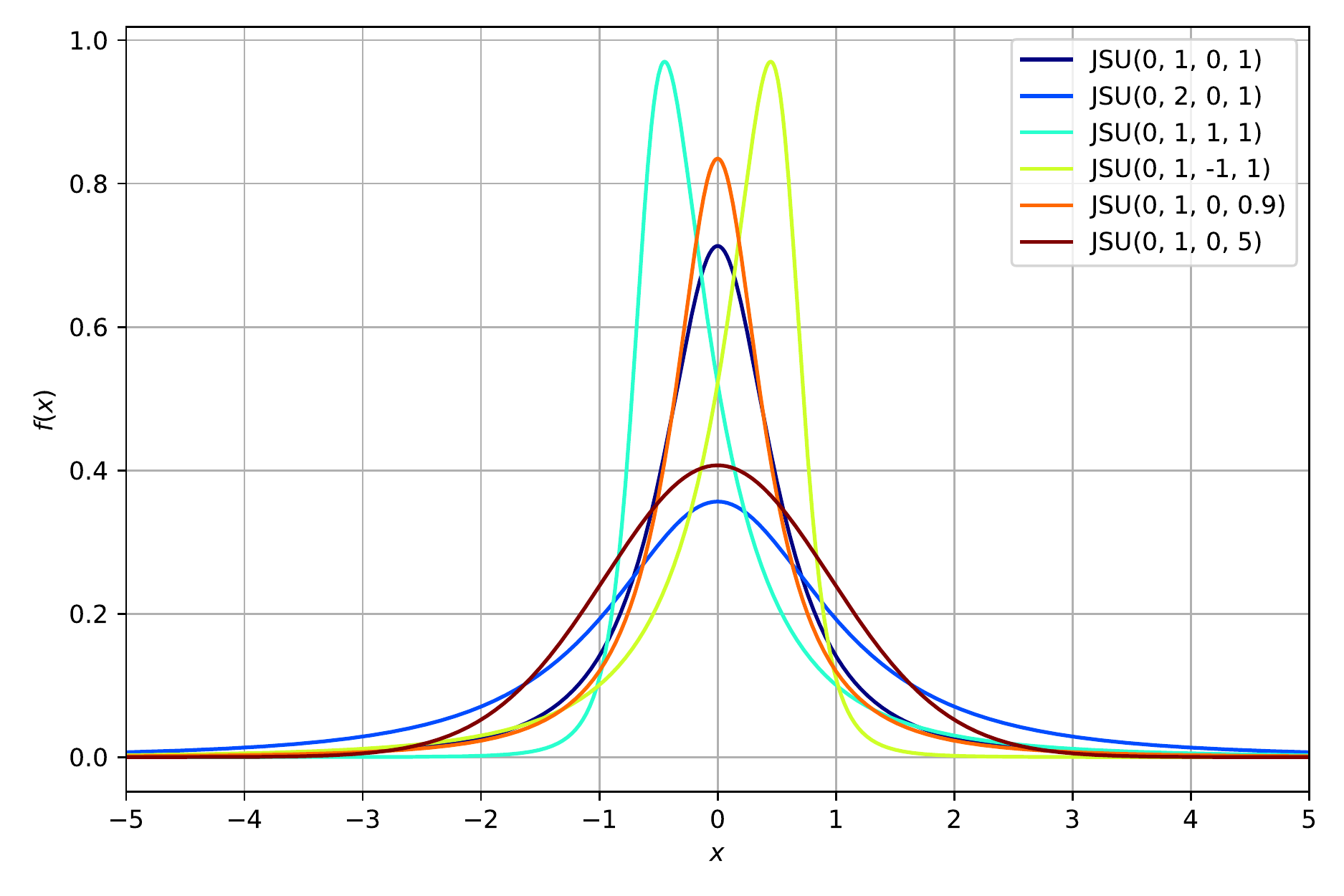}
\subcaption{Untransformed $y$-axis.}
\end{subfigure} %
\begin{subfigure}[c]{0.5\textwidth}
\includegraphics[width=\textwidth]{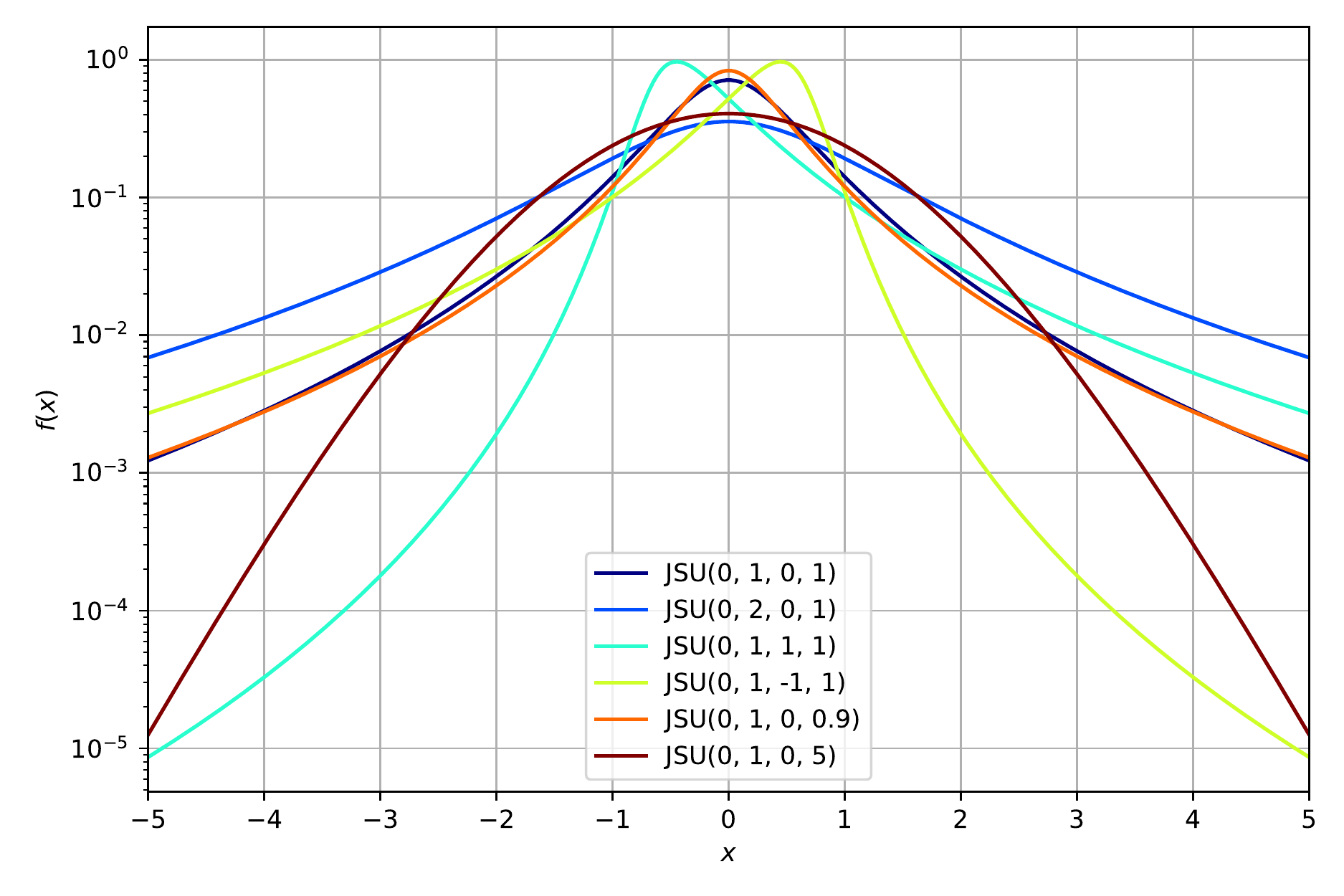}
\subcaption{Log-scale $y$-axis.}
\end{subfigure}
\caption{The PDF for Johnson's $S_U$ for different parametrisations.} \label{fig:JSU_paramterisation}
\end{figure}

The \gls{PDF} of the original $S_U$ distribution can be written as follows:
\begin{equation}
f(y \mid \mu, \sigma, \nu, \tau) = \frac{\tau}{\sigma} \frac{1}{(z^2 +1)^\frac{1}{2}} \frac{1}{\sqrt{2 \pi}} \text{exp}\left(- \frac{1}{2} r^2\right),
\end{equation}
where $z = \frac{y - \mu}{\sigma}$ and $r = \nu + \tau \text{sinh}^{-1}(z)$ for $-\infty < y < \infty$, $\mu = (-\infty, \infty)$, $\sigma > 0$, $\nu = (-\infty, \infty)$ and $\tau > 0$. The implementation in the \texttt{\gls{GAMLSS}}-package is parametrized such that $\mu$ is the mean and $\sigma$ is the standard deviation of the distribution \citep{rigby2017, rGamlssDist}. If $\tau$ approaches $\infty$ and $\nu=0$ the distribution equals the normal distribution. $\nu > 0$ indicates a positive or right-sided skewness, $\nu < 0$ indicates negative of left-sided skewness. The $S_U$-distribution has already been successfully used to model day-ahead prices by \cite{serinaldi2011} and \cite{bunn2018}. Figure \ref{fig:JSU_paramterisation} gives an intuition of the different parameters. The skewness of the $S_U$ for different values of $\tau$ is clearly visible.

\begin{figure}
\begin{subfigure}[c]{0.5\textwidth}
\includegraphics[width=\textwidth]{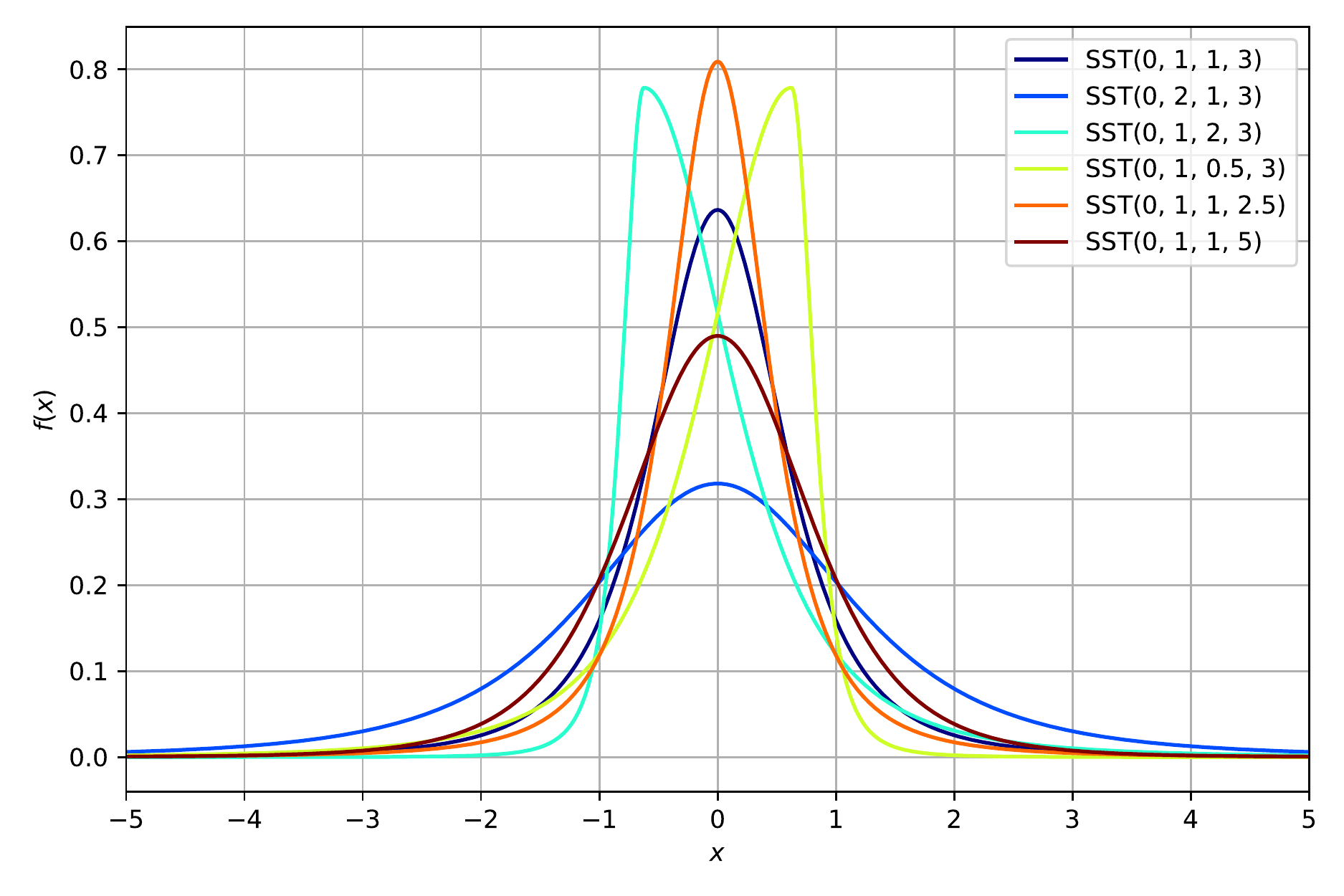}
\subcaption{Untransformed $y$-axis.}
\end{subfigure} %
\begin{subfigure}[c]{0.5\textwidth}
\includegraphics[width=\textwidth]{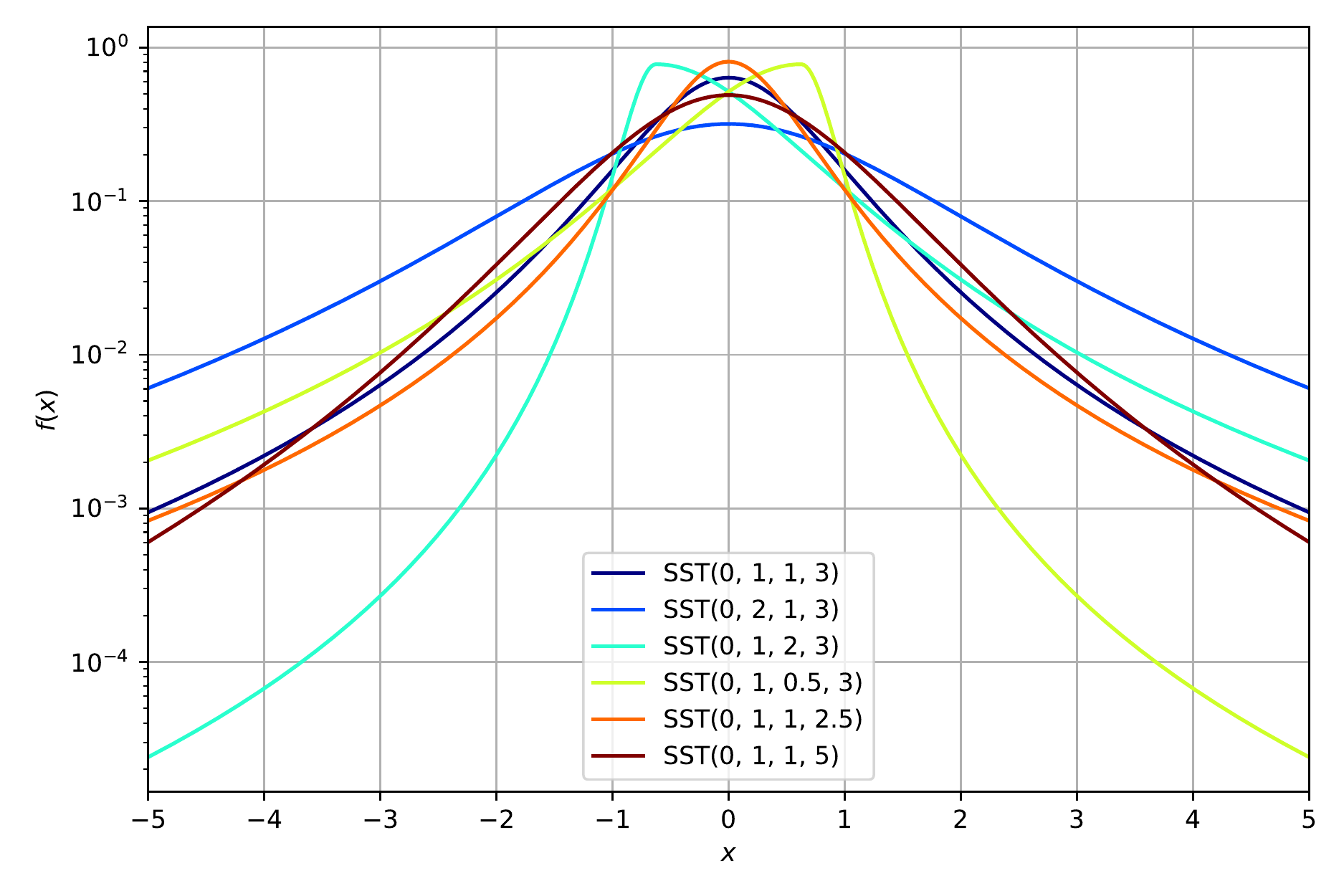}
\subcaption{Log-scale $y$-axis.}
\end{subfigure}
\caption{The PDF for skew-$t$ distribution for different parametrisations. Note that while for Johnson's $S_U$, the skewness is symmetric around 0, for the skew-$t$ the skew is symmetric around 1 for $1/\nu$.} \label{fig:SST_paramterisation}
\end{figure}

\paragraph{Skew-$t$ Distribution:}  The PDF for the 4-parameter skew-$t$ distribution follows the form of \cite{wurtz2006} and \cite{fernandez1998}, so that $\mu$ is the mean and $\sigma$ is the standard deviation. The following definition follows \cite{rigby2017} and is consistent to the implementation in GAMLSS.

\begin{equation}
f(y \mid \mu, \sigma, \nu, \tau) = 
	\begin{cases}
	\frac{c}{\sigma_0}\left( 1+ \frac{v^2 z^2}{\tau} \right)^{-(\tau+1)/2} & \text{if} \; y \leq \mu_0  \\
	\frac{c}{\sigma_0}\left( 1+ \frac{z^2}{v^2 \tau} \right)^{-(\tau+1)/2} & \text{if} \; y \geq \mu_0  \\
	\end{cases}
\end{equation} 
for $-\infty < y < \infty$, where $-\infty < \mu < \infty$, $\sigma > 0$, $\nu > 0$ and $\tau > 2$ and where $\mu_0 = \mu - \sigma m / s$ and $\sigma_0 = \sigma / s$ and $z = (y-\mu_0) / \sigma_0$, $c = 2\nu [(1+\nu^2)B(1/2, \tau/2)\tau^{1/2}]^{-1}$. $m$ and $s$ are obtained from equations (14.32) and (14.33) on pp. 261f in \cite{rigby2017}. Figure \ref{fig:SST_paramterisation} gives an intuition of the different parameters. 

\section{Supplementary Figures}\label{app:plots}

\begin{figure}
\begin{subfigure}[c]{\textwidth}
\includegraphics[width=\textwidth]{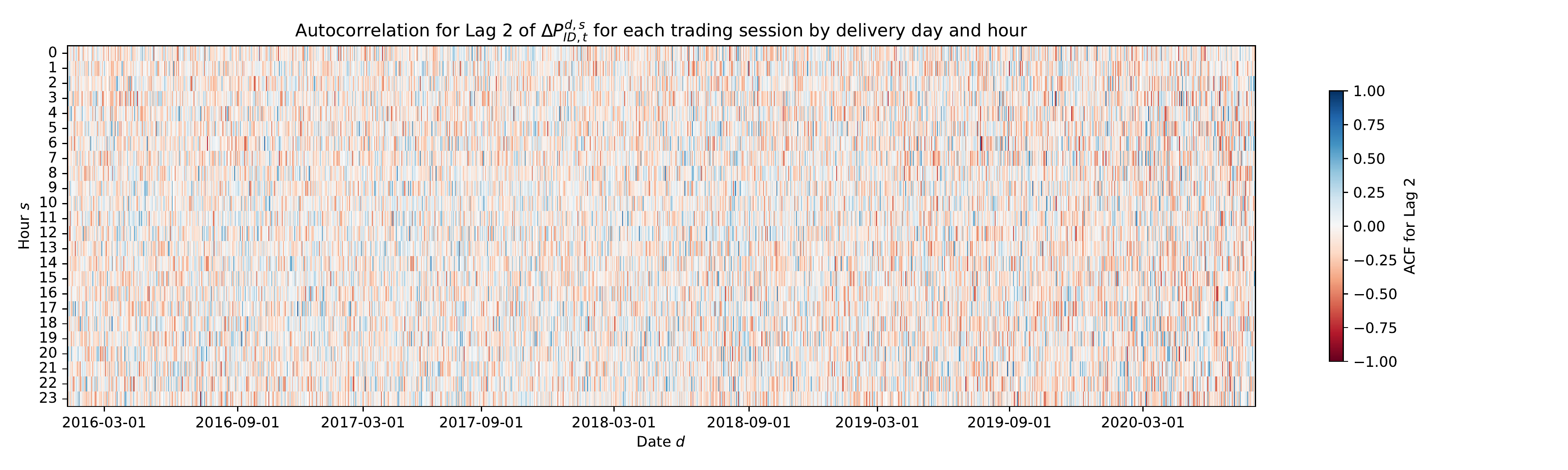}
\subcaption{Lag 2 autocorrelation of $\deltaPID{t}$.} 
\end{subfigure}
\begin{subfigure}[c]{\textwidth}
\includegraphics[width=\textwidth]{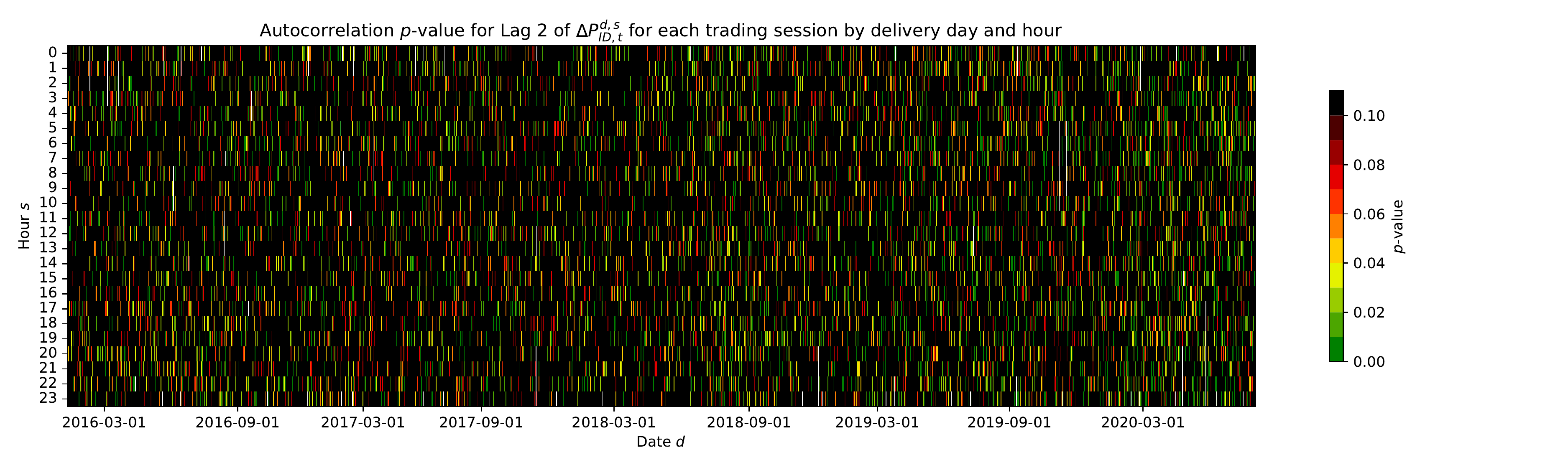}
\subcaption{$p$-values for lag 2 autocorrelation of $\deltaPID{t}$.} 
\end{subfigure}
\caption{Autocorrelation of $\deltaPID{t}$ per trading window for lag 2 and according $p$-values. The first heat maps show the size of the correlation coefficient by delivery day $d$ and hour $s$, second shows according $p$-values.} \label{fig:autocorrelation_lag2}
\end{figure}

\begin{figure}
\begin{subfigure}[c]{\textwidth}
\includegraphics[width=\textwidth]{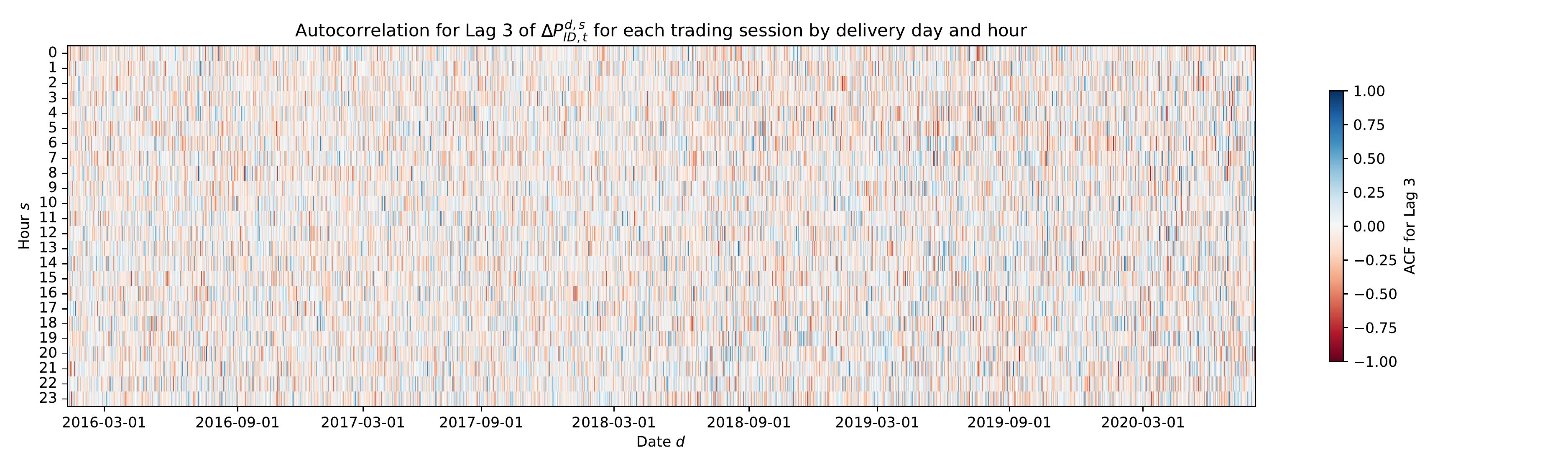}
\subcaption{Lag 3 autocorrelation of $\deltaPID{t}$.} 
\end{subfigure}
\begin{subfigure}[c]{\textwidth}
\includegraphics[width=\textwidth]{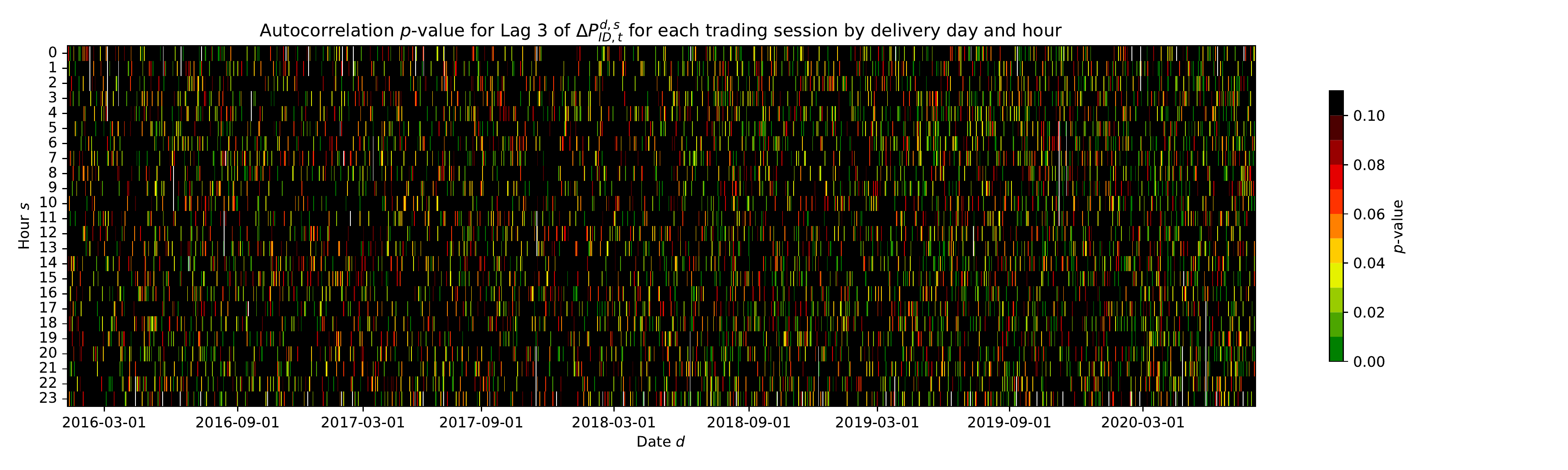}
\subcaption{$p$-values for lag 3 autocorrelation of $\deltaPID{t}$.} 
\end{subfigure}
\caption{Autocorrelation of $\deltaPID{t}$ per trading window for lag 3 and according $p$-values. The first heat maps show the size of the correlation coefficient by delivery day $d$ and hour $s$, second shows according $p$-values.} \label{fig:autocorrelation_lag3}
\end{figure}

\begin{figure}
\begin{subfigure}[c]{\textwidth}
\includegraphics[width=\textwidth]{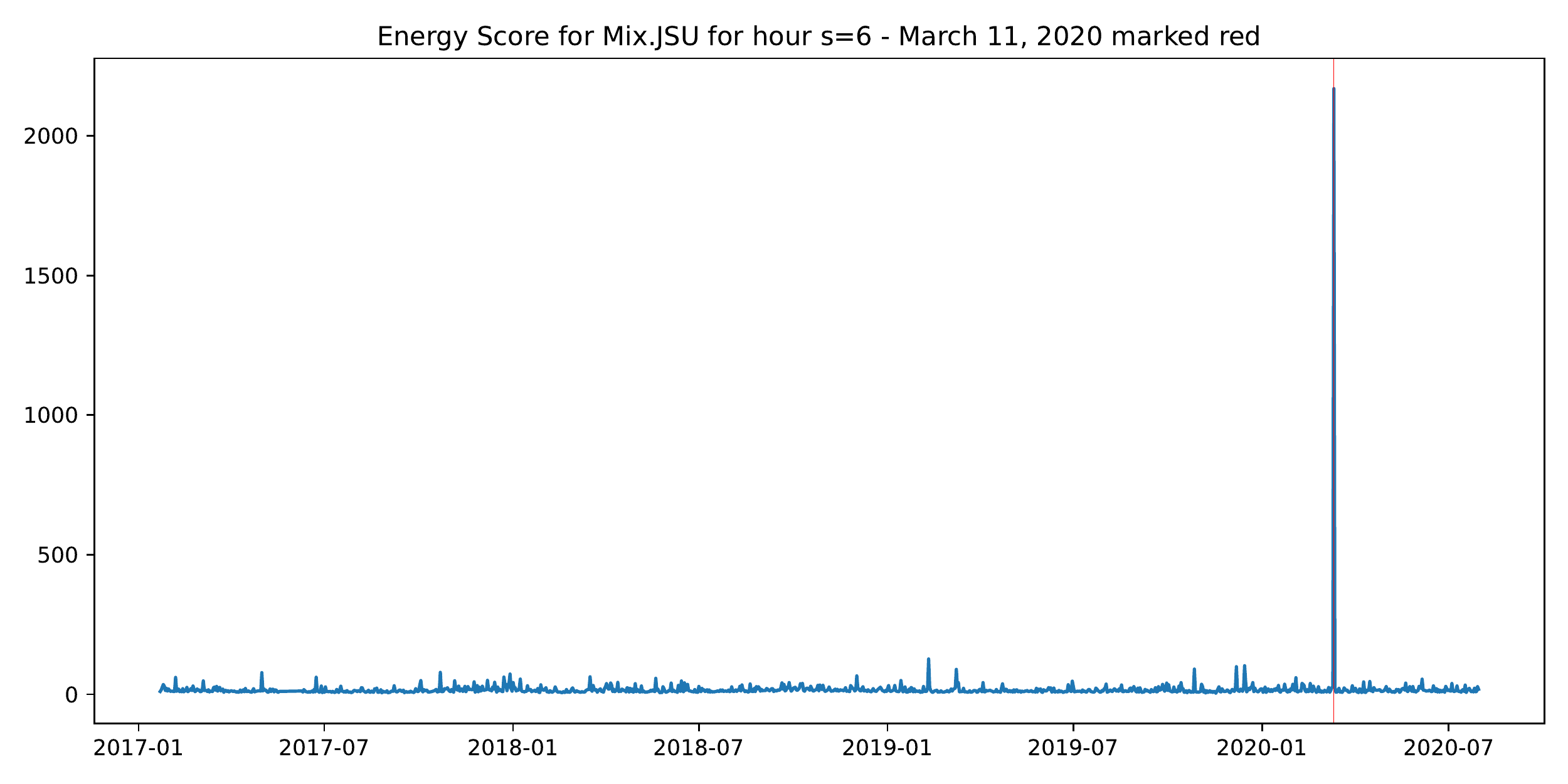}
\subcaption{Energy score for the \textbf{Mix.JSU} for hour $s=6$.} 
\end{subfigure}
\begin{subfigure}[c]{\textwidth}
\includegraphics[width=\textwidth]{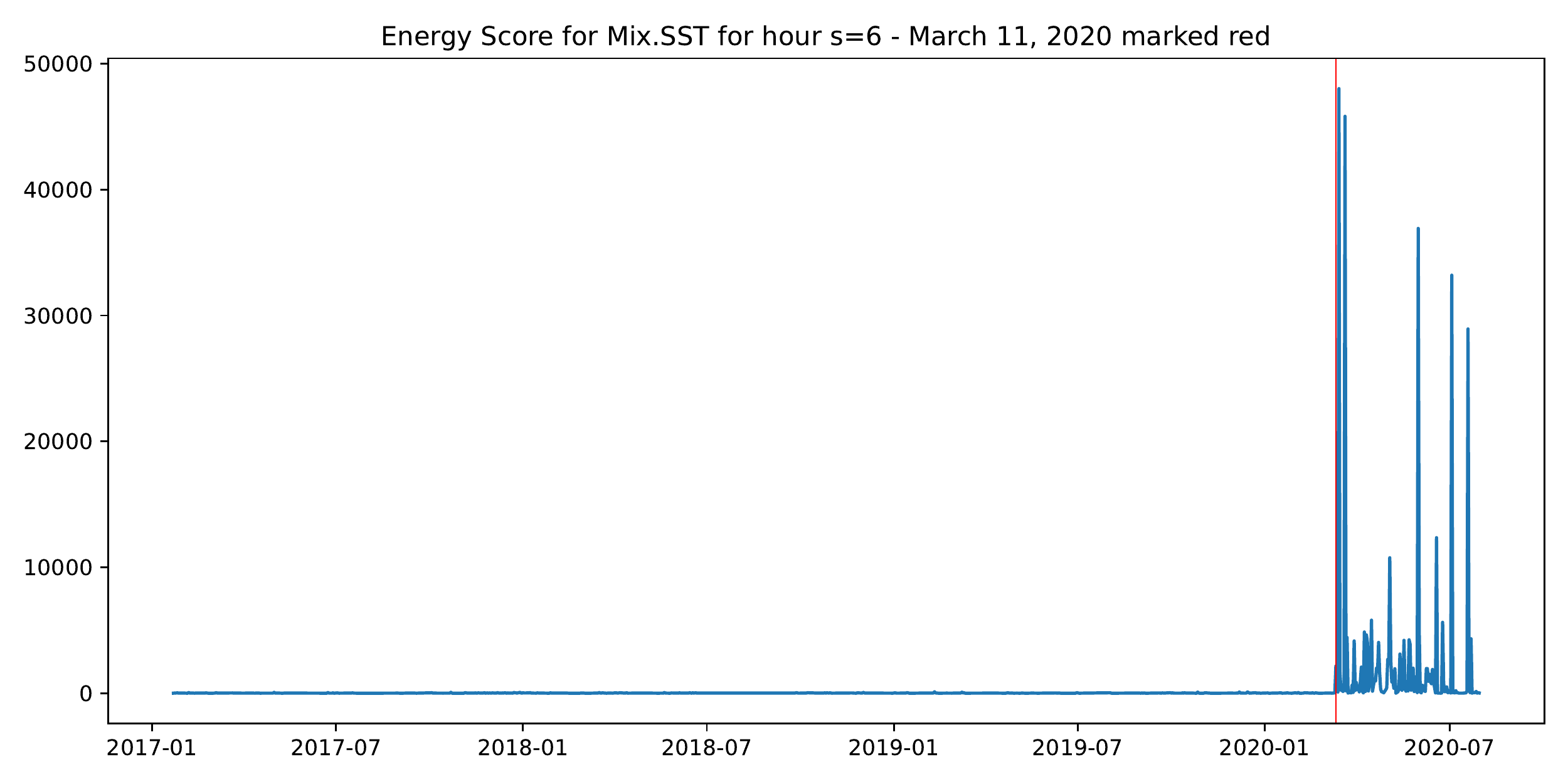}
\subcaption{Energy score for the \textbf{Mix.SST} for hour $s=6$.}  
\end{subfigure}
\caption{Energy scores for \textbf{Mix.JSU} and \textbf{Mix.SST} for hour $s=6$. March 11th, 2020, where price changes $\deltaPID{t}$ where larger than 2000 EUR/MWh, is marked in red. Note that the $y$-axis are different by a factor of more than 20.} \label{fig:error_outliers}
\end{figure}

\end{document}